\documentclass[prd,showpacs,preprintnumbers,amsmath,amssymb]{revtex4}
\usepackage{graphics,color,subfigure}
\usepackage{epsfig}
\usepackage{dcolumn}
\usepackage{bm}

\begin{document}
\title{$\Lambda_c N$ bound states revisited}
\author{Yan-Rui Liu}\email{yrliu@th.phys.titech.ac.jp}
\affiliation{Department of Physics, H-27, Tokyo Institute of Technology, Meguro,
Tokyo 152-8551, Japan}
\author{Makoto Oka}\email{oka@th.phys.titech.ac.jp}
\affiliation{Department of Physics, H-27, Tokyo Institute of Technology, Meguro,
Tokyo 152-8551, Japan}

\date{\today}

\begin{abstract}

The one-boson exchange potential model for $\Lambda_cN$ is constructed and the possibility of $\Lambda_cN$ bound states is examined. We consider an effective Lagrangian for the charmed baryons, $\Lambda_c$, $\Sigma_c$, and $\Sigma_c^*$, reflecting the heavy quark symmetry, chiral symmetry, and hidden local symmetry. We determine the coupling constants using various methods. With the derived nonrelativistic potentials, we study the bound state problem of the $\Lambda_cN$ system and relevant coupled channel effects. It is found that molecular bound states of $\Lambda_cN$ are plausible, for which the channel couplings of $\Sigma_cN$ and $\Sigma_c^*N$ are essential.
\end{abstract}


\maketitle

\section{Introduction}\label{sec1}

Heavy quark hadrons attract great interests since the observation of the exotic $D_{sJ}(2317)$ and $X(3872)$. One of their prominent features is the proximity to one or several thresholds. These mesons, followed by the observation of other exotica \cite{Hmesons,charmonium,XYZ,newHadron,rev-new}, triggered lots of discussions about the bound state problem of heavy quark hadrons. The case of hidden charm mesons is a major concern because they are relatively easy to produce in the experimental facilities.

There are several types of two-body hidden charm systems: $(c\bar{q})(\bar{c}q)$, $(\bar{c}\bar{q}\bar{q})(cqq)$, $(\bar{c}q)(cqq)$, $(c\bar{c})(q\bar{q})$, and $(c\bar{c})(qqq)$. For the first three types, the components are heavy and may be treated nonrelativistically. Since the interactions may be mediated by light meson exchanges, each type had been proposed to have bound states \cite{Tornqvist1994,Wong2004,Zhang2006,Chiu2007,Gamermann2007,Liu2008,Ding2009,Lee2009,Liu2010,Qiao2006,Wu2010,Wang2010}. For the last two types, the two-body attraction is not so strong and the formation of a molecule may be difficult. However, it is possible that the in-medium effects may drive the formation of bound states because of the van der Waals type force. Such quarkonia are called nuclear-bound charmonium \cite{Brodsky1990} or hadro-charmonium \cite{Dubynskiy2008}.

More exotic bound states are possible in the heavy quark case due to the relatively small kinetic term. The charmed meson-nucleus systems are interesting in that they are easy to tag. One may find works about $DN$ (or $\bar{D}N$) interaction and the charmed mesic nuclei in various methods in Refs. \cite{Tsushima1999,Mishra2004,Lutz2004,Mizutani2006,Haidenbauer2007,Yasui2009,Tejero2009,Gamermann2010,Recio2010,He2010}. The calculations favor the existence of such bound states.

The formation of the $\Lambda_c$ and $\Sigma_c$ hypernuclei was predicted more than 30 years ago \cite{Dover1977}. There were further studies about such hypernuclei \cite{Bhamathi1981,Bando1982,Gibson1983}. The observation of $\Lambda_c$-nuclei had also been claimed \cite{Batusov1981}. Recently, this problem was revisited with the quark-meson coupling model \cite{Tsushima2003} and it was found that the charmed hypernuclei are still probable. However, it is inconclusive whether the two-body bound states in the $\Lambda_cN$ channel exist or not \cite{Dover1977}. In Refs. \cite{Fromel2005,Diaz2005}, various heavy quark deuteronlike bound states, $N\Sigma_c$, $N\Xi_c^\prime$, $N\Xi_{cc}$, $\Xi\Xi_{cc}$, and so on, were discussed by scaling the strengths of the nucleon-nucleon interaction models. But the interactions between $\Lambda_c$ and $N$ could not be approximated by the rescaled $NN$ interaction. As $\Lambda_cN$ is the most fundamental charmed nucleon state, it is worthwhile to give it a serious study with the modern heavy quark effective theory and channel couplings to $\Sigma_cN$ and $\Sigma_c^*N$.

With the forthcoming facilities such as GSI-FAIR and J-PARC \cite{Riedl2007}, it would be possible to draw a conclusion whether the charmed hypernuclei exist or not. To understand further this problem theoretically, we revisit the two-body bound states in the $\Lambda_c N$ channel at the hadron level in this paper. Since the one-pion-exchange interaction is forbidden in the $\Lambda_c N$ channel, we consider scalar and vector meson exchange contributions. We also include the coupled channel effects due to the nearby $\Sigma_cN$ and $\Sigma_c^*N$ thresholds. The coupled channel effects in the heavy quark scenario probably have sizable contributions even though the thresholds are not very close to each other, which may be realized from the discussions about the resonance effects for the $D\bar{D}$ production \cite{Li2010,Zhang2010}. It is interesting to investigate how important  the coupled channel effects are for the formation of the $\Lambda_cN$ bound states.

In the early studies of the $\Lambda_cN$ interaction \cite{Dover1977}, the SU(4) symmetry was used to extend the one-boson-exchange model for $NN$ and $YN$ interactions. At that time, the heavy quark symmetry had not yet been explored. With this symmetry, one does not need to consider the largely broken SU(4) any more. As a result of the heavy quark symmetry, parts of interacting terms are related and the Lagrangian can be written in a compact form. Because of the spontaneous symmetry breaking of QCD, it is natural to consider chiral symmetry for pions. To include the light vector mesons, we turn to the hidden symmetry approach where the vector mesons are treated as gauge fields of a hidden local symmetry. This approach has been successful in explaining the KSRF relation $m_\rho^2=2f_\pi^2g_{\rho\pi\pi}$, the universality of $\rho$ coupling, and $\rho$ dominance of the photon coupling to pions \cite{Bando1985}. Here, when deriving the one-boson-exchange potentials, we use effective Lagrangians satisfying the heavy quark symmetry for heavy baryons, chiral symmetry for pions, and hidden local symmetry for vector mesons. The Lagrangian is analogous to the heavy meson version presented in Ref. \cite{Casalbuoni1992}. As for the unknown coupling constants, we turn to the quark model, chiral multiplets, vector meson dominance, and QCD sum rule.

Extension of the study to the bottom case is also interesting, but we would not explore it here. Our paper is organized as follows. After the introduction, we present the effective Lagrangians in Sec. \ref{sec2}. Then we determine the coupling constants with various methods in Sec. \ref{sec3}. In Sec. \ref{sec4}, we present the derived potentials. The numerical results for the $S$-wave spin-singlet and triplet $\Lambda_cN$ are given in Sec. \ref{sec5} and Sec. \ref{sec6}, respectively. Finally, we present our discussions and conclusions in Sec. \ref{sec7}.

\section{The Lagrangian}\label{sec2}

Here, we consider the $S$-wave $\Lambda_cN$ states and the channels which couple to them, i.e., the  $I=\frac12$ and $J^P=0^+$ or $1^+$ two-baryon states. We consider the contributions from the $\Sigma_cN$ and $\Sigma_c^*N$ channels. It is a 3-channel problem for $J^P=0^+$ and 7-channel problem for $J^P=1^+$. The labels are listed in Table \ref{chn-lab}. Because of the higher mass, we assume that the $\Delta(1232)$ contributions are negligible.

\begin{table}[htb]
\begin{tabular}{cccccccc}\hline
Channels & 1 & 2 & 3 & 4 & 5 & 6 & 7\\\hline
$J^P=0^+$&$\Lambda_cN(^1S_0)$&$\Sigma_cN (^1S_0)$&$\Sigma_c^*N (^5D_0)$\\
$J^P=1^+$&$\Lambda_cN(^3S_1)$&$\Sigma_cN(^3S_1)$&$\Sigma_c^*N(^3S_1)$&$\Lambda_cN(^3D_1)$&$\Sigma_cN(^3D_1)$&$\Sigma_c^*N(^3D_1)$&$\Sigma_c^*N(^5D_1)$ \\\hline
\end{tabular}
\caption{The $S$-wave $\Lambda_cN$ states and the channels which couple to them.}\label{chn-lab}
\end{table}

In the heavy quark limit, the ground state heavy baryons $Qqq$ form $SU(3)$ antitriplet with $J^P=\frac12^+$ and two degenerate sextets with $J^P=(\frac12,\frac32)^+$. As in Ref. \cite{Yan1992}, we use $B_{\bar 3}$, $B_6$, and $B_6^*$ to denote these multiplets. To write down the compact form of the Lagrangians, we use the notation of the superfield $S_\mu$, which is defined by
\begin{eqnarray}\label{superfield}
S_\mu=B_{6\mu}^*+\delta\frac{1}{\sqrt3}(\gamma_\mu+v_\mu)\gamma^5 B_6
\end{eqnarray}
where $v_\mu$ is the 4-velocity of the heavy baryon and $\delta$ is an arbitrary phase factor. One finds that an appropriate choice of $\delta$ is $-1$ (see Appendix \ref{app1}), which is different from the one often used in the literature. Actually, this phase does not affect the final results because it appears only in the transition potentials. In the multichannel case, it is easy to prove that the binding energy does not change with the following replacement for the Hermitian Hamiltonian
\begin{eqnarray}\label{phConvention}
\left(\begin{array}{cccccc}H_{11}&H_{12}&H_{13}&H_{14}&H_{15}&\cdots\\
&H_{22}&H_{23}&H_{24}&H_{25}&\cdots\\
&&H_{33}&H_{34}&H_{35}&\cdots\\
&&&H_{44}&H_{45}&\cdots\\
&&&&H_{55}&\cdots\\&&\cdots\end{array}\right)
\Rightarrow
\left(\begin{array}{cccccc}H_{11}&\delta_2H_{12}&\delta_3H_{13}&\delta_4H_{14}&\delta_5H_{15}&\cdots\\
&H_{22}&(\delta_2^*\delta_3)H_{23}&(\delta_2^*\delta_4)H_{24}&(\delta_2^*\delta_5)H_{25}&\cdots\\
&&H_{33}&(\delta_3^*\delta_4)H_{34}&(\delta_3^*\delta_5)H_{35}&\cdots\\
&&&H_{44}&(\delta_4^*\delta_5)H_{45}&\cdots\\
&&&&H_{55}&\cdots\\&&\cdots\end{array}\right),
\end{eqnarray}
where the subscripts of $H$ denote the channels and $\delta_i$ ($i$=2,3,...) are arbitrary phases. Because of the same reason, the convention of the relative phase between the sextet and the antitriplet will also not matter.

One constructs the effective Lagrangian according to the chiral symmetry, heavy quark symmetry, and hidden local symmetry \cite{Bando1985,Meissner1986,Bando1988,Casalbuoni1992}. The coupling terms are constrained by the couplings at the quark level: $0^+-0^+-\mathfrak{M}$, $1^+-1^+-\mathfrak{M}$, and $1^+-0^+-\mathfrak{M}$. Here $0^+$ and $1^+$ are the spin-parity quantum numbers of the light diquark inside the baryons and $\mathfrak{M}$ represents the pseudoscalar, scalar, or vector meson. The constructed heavy quark baryon Lagrangian is
\begin{eqnarray}\label{totLag}
{\cal L_B}&=&{\cal L}_{B_{\bar{3}}}+{\cal L}_{S}+{\cal L}_{int},\\
{\cal L}_{B_{\bar{3}}}&=&\frac12{\rm tr}[\bar{B}_{\bar{3}}(iv\cdot D)B_{\bar{3}}]+ i\beta_B{\rm tr}[\bar{B}_{\bar{3}}v^\mu(\Gamma_\mu-V_\mu) B_{\bar{3}}]+\ell_B{\rm tr}[\bar{B}_{\bar{3}}{\sigma} B_{\bar{3}}]\\
{\cal L}_{S}&=&-{\rm tr}[\bar{S}^\alpha(i v\cdot D-\Delta_B)S_\alpha]+\frac32 g_1(iv_\kappa)\epsilon^{\mu\nu\lambda\kappa}{\rm tr}[\bar{S}_\mu
A_\nu S_\lambda]\nonumber\\
&&+i\beta_S{\rm tr}[\bar{S}_\mu v_\alpha (\Gamma^\alpha-V^\alpha) S^\mu] + \lambda_S{\rm tr}[\bar{S}_\mu F^{\mu\nu}S_\nu]+\ell_S{\rm tr}[\bar{S}_\mu \sigma S^\mu]\\
{\cal L}_{int}&=&g_4 {\rm tr}[\bar{S}^\mu A_\mu B_{\bar{3}}]+i\lambda_I \epsilon^{\mu\nu\lambda\kappa}v_\mu{\rm tr}[\bar{S}_\nu F_{\lambda\kappa} B_{\bar{3}}]+h.c.,
\end{eqnarray}
where we use the notations, Eqs. (\ref{notation-7})-(\ref{notation-10}), $\Delta_B=M_6-M_{\bar 3}$ is the mass difference between the sextet and the antitriplet, and $\sigma$ is the scalar singlet meson. Note that $B_6$ $(\frac12^+)$ and $B_6^*$ $(\frac32^+)$ are degenerate because of the heavy quark spin symmetry in this effective theory. In the coupled channel calculation of the $\Lambda_cN-\Sigma_cN-\Sigma_c^*N$ system, we will use the empirical values of the masses of $\Lambda_c$, $\Sigma_c$ and $\Sigma_c^*$ and therefore the $\Sigma_c-\Sigma_c^*$ splitting is properly included. The coupling constants $g_1$ and $g_4$ are the same as those in Ref. \cite{Yan1992}. Other definitions are given below.
\begin{eqnarray}
&B_{\bar{3}}=\left(\begin{array}{ccc}
0& \Lambda_c^+ & \Xi_c^{+}\\
-\Lambda_c^+ & 0 & \Xi_c^{0}\\
-\Xi_c^{+}& -\Xi_c^{0} & 0 \end{array}\right), \quad
B_6=\left(\begin{array}{ccc}
\Sigma_c^{++}& \frac{1}{\sqrt2}\Sigma_c^+ & \frac{1}{\sqrt2}\Xi_c^{\prime+}\\
\frac{1}{\sqrt2}\Sigma_c^+ & \Sigma_c^{0} & \frac{1}{\sqrt2}\Xi_c^{\prime0}\\
\frac{1}{\sqrt2}\Xi_c^{\prime+}& \frac{1}{\sqrt2}\Xi_c^{\prime0} & \Omega_c^0 \end{array}\right),\quad
B^*_6=\left(\begin{array}{ccc}
\Sigma_c^{*++}& \frac{1}{\sqrt2}\Sigma_c^{*+} & \frac{1}{\sqrt2}\Xi_c^{*+}\\
\frac{1}{\sqrt2}\Sigma_c^{*+} & \Sigma_c^{*0} & \frac{1}{\sqrt2}\Xi_c^{*0}\\
\frac{1}{\sqrt2}\Xi_c^{*+}& \frac{1}{\sqrt2}\Xi_c^{*0} & \Omega_c^{*0} \end{array}\right),&\label{notation-7}\\
&\Pi={\sqrt2}\left(
\begin{array}{ccc}
\frac{\pi^0}{\sqrt2}+\frac{\eta}{\sqrt6}&\pi^+&K^+\\
\pi^-&-\frac{\pi^0}{\sqrt2}+\frac{\eta}{\sqrt6}&K^0\\
K^-&\bar{K}^0&-\frac{2}{\sqrt6}\eta
\end{array}\right),\quad
V^\mu=i\frac{g_V}{\sqrt2}\left(\begin{array}{ccc}
\frac{\rho^{0}}{\sqrt{2}}+\frac{\omega}{\sqrt{2}}&\rho^{+}&K^{*+}\\
\rho^{-}&-\frac{\rho^{0}}{\sqrt{2}}+\frac{\omega}{\sqrt{2}}&
K^{*0}\\
K^{*-} &\bar{K}^{*0}&\phi
\end{array}\right)^\mu,&\\
&A_\mu=\frac{i}{2}[\xi^\dag(\partial_\mu\xi)+(\partial_\mu\xi)\xi^\dag],\quad \Gamma_\mu=\frac{1}{2}[\xi^\dag(\partial_\mu\xi)-(\partial_\mu\xi)\xi^\dag], \quad  \xi=\exp[\frac{i\Pi}{2f}], \quad F_{\mu\nu}=\partial_\mu V_\nu-\partial_\nu V_\mu+[V_\mu,V_\nu],&\\
&D_\mu B_{\bar 3}=\partial_\mu B_{\bar 3}+\Gamma_\mu B_{\bar 3}+B_{\bar 3}\Gamma_\mu^T,\quad
D_\mu S_\nu=\partial_\mu S_\nu+\Gamma_\mu S_\nu+S_\nu\Gamma_\mu^T.&\label{notation-10}
\end{eqnarray}
We use $f=92.3$ MeV for the pion decay constant. The constant $g_V=m_\rho/(\sqrt2f_\pi)=5.8$ is derived with the vector meson dominance (VMD) \cite{Bando1985,Meissner1986,Casalbuoni1992}.
For the nucleon-nucleon interaction part, we use the following SU(2) Lagrangian,
\begin{eqnarray}
{\cal L}_N&=&-\frac{g_A}{2f}\bar{N}\gamma^\mu\gamma^5 \partial_\mu (\pi^i\tau^i) N-h_\sigma \bar{N}\sigma N-h_V\bar{N}\gamma^\mu(\tau^i\rho^i_\mu+\omega_\mu)N -h_T\bar{N}\sigma^{\mu\nu} \partial_\mu(\tau^i\rho^i_\nu+\omega_\nu)N,
\end{eqnarray}
where $\tau^i$ is the Pauli matrix, representing the isospin.

\section{The potentials}\label{sec3}

\begin{figure}[htb]
\includegraphics[scale=0.7]{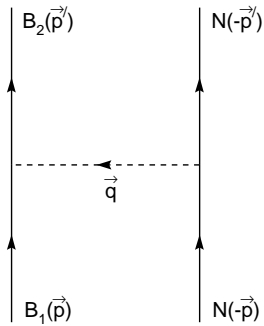}
\caption{The considered interaction in deriving the potentials. $B_1$ ($B_2$) may be $\Lambda_c$, $\Sigma_c$, or $\Sigma_c^*$. The dashed line denotes the exchanged meson $\pi$, $\sigma$, $\rho$, or $\omega$.}\label{int_diag}
\end{figure}

The one-boson-exchange diagram in Fig. \ref{int_diag} is considered, where we use $B_1$ and $B_2$ to denote the charmed baryons. When deriving the potentials, we use the heavy quark limit for the charmed baryons, i.e. ignoring the $1/M_{\bar{3},6}$ corrections but keeping up to $1/M_N^2$ corrections for the nucleon. At each interacting vertex, we introduce a cutoff $\Lambda$ through the monopole type form factor
\begin{eqnarray}
F(q)=\frac{\Lambda^2-m^2}{\Lambda^2-q^2},
\end{eqnarray}
where $m$ is the mass of the exchanged meson and $q$ is its 4-momentum. Because the meson exchange occurs between light quarks, we use the same cutoff for $B_1B_2\pi$ and $NN\pi$ vertices. The cutoffs are taken to be around 1 GeV, while they may be different for the scalar and the vector meson exchanges.

For convenience, we define some functions for the final potentials after the Fourier transformation,
\begin{eqnarray}
\frac{[F(q)]^2}{q^2-m^2}&\to&-\frac{1}{4\pi}mY_1(m,\Lambda,r),\nonumber\\
\frac{[F(q)]^2}{q^2-m^2}\vec{q}^{\,2}&\to&\frac{1}{4\pi}m^3Y_3(m,\Lambda,r),\nonumber\\
\frac{[F(q)]^2}{q^2-m^2}[i\vec{\cal O}_1\cdot(\vec{k}\times\vec{q})]&\to&-\frac{\vec{L}\cdot\vec{\cal O}_1}{4\pi}m^3Z_3(m,\Lambda,r),\nonumber\\
\frac{[F(q)]^2}{q^2-m^2}[(\vec{\cal O}_1\cdot\vec{q})(\vec{\cal O}_2\cdot\vec{q})]&\to&\frac{\vec{\cal O}_1\cdot\vec{\cal O}_2}{12\pi}m^3Y_3(m,\Lambda,r)+\frac{{\cal O}_{ten}}{12\pi}m^3H_3(m,\Lambda,r),
\end{eqnarray}
where $\vec{k}=\vec{p}+\vec{p}^{\,\prime}$ with the notation in Fig. \ref{int_diag}, $\vec{\cal O}_1$ is the (transition) spin operator of the charmed baryon, $\vec{\cal O}_2$ is the spin operator of the nucleon, and ${\cal O}_{ten}$ is the tensor operator defined by ${\cal O}_{ten}=\frac{3(\vec{\cal O}_1\cdot\vec{r})(\vec{\cal O}_2\cdot\vec{r})}{r^2}-(\vec{\cal O}_1\cdot\vec{\cal O}_2)$. The other definitions are
\begin{eqnarray}
Y(x)&=&\frac{e^{-x}}{x},\quad Z(x)=(\frac{1}{x}+\frac{1}{x^2})Y(x),\quad H(x)=(1+\frac{3}{x}+\frac{3}{x^2})Y(x),\nonumber\\
Y_1(m,\Lambda,r)&=&Y(mr)-\left(\frac{\Lambda}{m}\right)Y(\Lambda r)-\frac{\Lambda^2-m^2}{2m\Lambda}e^{-\Lambda r},\nonumber\\
Y_3(m,\Lambda,r)&=&Y(mr)-\left(\frac{\Lambda}{m}\right)Y(\Lambda r)-\frac{(\Lambda^2-m^2)\Lambda}{2m^3}e^{-\Lambda r},\nonumber\\
Z_3(m,\Lambda,r)&=&Z(mr)-\left(\frac{\Lambda}{m}\right)^3Z(\Lambda r)-\frac{(\Lambda^2-m^2)\Lambda}{2m^3}Y(\Lambda r),\nonumber\\
H_3(m,\Lambda,r)&=&H(mr)-\left(\frac{\Lambda}{m}\right)^3H(\Lambda r)-\frac{(\Lambda^2-m^2)\Lambda}{2m^3}Y(\Lambda r)-\frac{(\Lambda^2-m^2)\Lambda}{2m^3}e^{-\Lambda r}.
\end{eqnarray}
When getting these formulas, we have adopted the approximation $q^0=0$. The lowest excited state of  $\Lambda_cN$ is the three-body $\Lambda_cN\pi$ state. But in this study we do not consider the coupling of this channel because we need to excite the system to $L=1$ in order to get the correct parity. Such excitation will hinder the channel coupling.

In the multichannel case, we have transition potentials, e.g. $V(\Lambda_cN\to\Sigma_c^*N)$, and we need to define the transition spin. Here we define $u^\mu=S_t^\mu\Phi$, where $u^\mu$ is the Rarita-Schwinger field, $S_t^\mu$ is the transition spin matrix ($2\times4$), and the spin wave functions of $\Sigma_c^*$ are
\begin{eqnarray}
\Phi(3/2)=(1,0,0,0)^T,\quad \Phi(1/2)=(0,1,0,0)^T,\quad \Phi(-1/2)=(0,0,1,0)^T,\quad \Phi(-3/2)=(0,0,0,1)^T.
\end{eqnarray}
We also need the spin operator for $\Sigma_c^*$, which is defined through $\vec{\sigma}_{rs}\equiv-S_{t\mu}^\dag\vec{\sigma}S_t^\mu$, $\vec{S}_{rs}\equiv\frac32\vec{\sigma}_{rs}$. One may check $\vec{S}_{rs}^2=\frac32(\frac32+1)=\frac{15}{4}$.

Now we may write down the general form of the potentials,
\begin{eqnarray}
V_\pi(i,j)&=&C_\pi(i,j)\frac{m_\pi^3}{24\pi f_\pi^2}\Big\{\vec{\cal O}_1\cdot\vec{\cal O}_2Y_1(m_\pi,\Lambda,r)+{\cal O}_{ten}H_3(m_\pi,\Lambda,r)\Big\},\nonumber\\
V_\sigma(i)&=&C_\sigma(i)\frac{m_\sigma}{16\pi}\Big\{4Y_1(m_\sigma,\Lambda,r)+\vec{L}\cdot\vec{\sigma}_2\left(\frac{m_\sigma}{M_N}\right)^2Z_3(m_\sigma,\Lambda,r)\Big\},\nonumber\\
V_\rho(i,j)&=&C_{\rho1}(i,j)\frac{m_\rho h_V}{32\pi}\Big\{8Y_1(m_\rho,\Lambda,r)+(1+\frac{4M_Nh_T}{h_V})\frac{m_\rho^2}{M_N^2}\Big[Y_1(m_\rho,\Lambda,r)-2\vec{L}\cdot\vec{\sigma}_2 Z_3(m_\rho,\Lambda,r)\Big]\Big\}\nonumber\\
&+&C_{\rho2}(i,j)\frac{m_\rho^3h_V}{36\pi M_N}\Big\{(1+\frac{2M_Nh_T}{h_V})\Big[2\vec{\cal O}_1\cdot\vec{\cal O}_2Y_1(m_\rho,\Lambda,r)-{\cal O}_{ten}H_3(m_\rho,\Lambda,r)\Big]-6\vec{L}\cdot\vec{\cal O}_1 Z_3(m_\rho,\Lambda,r)\Big\},
\end{eqnarray}
where $i$ and $j$ are labels of the channels, $C(i,j)$ and $C(i)$ are channel-dependent coefficients, the operator $\sigma_2$ or ${\cal O}_2$ is the Pauli matrix of the nucleon spin, and ${\cal O}_1$ is the corresponding spin matrix for the charmed baryon $\vec{\sigma}_1$ ($\vec{\sigma}_{rs}$) or the transition spin $\vec{S}_t$ or $\vec{S}_t^\dag$. The $\omega$ exchange potential is similarly defined as the $\rho$ potential.

It should be noted that we omit the contact ($\delta$-functional) part of the potential, which appears in $Y_3$ in our model. The main reason is that the interacting two baryons should get small contributions from such terms when they form a molecule-type bound state, in which two baryons are well separated. The difference between the defined functions $Y_3$ and $Y_1$ is the contact part. So the $Y_3$ does not appear in our model potentials.

For the calculation with the potentials, one may use any consistent phase convention of the coefficients $C(i,j)$. The correct result is ensured by Eq. (\ref{phConvention}). One should also note the possible convention problem when calculating the matrix elements $\langle\vec{\cal O}_1\cdot\vec{\cal O}_2\rangle$, $\langle{\cal O}_{ten}\rangle$, and $\langle{\vec L}\cdot\vec{\cal O}\rangle$ with the Wigner-Eckart theorem. The conventions must also be consistent.

\section{The coupling constants}\label{sec4}

The next task is to determine the coupling constants in the Lagrangians. We use several methods to constrain the values: (1) the strong decay of the baryons, (2) quark model estimation, (3) vector meson dominance assumption, (4) chiral multiplet assumption, and (5) QCD sum rule calculations. In Ref. \cite{Yan1992}, the coefficients of $B_6B_6\pi$, $B_6B_{\bar 3}\pi$, $B_6^*B_6\pi$, $B_6^*B_{\bar 3}\pi$, and $B_6^*B_6^*\pi$ coupling terms are called $g_1$, $g_2$, $g_3$, $g_4$, and $g_5$, respectively. In the following parts, we also use these notations.

The available strong decays are only $\Sigma_c\to\Lambda_c\pi$ and $\Sigma_c^*\to\Lambda_c\pi$ \cite{PDG2010}. The coupling constants $g_2$ and $g_4$ may be derived from the decay widths
\begin{eqnarray}
\Gamma(\Sigma_c\to\Lambda\pi)=\frac{g_2^2}{4\pi f_\pi^2}\frac{M_{\Lambda_c}}{M_{\Sigma_c}}|\vec{p}_\pi^{\,3}|,\qquad \Gamma(\Sigma_c^*\to\Lambda\pi)=\frac{g_4^2}{12\pi f_\pi^2}\frac{M_{\Lambda_c}}{M_{\Sigma_c^*}}|\vec{p}_\pi^{\,3}|,
\end{eqnarray}
where $|\vec{p}_\pi|=94$ (180) MeV for $\Sigma_c\to\Lambda_c\pi$ ($\Sigma_c^*\to\Lambda_c\pi$) is the c.m. momentum of the pion. In the numerical evaluation, we will use the averaged values derived from different decay modes: $|g_2|=0.598$ and $|g_4|=0.999$.

All the coupling constants may be estimated with the quark model. The basic procedure is to calculate the average of the coupling vertex twice, at hadron level and at quark level, and to equate them. The effective Lagrangian for the light constitute $u$, $d$ quark fields, $\psi$ is
\begin{eqnarray}
{\cal L}_q=-\frac{g_A^q}{2f}\bar{\psi}\gamma^\mu\gamma^5\partial_\mu(\pi^i\tau^i)\psi -g_\sigma^q \bar{\psi}\sigma\psi -g_\rho^q \bar{\psi}\gamma^\mu(\rho_\mu^i\tau^i+\omega_\mu) \psi-f_\rho^q\bar{\psi}\sigma^{\mu\nu}\partial_\mu(\rho_\nu^i\tau^i+\omega_\nu)\psi.
\end{eqnarray}
We assume that the heavy quark does not couple to the light mesons. For the pion coupling with heavy baryons, we just need to estimate two independent coupling constants, e.g. $g_1$ and $g_2$, and relate others to them with the heavy quark symmetry relations $g_3=\frac{\sqrt3}{2}g_1$, $g_4=-\sqrt3g_2$, and $g_5=-\frac32g_1$ \cite{Yan1992}. The obtained values are $g_1=\frac43g_A^q$ and $g_2=-\sqrt{\frac23}g_A^q$. For the scalar meson coupling, we have $\ell_B=-g_\sigma^q$ and $\ell_S=2g_\sigma^q$. For the vector meson coupling, one gets $(\beta_B g_V)=-2g_\rho^q$, $(\beta_Sg_V)=4g_\rho^q$, $(\lambda_Sg_V)=\frac{2g_\rho^q}{m_q}+4f_\rho^q$, and $(\lambda_Ig_V)=-\frac{\sqrt2}{4}\left(\frac{2g_\rho^q}{m_q}+4f_\rho^q\right)$, where $m_q$ is the quark mass. We have neglected the $1/M_{B_{\bar{3},6}}$ terms in deriving the above relations. For the nucleon-nucleon sector, we have $g_A=\frac53g_A^q$, $h_\sigma=3g_\sigma^q$, $h_V=g_\rho^q$, and $h_T=\frac{5g_\rho^q}{6M_N}(\frac{M_N}{m_q}-\frac35)+\frac53f_\rho^q$.

With $g_A^q=0.75$ in the chiral quark model \cite{Manohar1984}, one obtains $g_1=1.00$, $g_2=-0.61$, $g_3=0.87$, $g_4=1.06$, $g_5=-1.50$, and $g_A=1.25$. For the coupling constant $g_\sigma^q$, the value in a $\sigma$ model is 3.65 \cite{Riska1999}. We have $\ell_B=-3.65$, $\ell_S=7.30$, and $h_\sigma=10.95$. If we use the value $g_\sigma^q=2.621$ in a chiral quark model \cite{Zhang1994}, we have $\ell_B=-2.621$, $\ell_S=5.242$, and $h_\sigma=7.863$. For the vector meson-baryon coupling constants, we use the quark mass $m_q=313$ MeV \cite{Zhang1994} for the discussions. In the Nijmegen model D \cite{Nagels1975}, the authors used $h_V=2.1$ and $h_T=9.1$ GeV$^{-1}$, which correspond to $g_\rho^q=2.1$ and $f_\rho^q=2.8$ GeV$^{-1}$ and gives $(\beta_Bg_V)=-4.2$, $(\beta_Sg_V)=8.4$, $(\lambda_Sg_V)=24.6$ GeV$^{-1}$ and $(\lambda_Ig_V)=-8.7$ GeV$^{-1}$. In Ref. \cite{Riska2001}, the values $g_\rho^q=3.0$ and $f_\rho^q=0.0$ were used, which correspond to $h_V=3.0$ and $h_T=6.4$ GeV$^{-1}$ and give $(\beta_Bg_V)=-6.0$, $(\beta_Sg_V)=12.0$, $(\lambda_Sg_V)=19.2$ GeV$^{-1}$, $(\lambda_Ig_V)=-6.8$ GeV$^{-1}$.

In the heavy meson case, the relevant vector meson coupling constants have been estimated with VMD \cite{Isola2003}. One can also use this approach to estimate the coupling constants $\beta_B$ and $\beta_S$ here. For the heavy quark hadrons, VMD assumes that the coupling of the electromagnetic current to the light quarks is dominated by the vector mesons. One may find the electromagnetic interactions of heavy hadrons in Ref. \cite{Cheng1993}. To avoid the confusion due to the existence of the heavy quark spectator, one uses only the isovector $\rho$ meson dominance. The obtained values are $(\beta_Bg_V)=-\frac{\sqrt2m_\rho}{f_\rho}=-5.04$ and $(\beta_Sg_V)=\frac{2\sqrt2m_\rho}{f_\rho}=10.08$, where the decay constant $f_\rho=216$ MeV. Comparing these numbers with the quark model estimation, one finds that they are roughly consistent.

In Ref. \cite{Bardeen2003}, the chiral partner of the ground charmed meson multiplet was studied to interpret $D_{sJ}(2317)$. In that method, one may obtain an estimation of the coupling constant between the scalar meson and the charmed mesons. The authors also explored the doubly heavy baryons. Recall that the parity partner of the nucleon has also been explored \cite{Jido2001}. One may extend the previous studies to the charmed baryon case. Here, we derive the scalar meson coupling constants $\ell_B$ and $\ell_S$ with the chiral multiplet assumption.

For the antitriplet $B_{\bar{3}}$, its interpolating current is
\begin{eqnarray}
B_{\bar{3}}&\sim& (q^TC\gamma^5q)Q=(q_L^TC q_L)Q-(q_R^TC q_R)Q,
\end{eqnarray}
which transforms as $(\bar{3},1)\oplus(1,\bar{3})$ under $SU(3)_L\times SU(3)_R$. Here, we have used $\gamma_5 q_L=q_L$ and $\gamma_5 q_R=-q_R$. Its parity partner is
\begin{eqnarray}
\tilde{B}_{\bar{3}}&\sim& (q^TC q)Q=(q_L^TC q_L)Q+(q_R^TC q_R)Q.
\end{eqnarray}
Therefore, one may define the left field $B_{\bar{3}L}=\frac{1}{\sqrt2}(\tilde{B}_{\bar{3}}+ B_{\bar{3}})$ and the right field $B_{\bar{3}R}=\frac{1}{\sqrt2}(\tilde{B}_{\bar{3}}- B_{\bar{3}})$, which transform as $B_{\bar{3}L}\to LB_{\bar{3}L}L^T$ and $B_{\bar{3}R}\to RB_{\bar{3}R}R^T$. Then the lowest order chirally invariant Lagrangian is:
\begin{eqnarray}
{\cal L}_{LH}^{\bar{3}}=\frac12tr[\bar{B}_{\bar{3}L}(i\partial\!\!\!/-m_{\bar{3}})B_{\bar{3}L}]+\frac12tr[\bar{B}_{\bar{3}R}(i\partial\!\!\!/-m_{\bar{3}})B_{\bar{3}R}]+\frac{G_\pi}{4}[tr(\bar{B}_{\bar{3}L}\Sigma B_{\bar{3}R}\Sigma^T)+tr(\bar{B}_{\bar{3}R}\Sigma^\dag B_{\bar{3}L}\Sigma^{*})],
\end{eqnarray}
where $\Sigma=\xi\sigma\xi$, $\Sigma^T$ is its transposition, and $\Sigma$ transforms as $\Sigma\to L\Sigma R^\dag$. $\sigma$ can actually be the scalar nonet \cite{Bardeen2003}. In the chiral symmetric phase, the left and right fields have the degenerate mass $m_{\bar{3}}$. To get the linear representation, one redefines the fields $B_{\bar{3}}$ and $\tilde{B}_{\bar{3}}$,
\begin{eqnarray}
B_{\bar{3}L}=\frac{1}{\sqrt2}\xi(\tilde{B}_{\bar{3}}+ B_{\bar{3}})\xi^T,\quad B_{\bar{3}R}=\frac{1}{\sqrt2}\xi^\dag(\tilde{B}_{\bar{3}}-{B}_{\bar{3}})\xi^{*},
\end{eqnarray}
where the physical heavy field $B_{\bar{3}}$ or $\tilde{B}_{\bar{3}}$ transforms as $B\to UB U^T$. With the fact that $\langle \tilde{\sigma}\rangle\approx f_\pi$ in the chiral broken phase, we get
\begin{eqnarray}
{\cal L}_{LH}^{\bar{3}}&=&\frac12tr[\bar{B}_{\bar{3}} iD\!\!\!\!/ B_{\bar{3}} -m_{\bar{3}}\bar{B}_{\bar{3}} B_{\bar{3}}]+\frac12tr[\bar{\tilde{B}}_{\bar{3}}iD\!\!\!\!/\tilde{B}_{\bar{3}}-m_{\bar{3}}\bar{\tilde{B}}_{\bar{3}}\tilde{B}_{\bar{3}} ]+ tr[ \bar{\tilde{B}}_{\bar{3}}\gamma_\mu A^\mu B_{\bar{3}} + \bar{B}_{\bar{3}}\gamma_\mu A^\mu \tilde{B}_{\bar{3}}]\nonumber\\
&&+\frac{G_\pi f_\pi^2}{4}tr\{\bar{\tilde{B}}_{\bar{3}}\tilde{B}_{\bar{3}}- \bar{B}_{\bar{3}} B_{\bar{3}} \}+\frac{G_\pi f_\pi}{2}tr\{\bar{\tilde{B}}_{\bar{3}}\tilde{\sigma}_{phy}\tilde{B}_{\bar{3}}- \bar{B}_{\bar{3}}\tilde{\sigma}_{phy} B_{\bar{3}}\}+\ldots
\end{eqnarray}
Now the mass difference $\Delta M_{\bar{3}}$ between $B_{\bar{3}}$ and $\tilde{B}_{\bar{3}}$ appears. Comparing with Eq. (\ref{totLag}), one has
\begin{eqnarray}
\Delta M_{\bar{3}}=G_\pi f_\pi^2,\quad \ell_B=-\frac{\Delta M_{\bar{3}}}{2 f_\pi}=-\frac{G_\pi f_\pi}{2}.
\end{eqnarray}
To estimate the numerical value of $\ell_B$, we adopt the masses of the $\Lambda_c$ states. There are three P-wave fields: $\Lambda_{c1}$, $\tilde{\Lambda}_{c0}$ and $\tilde{\Lambda}_{c1}$ \cite{Cheng2007}. The parity partner of the S-wave $\Lambda_c$ should be $\tilde{\Lambda}_{c0}$. Unfortunately, it has not been reported. We may use the $\Lambda$ baryons to estimate the mass splitting $\Delta M_{\bar 3}\approx \Lambda(1670)-\Lambda(1405)+\Lambda_c(2595)^+-\Lambda_c^+\approx 573$ MeV, which gives $\ell_B=-3.1$. This value and the sign are consistent with the quark model.

The chiral multiplet $\tilde{S}_\mu$ ($\frac12^-$, $\frac32^-$) of $S_\mu$ ($\frac12^+$, $\frac32^+$) has not been identified. But we can still give a rough estimate of the coupling constant $\ell_S$. Here, we discuss only the $J=\frac12$ parity partners $B_6$ and $\tilde{B}_6$. If we use the interpolating current $B_{6}\sim (q^TC\gamma_\mu q)\gamma^\mu_t \gamma^5 Q$ \cite{Liu2008a,Huang2009}, its partner's current is $\tilde{B}_6\sim (q^T C\gamma_\mu\gamma^5 q)\gamma^\mu_t\gamma^5 Q$, where $\gamma^\mu_t=\gamma^\mu-v^\mu v\!\!\!/$. We may define
\begin{eqnarray}
B_{6RL}\sim (q_L^TC\gamma_\mu q_R)\gamma^\mu_t\gamma^5 Q,\quad B_{6LR}\sim (q_R^TC\gamma_\mu q_L)\gamma^\mu_t\gamma^5 Q.
\end{eqnarray}
In fact, $(q_R^TC\gamma_\mu q_L)^T=(q_L^T\gamma_\mu^TC^T q_R)=(q_L^TC\gamma_\mu q_R)$, so that $B_{6RL}=B_{6LR}\sim(3,3)$,
\begin{eqnarray}
B_6&\sim& (q^TC\gamma_\mu q)\gamma^\mu_t \gamma^5 Q=(q_R^TC\gamma_\mu q_L)\gamma^\mu_t \gamma^5 Q+(q_L^TC\gamma_\mu q_R)\gamma^\mu_t \gamma^5 Q+ 0\sim(3,3),\\
\tilde{B}_6&\sim& (q^TC\gamma_\mu \gamma^5 q)\gamma^\mu_t \gamma^5 Q=(q_R^TC\gamma_\mu q_L)\gamma^\mu_t \gamma^5 Q-(q_L^TC\gamma_\mu q_R)\gamma^\mu_t \gamma^5 Q+ 0\sim 0.
\end{eqnarray}
This means the defined current $B_6$ with the transformation (3,3) does not have a parity partner. However, one may use the following operators for the argument:
\begin{eqnarray}
B_{6}&\sim& (q^TC\sigma_{\mu\nu} q)\sigma^{\mu\nu} Q=(q_L^TC\sigma_{\mu\nu} q_L)\gamma_5\sigma^{\mu\nu} Q-(q_R^TC\sigma_{\mu\nu} q_R)\gamma_5\sigma^{\mu\nu} Q,\\
\tilde{B}_6&\sim& (q^T C\sigma_{\mu\nu} q)\gamma^5\sigma^{\mu\nu} Q=(q_L^TC\sigma_{\mu\nu} q_L)\gamma_5\sigma^{\mu\nu} Q+(q_R^TC\sigma_{\mu\nu} q_R)\gamma_5\sigma^{\mu\nu} Q,
\end{eqnarray}
which transform as $(6,1)\oplus(1,6)$. Then we may identify:
\begin{eqnarray}
B_{6L}=\frac{1}{\sqrt2}(\tilde{B}_6+ B_6),\quad B_{6R}=\frac{1}{\sqrt2}(\tilde{B}_6- B_6).
\end{eqnarray}
They transform as $B_{6L}\to LB_{6L}L^T$, $B_{6R}\to RB_{6R}R^T$. With the similar procedure, we finally get
$\ell_S=\frac{\Delta M_6}{f_\pi}$. The parity partner of $\Sigma$ is denoted as $\tilde{\Sigma}_{c1}$ \cite{Cheng2007}. It has not been measured, either. However, if we assume the mass difference comes mainly from the excitation of the light diquark, one may estimate $\Delta M_6\thickapprox\Delta M_{\bar 3}\approx 573$ MeV, and thus $\ell_S\approx2\ell_B\approx 6.2$.

Now we turn to some QCD sum rule calculations about the coupling constants. For the $NN\sigma$ coupling constant, there is a calculation $h_\sigma=14.6$ \cite{Erkol2006}, which is a little larger than the estimation in the quark model. From the pion couplings $\Sigma_c^*\Sigma_c\pi$ and $\Sigma_c^*\Lambda_c\pi$ in Ref. \cite{Zhu1998}, one gets $g_3=0.90\pm0.17\pm0.17$, $g_4=0.94\pm0.06\pm0.20$, and therefore $g_1\approx 1.04$. From the values $g_{\Sigma_c^*\Sigma_c\pi}=4.2\pm0.5$ GeV$^{-1}$ and $g_{\Sigma_c^*\Lambda_c\pi}=7.8\pm1.0$ GeV$^{-1}$ in Ref. \cite{Aliev2010}, one has $g_3=0.78\pm 0.09$ and $g_4=1.02\pm 0.13$. These numbers are all consistent with the quark model results. In Ref. \cite{Azizi2010}, however, the authors use a different definition of the coupling constant $g_{\Sigma_c\Sigma_c\pi}=-8.0\pm1.7$. With the help of the Goldberger-Treiman relation (GTR) \cite{Yu1993}, one obtains a value $g_1\sim 0.3$, which is much smaller than the quark model estimation. Ref. \cite{Huang2009} presents a calculation of the couplings of the light vector mesons and heavy baryons. To compare their results with our quark model estimation, we again use GTR and get $(\beta_Bg_V)\approx -43.4$ for $g_{\Lambda_c\Lambda_c\omega}=-1.85\pm 0.06$, $(\beta_Sg_V)\approx123.6$ for $g_{\Sigma_c^*\Sigma_c^*\rho}^{p0}=2.65\pm0.20$, and $(\beta_Sg_V)\approx78.0$ for $g_{\Sigma_c^*\Sigma_c^*\omega}^{p0}=1.51\pm0.18$. These results are much larger than the quark model estimation. But, for another coupling constant, we do not need GTR and we have $(\lambda_Sg_V)\approx21.0$ GeV$^{-1}$ for $g_{\Sigma_c^*\Sigma_c^*\rho}^{p1}=2.27\pm0.20$ and $(\lambda_Sg_V)\approx13.5$ GeV$^{-1}$ for $g_{\Sigma_c^*\Sigma_c^*\omega}^{p1}=1.32\pm0.18$, which are not far from the quark model estimation. It is an interesting problem for the QCD sum rule calculation that we get inconsistent results when comparing different definitions of the coupling constants with the help of GTR in the heavy quark baryon case. Because of this reason, we do not use these values. By the way, one may conclude that the hidden local symmetry is supported by taking a glimpse at the small coupling constants $g_{\Sigma_c^*\Sigma_c^*\rho}^{p2}$ and $g_{\Sigma_c^*\Sigma_c^*\omega}^{p2}$ in Ref. \cite{Huang2009}.

\begin{table}[htb]\centering
\begin{tabular}{c|c|c|c|c|c}\hline
Couping & Quark Model & Chiral Multiplet & VMD &QSR& Decay\\\hline
$g_1$ & $\frac43g_A^q=1.00$&&&&\\
$g_2$ & $-\sqrt{\frac23}g_A^q=-0.61$ & &   & & $ -0.598$\\
$g_3$ & $\frac{2}{\sqrt3} g_A^q=0.87$& & &$0.90\pm0.17\pm0.17$ \cite{Zhu1998}\\
      &                              & & &$0.78\pm 0.09$ \cite{Aliev2010}\\
$g_4$ & $\sqrt{2} g_A^q=1.06$ & &   &$0.94\pm0.06\pm0.2$ \cite{Zhu1998}& $0.999$\\
      &                              & & &$1.02\pm 0.13$ \cite{Aliev2010}\\
$g_5$ & $-2g_A^q=-1.50$&&&&\\\hline
$\ell_B$&$-g_\sigma^q=-3.65$& $-\frac{\Delta M}{2f_\pi}\approx -3.1$&&\\
$\ell_S$&$2g_\sigma^q=7.30$&$\frac{\Delta M}{f_\pi}\approx 6.2$&&\\\hline
$(\beta_Bg_V)$&$-2g_\rho^q=-6.0$ &         & $-\frac{m_V}{f_V}\approx-5.04$&\\
$(\beta_Sg_V)$&$4g_\rho^q=12.0$  &         & $\frac{2m_V}{f_V}\approx10.08$&\\
$(\lambda_Sg_V)$& $2\left(\frac{g_\rho^q}{m_q}+2f_\rho^q\right)=19.2$ GeV$^{-1}$ &&&$21.0$ GeV$^{-1}$, $13.5$ GeV$^{-1}$ \cite{Huang2009}&\\
$(\lambda_I g_V)$& $-\frac{1}{\sqrt2}\left(\frac{g_\rho^q}{m_q}+2f_\rho^q\right)=-6.8$ GeV$^{-1}$&&&&\\\hline
$g_A$&$\frac53g_A^q=1.25$&&&&\\
$h_\sigma$&$3g_\sigma^q=10.95$&&&14.6 \cite{Erkol2006}&\\
$h_V$& $g_\rho^q=3.0$&&&&\\
$h_T$& $-\frac{g_\rho^q}{2M_N}+\frac56\left(\frac{g_\rho^q}{m_q}+2 f_\rho^q\right)=6.4$ GeV$^{-1}$ &&&& \\
\hline
\end{tabular}
\caption{The coupling constants in different methods. For the quark model estimation, we present the results with $g_A^q=0.75$, $g_\sigma^q=3.65$ \cite{Riska1999}, $g_\rho^q=3.0$, and $f_\rho^q=0.0$ \cite{Riska2001}. We have used the phase convention of $g_{2,4}$ from the decay being consistent with the quark model.}\label{couplings}
\end{table}

For comparison, we summarize in Table \ref{couplings} the values of the coupling constants given by different methods. There we have taken our convention for the phases. In our numerical evaluation, we will choose the following values and relations,
\begin{eqnarray}
&g_2=-0.598,\quad g_4=0.999,\quad g_1=\frac{\sqrt8}{3}g_4,\quad g_3=\sqrt{\frac23}g_4,\quad g_5=-\sqrt{2}g_4,\quad g_A=1.25,&\nonumber\\
&\ell_B=-3.1,\quad \ell_S=-2\ell_B,\quad h_\sigma=10.95,&\nonumber\\
&(\beta_Bg_V)=-6.0, \quad (\beta_Sg_V)=-2(\beta_Bg_V),\quad (\lambda_Sg_V)=19.2 \text{ GeV}^{-1},\quad (\lambda_Ig_V)=-(\lambda_Sg_V)/\sqrt{8},&\nonumber\\
&h_V=3.0,\quad h_T=6.4 \text{ GeV}^{-1}.&
\end{eqnarray}

\section{$J^P=0^+$ case}\label{sec5}

We adopt the variational method \cite{hiyama} to solve the bound state problem. For the hadron masses, we use $m_\pi=137.27$ MeV, $m_\rho=775.49$ MeV, $m_\omega=782.65$ MeV, $m_N=938.92$ MeV, $m_{\Lambda_c}=2286.46$ MeV, $m_{\Sigma_c}=2453.56$ MeV, and $m_{\Sigma_c^*}=2517.97$ MeV \cite{PDG2010}. The scalar $\sigma$ meson is a broad state with strong coupling to $\pi\pi$ S-wave scattering states and its mass has not been accurately given. The recent investigations indicate that the pole mass is around $400\sim600$ MeV \cite{Ishida1996,Igi1999,Colangelo2006}. In the following calculation, we just adopt a large value $m_\sigma=600$ MeV. If we take a smaller value, then it will enhance attraction and will result in deeper bound states. Another parameter we cannot determine is the cutoff. In fact, the results are sensitive to it. We treat it as a free parameter and discuss the effects on the binding energy. In this section, we investigate the spin-singlet case.

\begin{figure}[htb]
\centering
\begin{tabular}{cc}
\scalebox{0.6}{\includegraphics{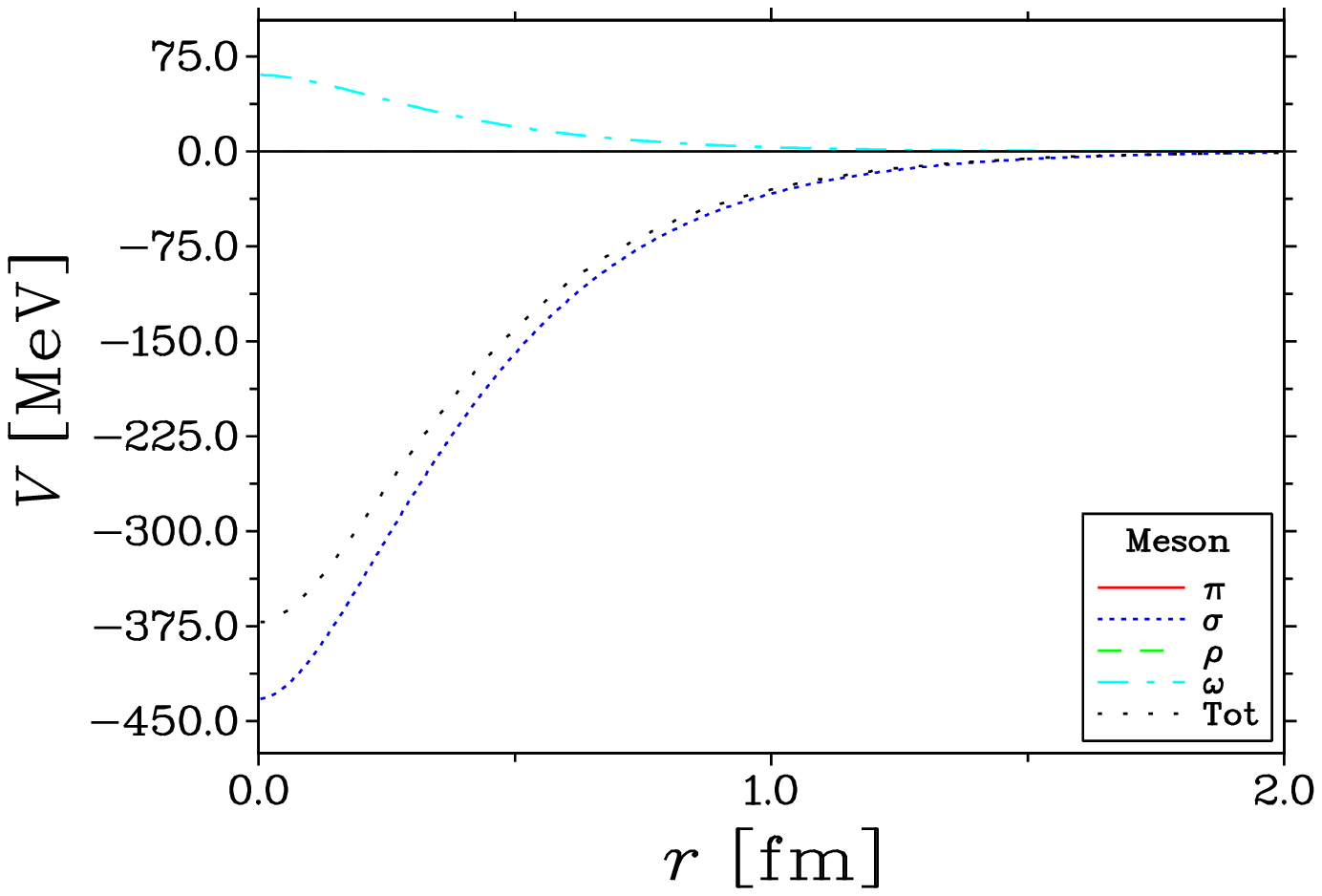}}&
\scalebox{0.6}{\includegraphics{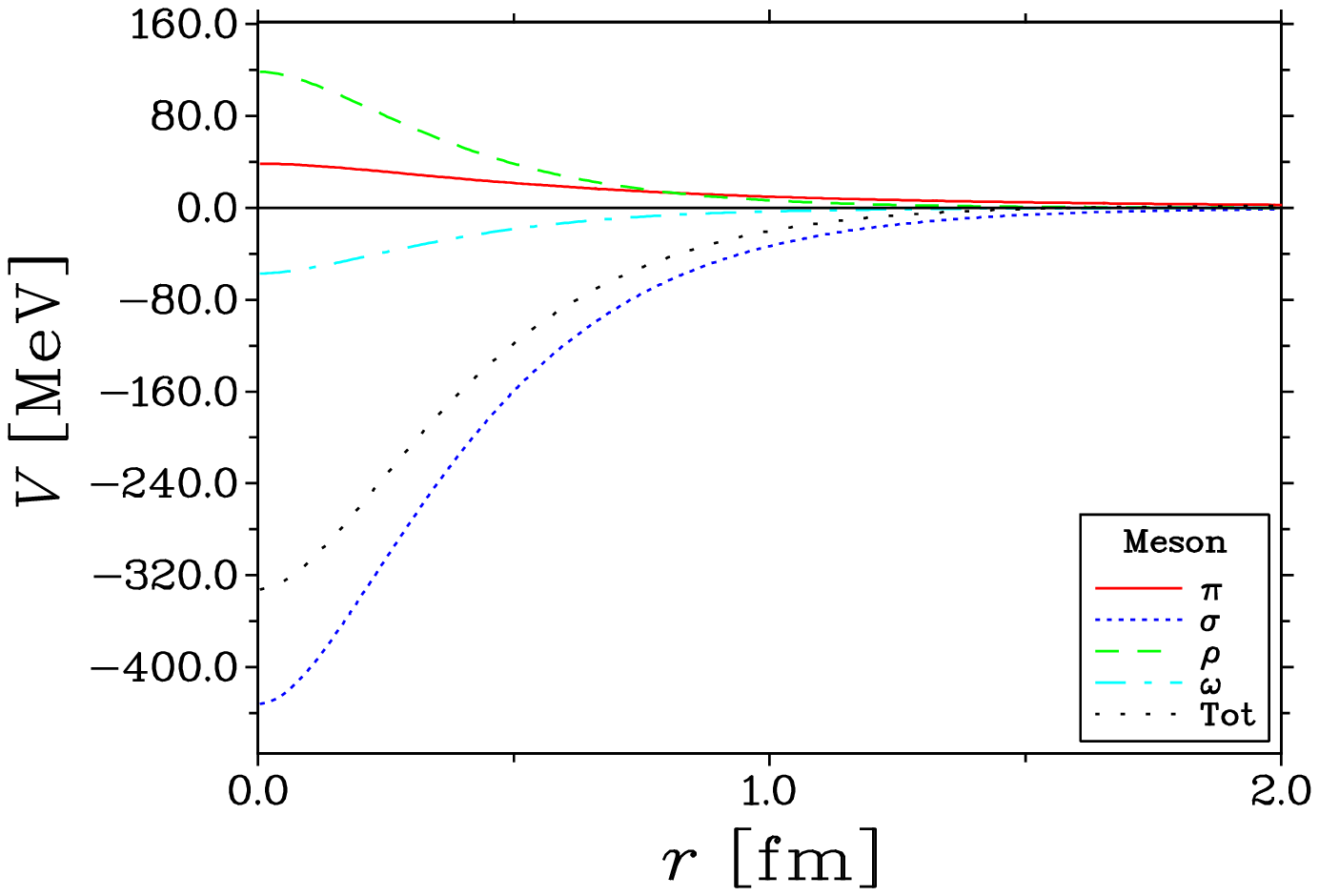}}\\
(11): $\Lambda_cN(^1S_0)\leftrightarrow\Lambda_cN(^1S_0)$&
(22): $\Sigma_cN(^1S_0)\leftrightarrow\Sigma_cN(^1S_0)$\\
\scalebox{0.6}{\includegraphics{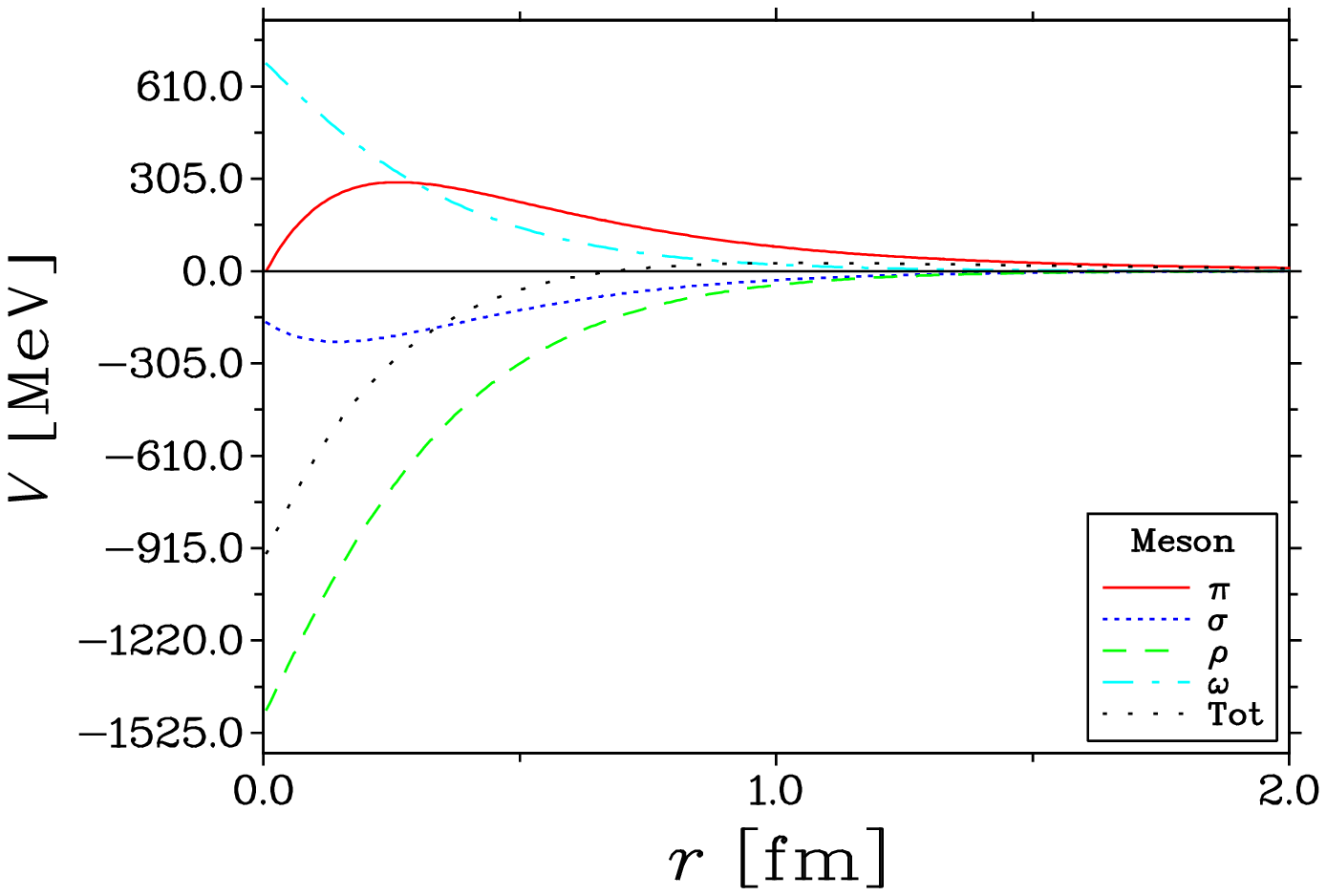}}&
\scalebox{0.6}{\includegraphics{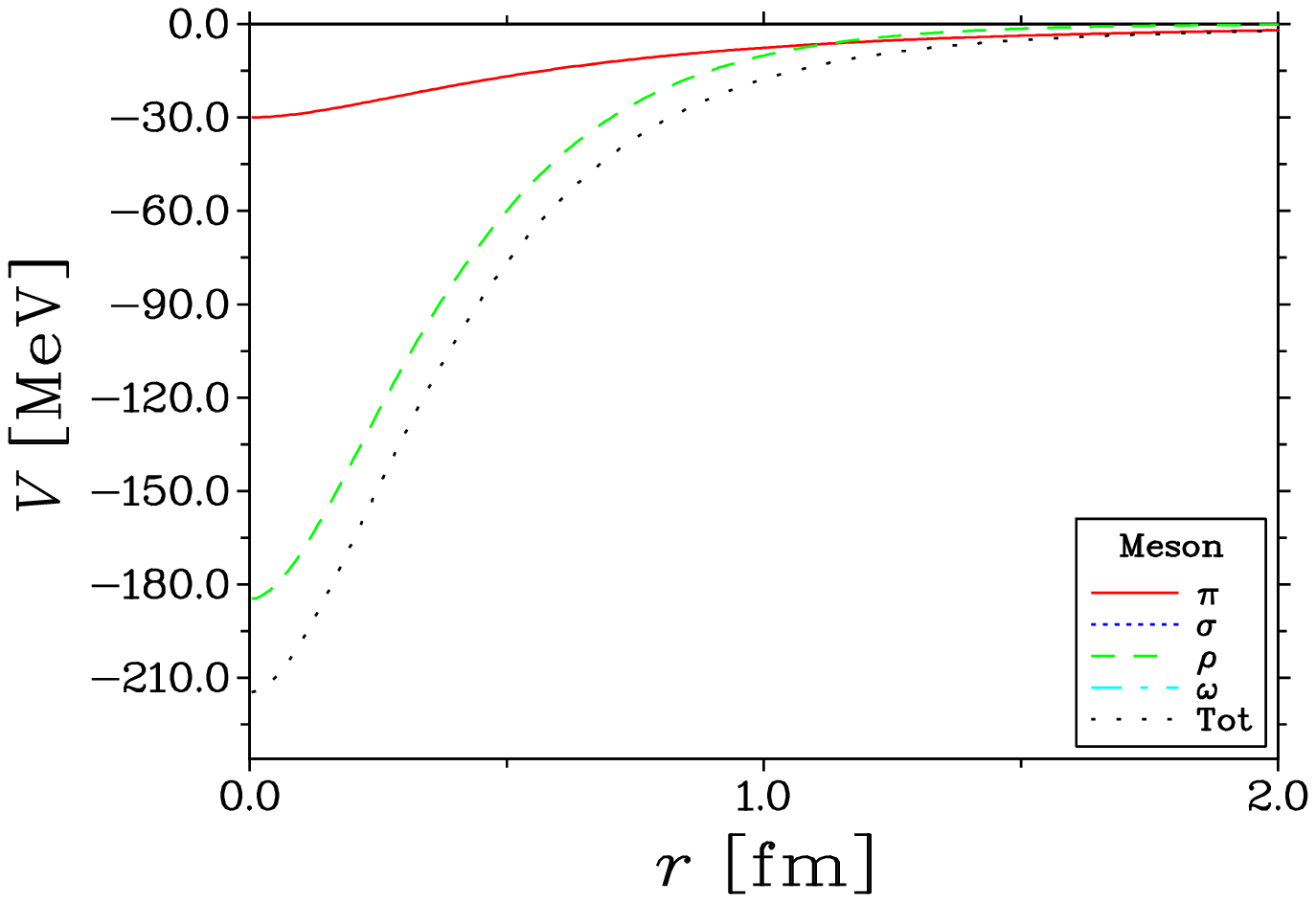}}\\
(33): $\Sigma_c^*N(^5D_0)\leftrightarrow\Sigma_c^*N(^5D_0)$&
(12): $\Lambda_cN(^1S_0)\leftrightarrow\Sigma_cN(^1S_0)$\\
\scalebox{0.6}{\includegraphics{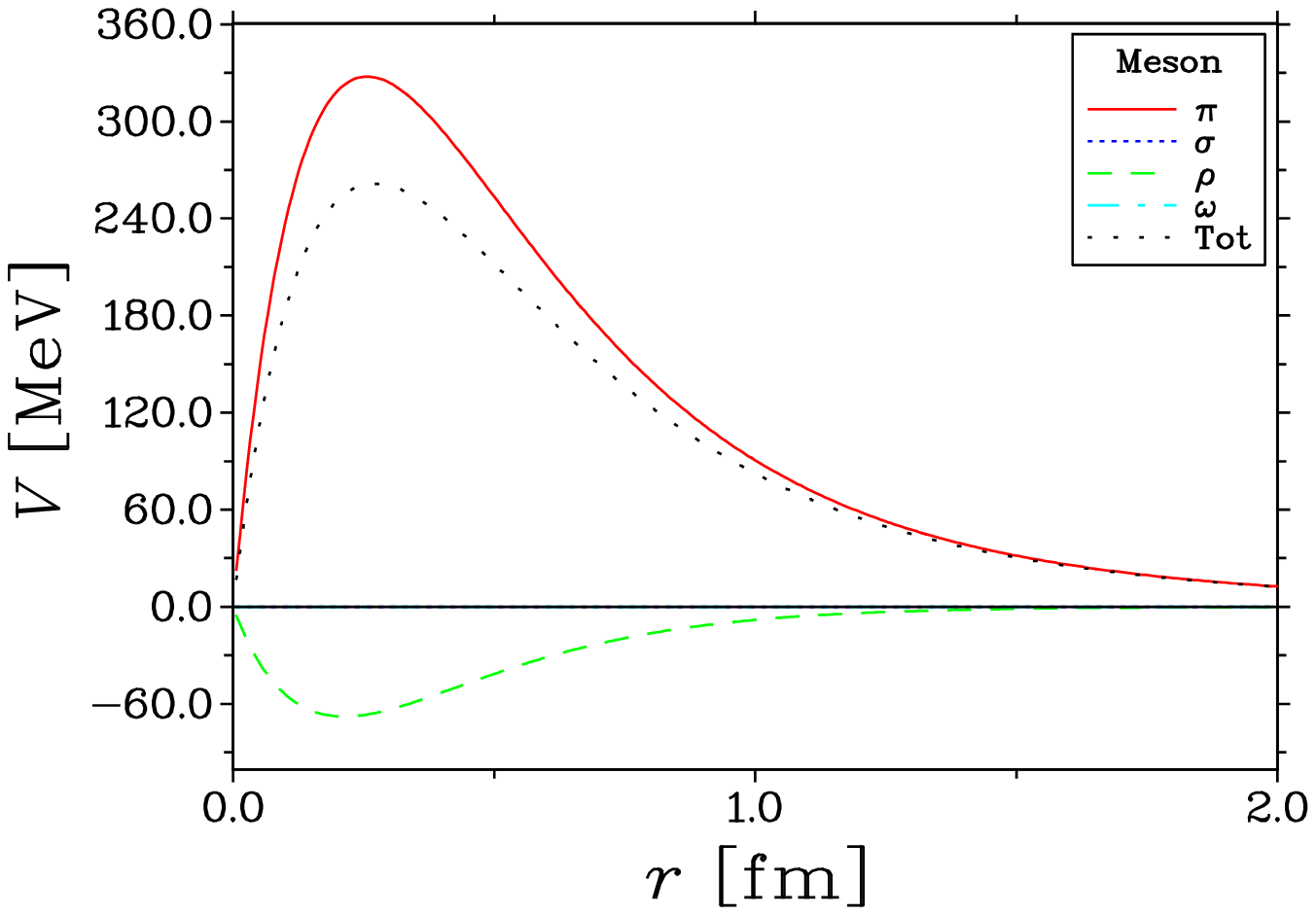}}&
\scalebox{0.6}{\includegraphics{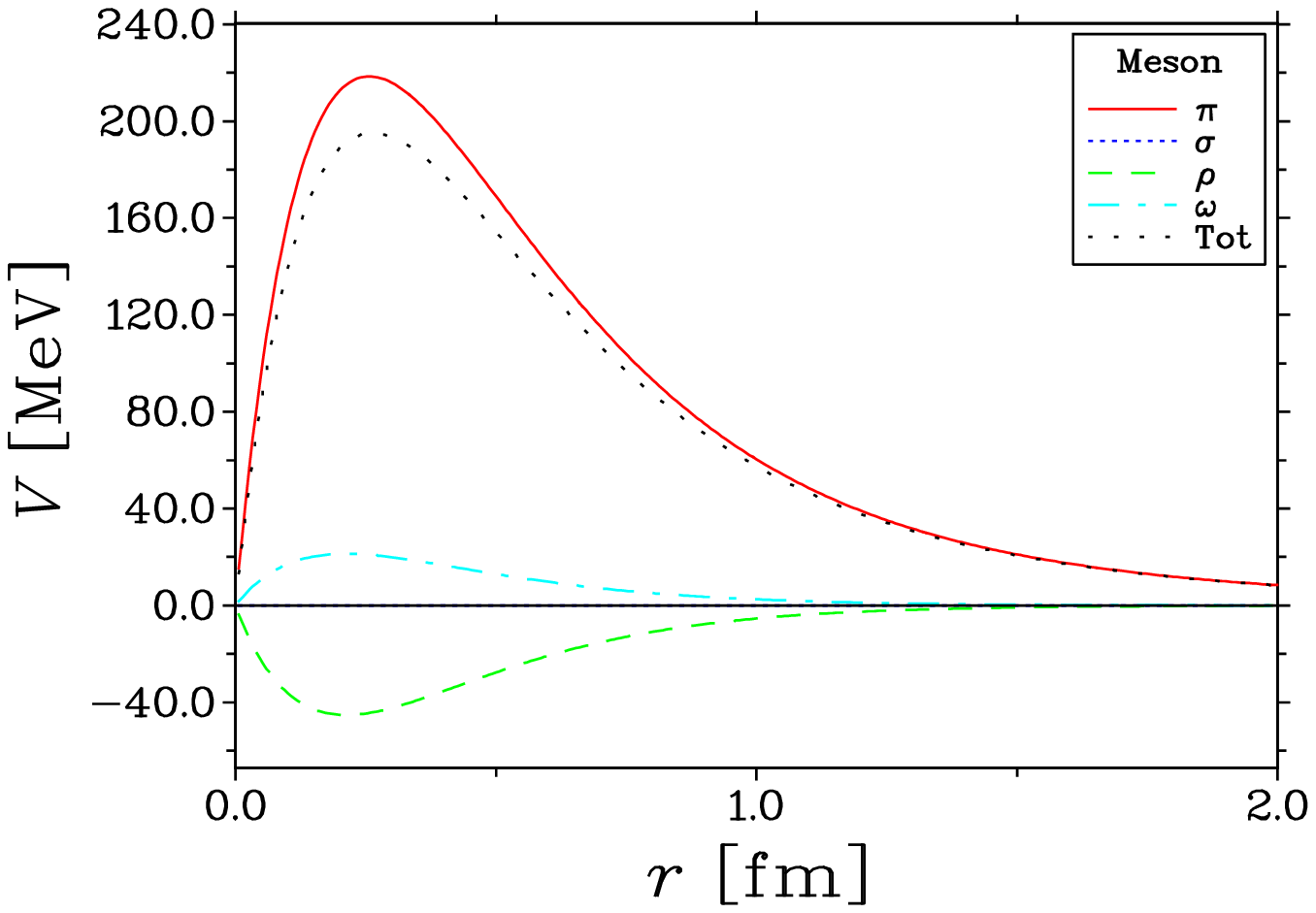}}\\
(13): $\Lambda_cN(^1S_0)\leftrightarrow\Sigma_c^*N(^5D_0)$&
(23): $\Sigma_cN(^1S_0)\leftrightarrow\Sigma_c^*N(^5D_0)$
\end{tabular}
\caption{The potentials of different channels for the $J^P=0^+$ case with $\Lambda_\pi=\Lambda_\sigma=\Lambda_{\rm vec}=1$ GeV.}\label{potential-J0}
\end{figure}

It is useful to take a look at the potentials. We plot the potentials of different channels in Fig. \ref{potential-J0} with the cutoff $\Lambda_\pi=\Lambda_\sigma=\Lambda_{\rm vec}=1$ GeV. The pion potentials in the channels 2 and 3 are both repulsive while the transition potentials are strong. The central force of the $\rho$ potential dominates the transition process $\Lambda_cN(^1S_0)\leftrightarrow\Sigma_cN(^1S_0)$ while the tensor force of the $\pi$ potential dominates the processes $\Lambda_cN(^1S_0)\leftrightarrow\Sigma_c^*N(^5D_0)$ and $\Sigma_cN(^1S_0)\leftrightarrow\Sigma_c^*N(^5D_0)$. We study first whether only one-pion-exchange interaction would lead to a bound state and then investigate the scalar and vector meson contributions. We call the corresponding potentials OPEP (one-pion-exchange potential) and OBEP (one-boson-exchange potential), respectively. Certainly the portions of the contributions from different mesons change if one uses other cutoffs $\Lambda_\pi\neq\Lambda_\sigma\neq\Lambda_{\rm vec}$. We will discuss the effects in the following calculation.

\subsection{OPEP model}

Let us first consider the case without channel coupling. For the channel $\Lambda_cN$, the direct one-pion-exchange is forbidden. For the channels $\Sigma_cN$ and $\Sigma_c^*N$, the potentials are both repulsive, therefore one cannot get a bound state.

\begin{table}[htb]
\begin{tabular}{c|cccccc}\hline
$\Lambda_\pi$ (GeV)                         &1.2     & 1.3  & 1.4   & 1.5    & 1.6 & 1.7\\\hline
$B.E. (J=0)$ (MeV)&0.64&6.16&18.51&38.88 & 68.29 & 107.64\\
$\sqrt{\langle r^2\rangle}$ (fm)& 5.2 & 1.9 & 1.2 & 0.9& 0.8 & 0.7\\
Prob. (\%)& (98.2/0.6/1.2)&(94.0/2.3/3.7)&(89.3/4.6/6.1)&(84.5/7.2/8.3)&(80.1/9.8/10.1)&(76.1/12.2/11.7)\\
\hline
\end{tabular}
\caption{Binding solutions for the $J^P=0^+$ case with channel coupling in the OPEP model. The binding energies (B.E.) are given relative to the $\Lambda_cN$ threshold. The probabilities correspond to $\Lambda_cN(^1S_0)$, $\Sigma_cN(^1S_0)$, and $\Sigma_c^*N(^5D_0)$, respectively. }\label{J0OpEP}
\end{table}

\begin{figure}[htb]\centering
\begin{tabular}{cc}
\scalebox{0.6}{\includegraphics{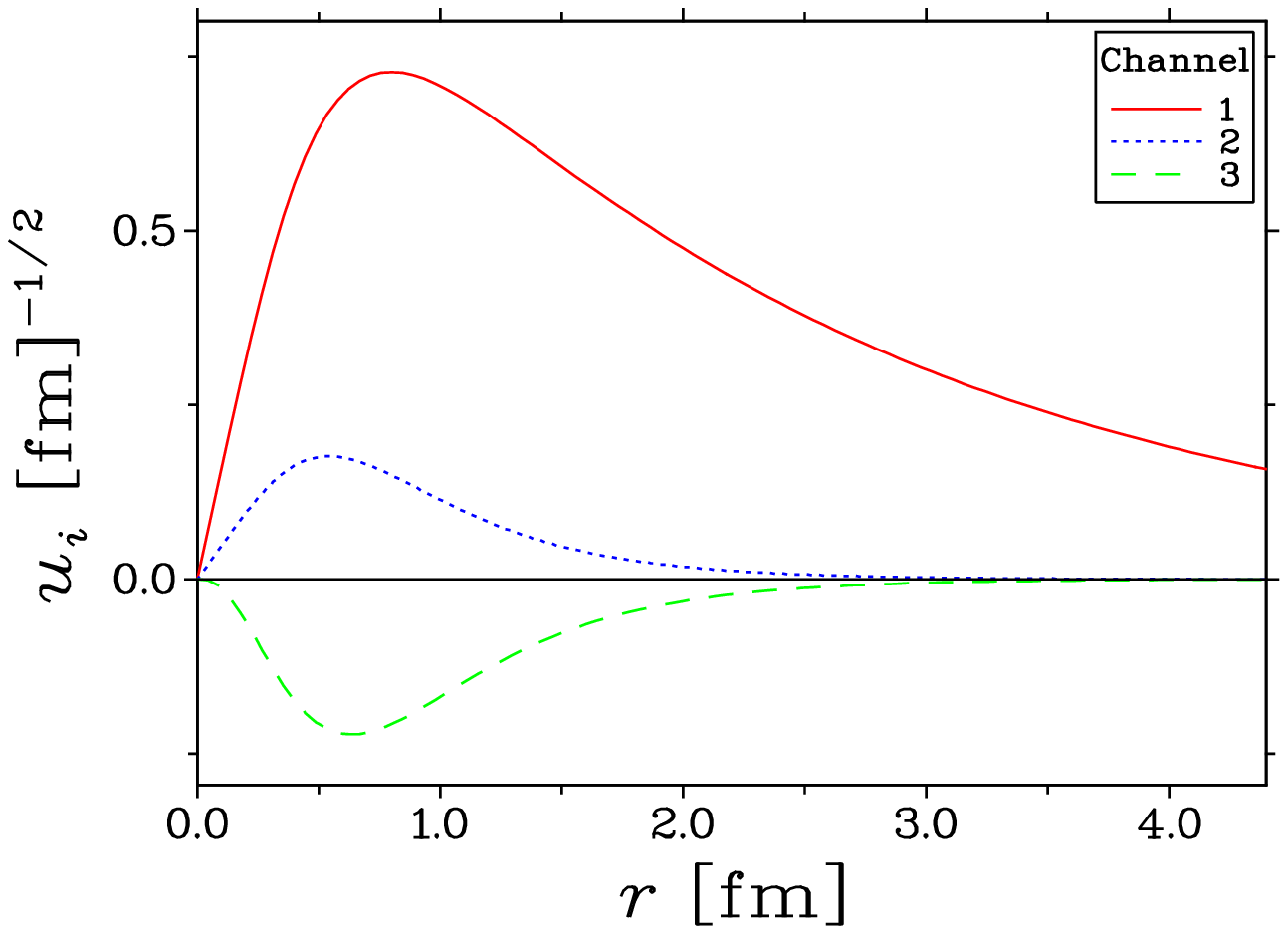}}&\scalebox{0.6}{\includegraphics{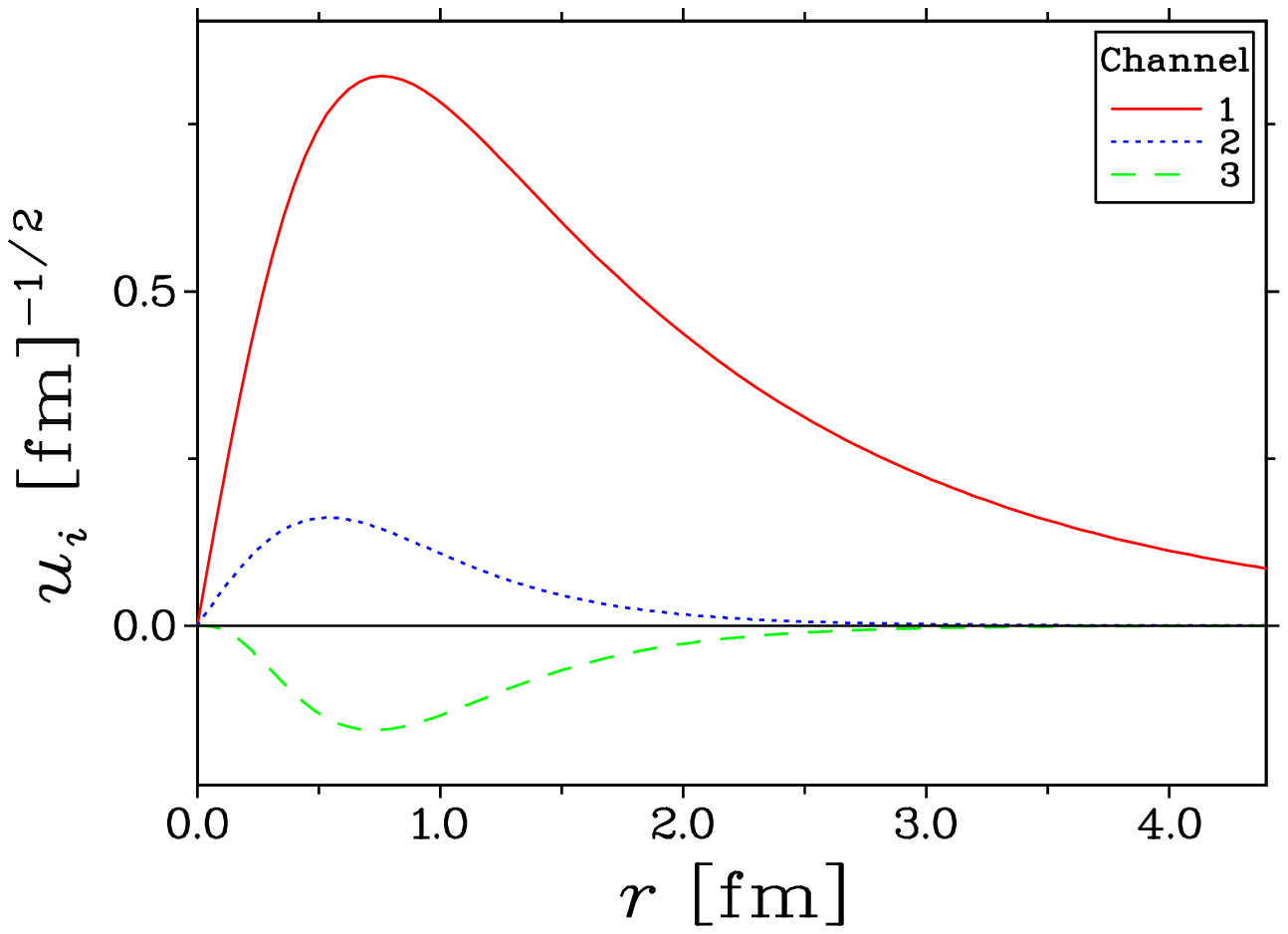}}\\
(a) & (b)\\
\scalebox{0.6}{\includegraphics{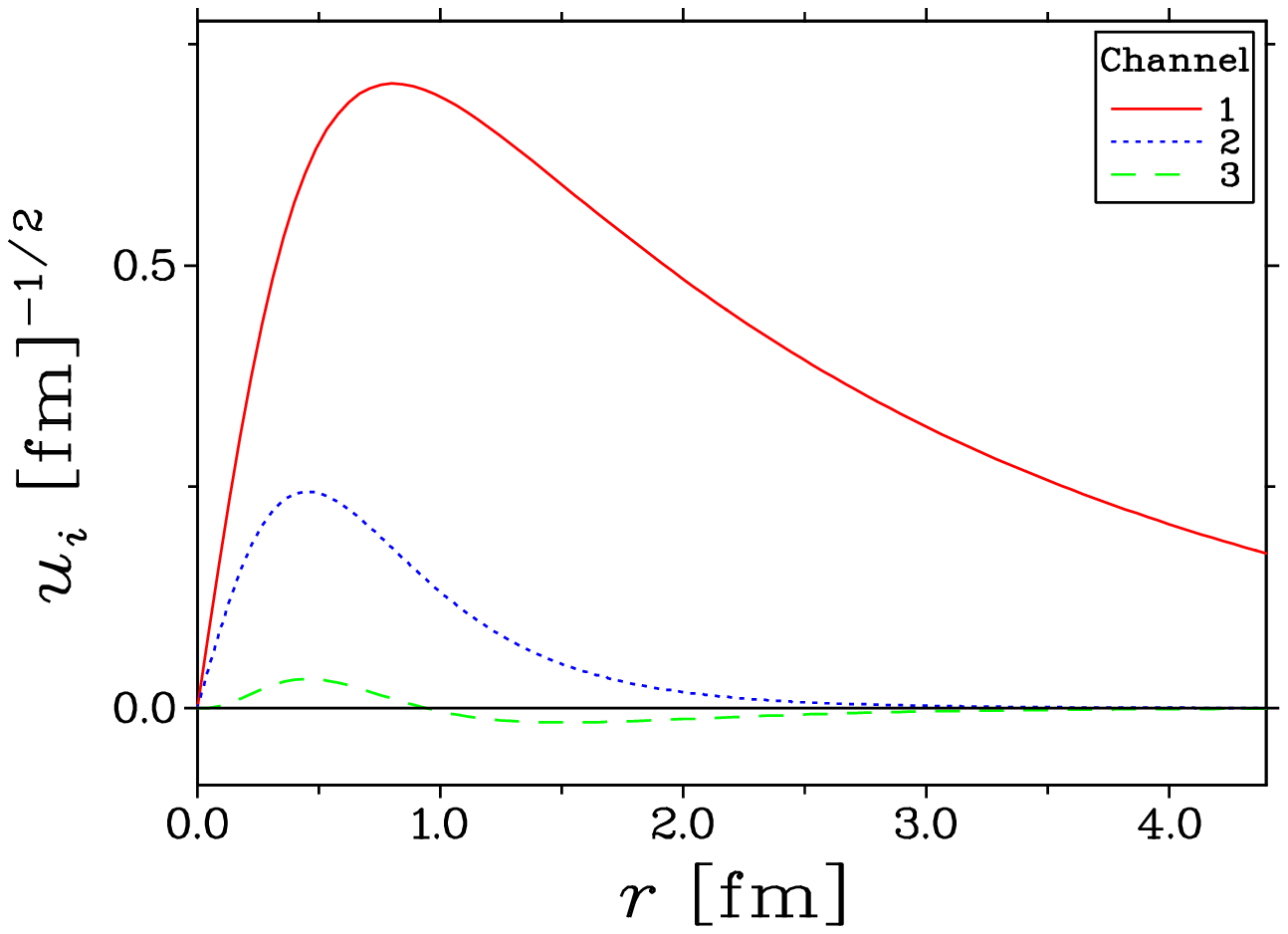}}&\scalebox{0.6}{\includegraphics{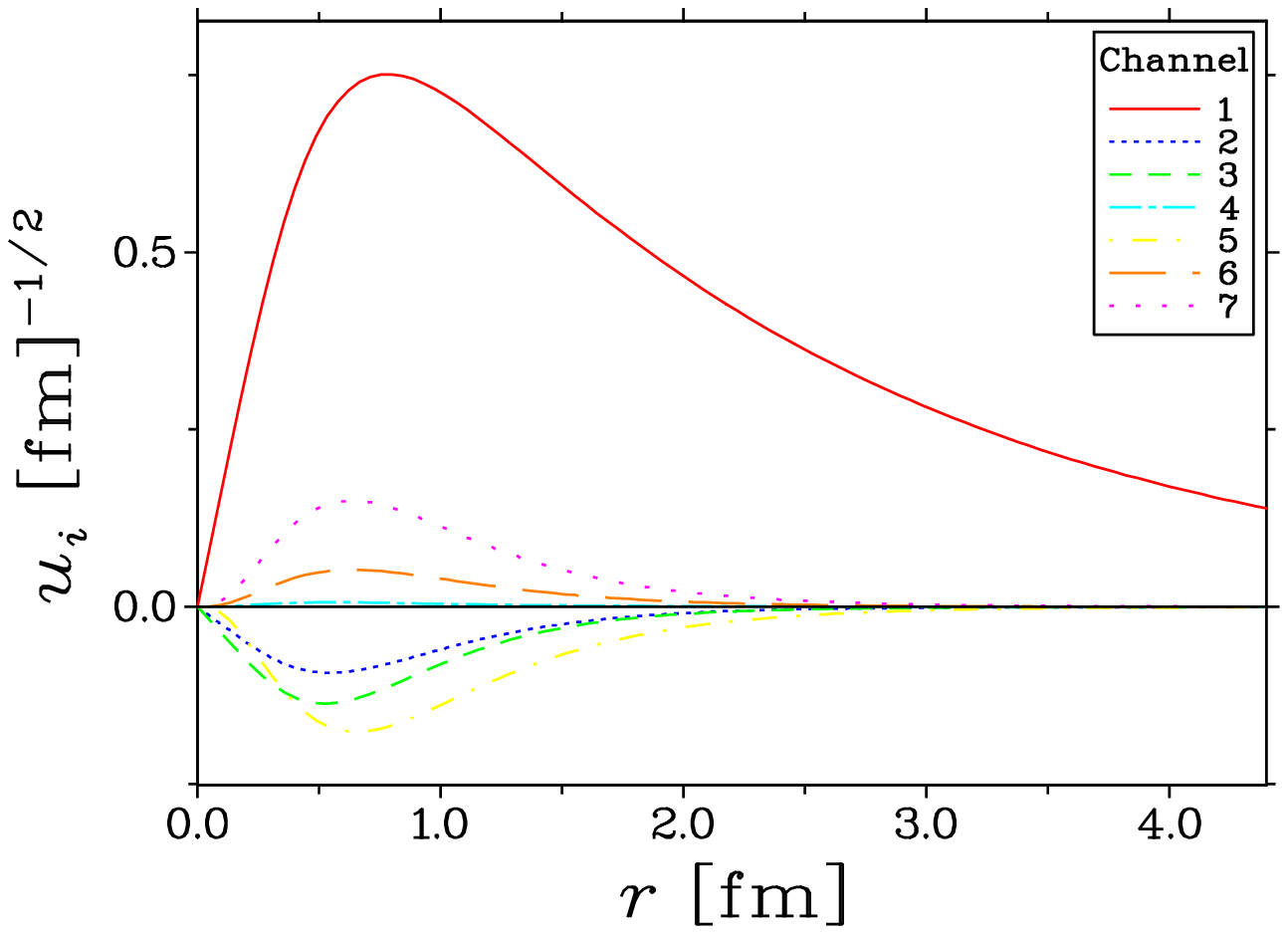}}\\
(c) & (d)\\
\scalebox{0.6}{\includegraphics{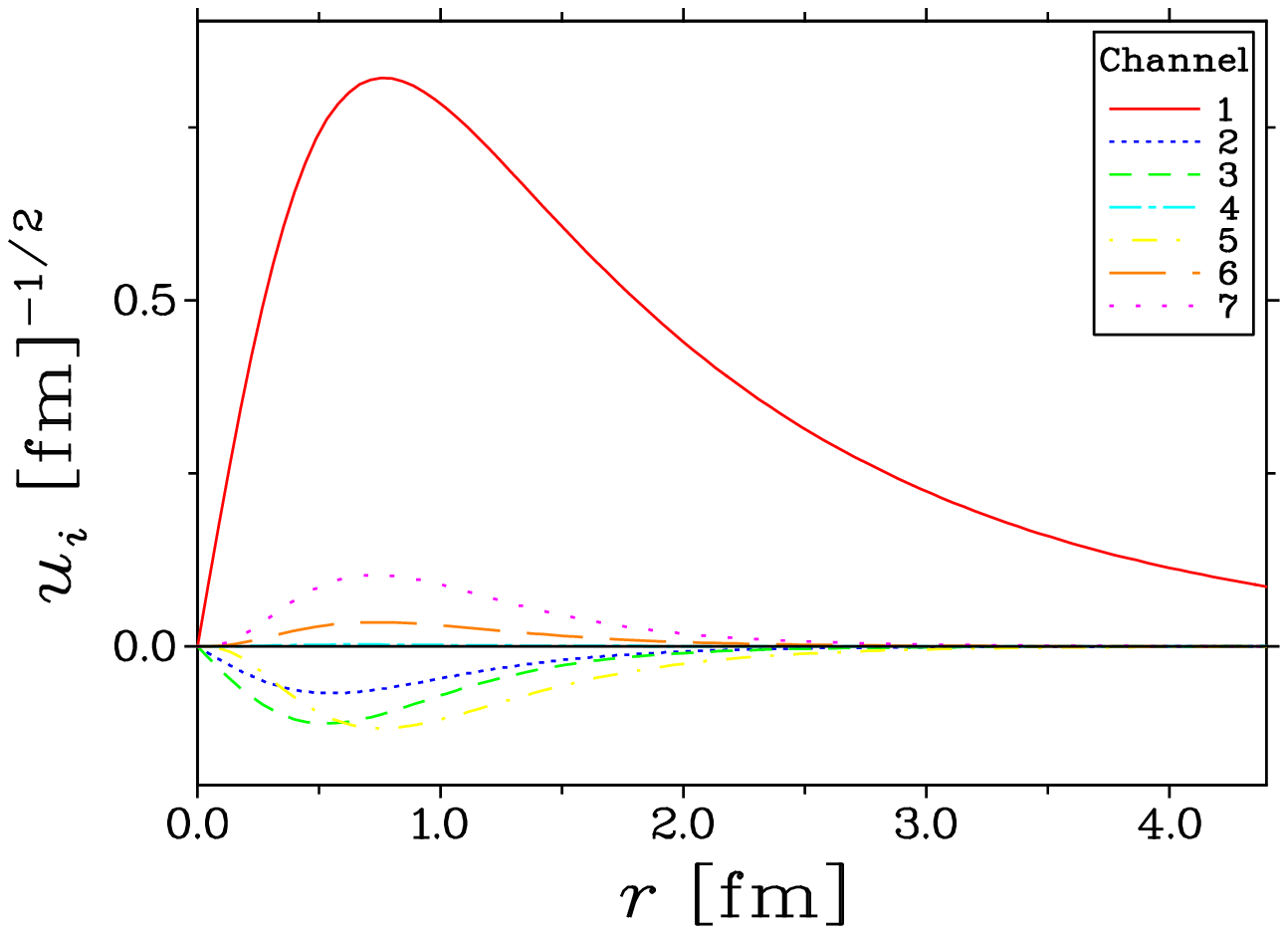}}&\scalebox{0.6}{\includegraphics{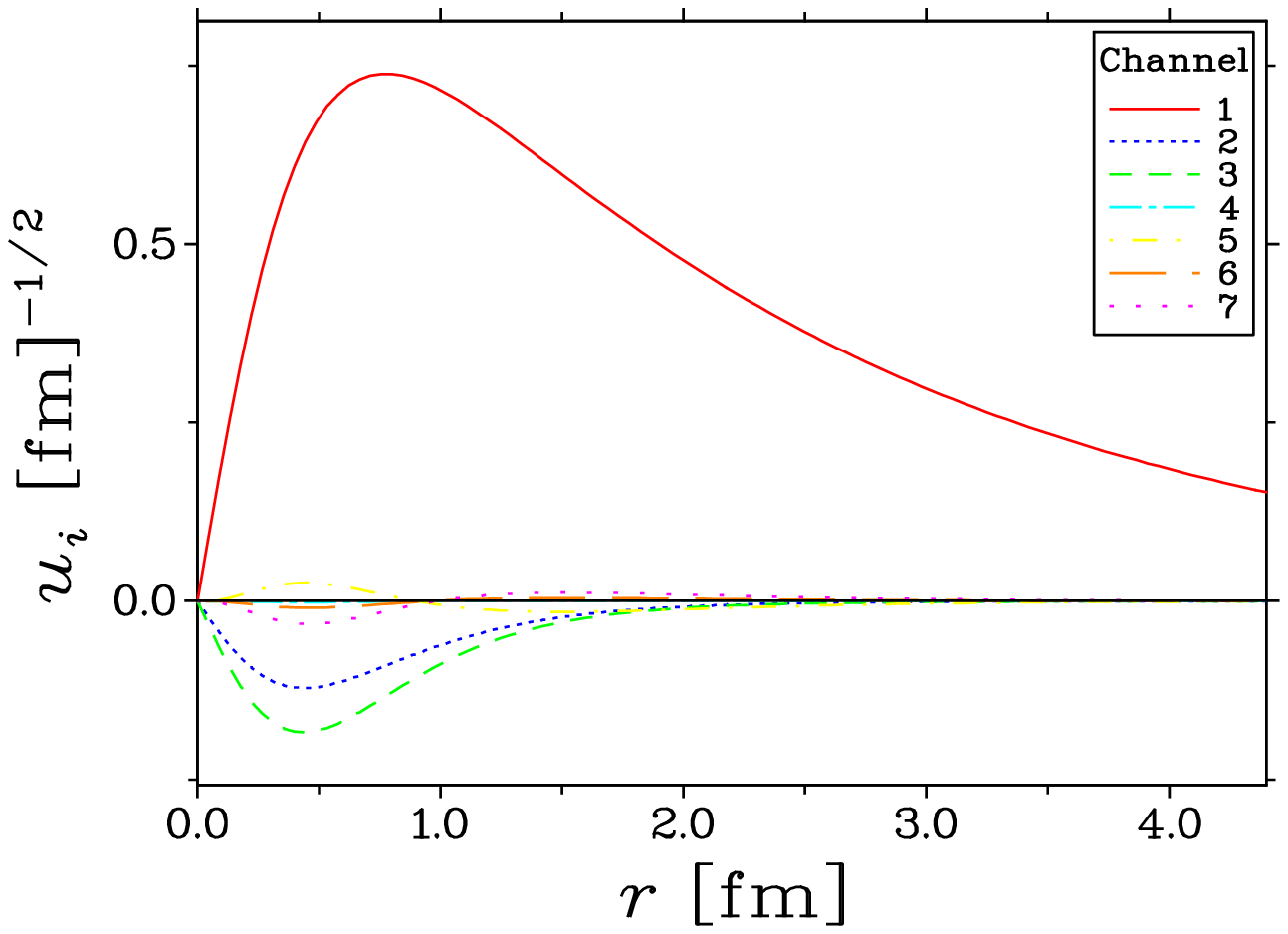}}\\
(e) & (f)\\
\end{tabular}
\caption{The wave functions $u_i$ ($i$=1,2, ..., 7) of different channels. The first three diagrams correspond to the results for the spin-singlet state: (a) OPEP case with the cutoff $\Lambda_\pi=1.3$ GeV; (b) OBEP case with the cutoff $\Lambda_\pi=\Lambda_\sigma=\Lambda_{\rm vec}=0.9$ GeV; (c) OBEP case with the parameter $\alpha=1.4$. The last three diagrams correspond to the results for the spin-triplet state: (d) OPEP case with the cutoff $\Lambda_\pi=1.3$ GeV; (e) OBEP case with the cutoff $\Lambda_\pi=\Lambda_\sigma=\Lambda_{\rm vec}=0.9$ GeV; (f) OBEP case with the parameter $\alpha=1.5$. }\label{WAVEu}
\end{figure}

For the case with channel coupling, we have one free parameter $\Lambda_\pi$. It is interesting that a bound state is obtained within the reasonable range of the cutoff. We present the binding energies, the root-mean-square (RMS) radius, and the probabilities of each channel in Table \ref{J0OpEP}. The binding energy is given with relative to the $\Lambda_cN$ threshold. As a deuteronlike molecule, the two baryons should not be very close. Thus we only list the results with the RMS radius larger than 0.7 fm. The values indicate that the reasonable binding energy should be no more than tens of MeV although it depends on the poorly known cutoff. As an example, we show the wave functions of different channels for $\Lambda_\pi=1.3$ GeV in Fig. \ref{WAVEu} (a). That the probability of the channel $\Sigma_c^*N (^5D_0)$ is larger than that of the channel $\Sigma_cN (^1S_0)$ indicates the importance of the tensor force in the model. As a check, we have calculated the two channel case: $\Lambda_cN (^1S_0)$ and $\Sigma_cN (^1S_0)$ and we do not find any binding solutions.

We have omitted the $\delta$-functional part in our potentials. Once that part is included, deeper molecular bound states are obtained. In that case, the $S$-wave $\Sigma_cN$ channel is dominant over to $D$-wave $\Sigma_c^*N$ channel. For example, if we use $\Lambda_\pi=0.8$ GeV, a 3-channel calculation gives the binding energy $B.E.=19.18$ MeV and a 2-channel calculation (without the $D$-wave channel) gives $B.E.=17.85$ MeV. However, the binding energy is more sensitive to the cutoff parameter. We get $B.E.=111.53$ MeV in the 3-channel calculation with a little larger cutoff $\Lambda_\pi=0.9$ GeV.

\subsection{OBEP model}

After we consider the contributions from the scalar meson $\sigma$ and the vector mesons $\rho$ and $\omega$, the binding energies will change accordingly. However, we have some freedoms to choose: the three cutoff parameters $\Lambda_\pi$, $\Lambda_\sigma$, and $\Lambda_{\rm vec}$. In the nucleon-nucleon case, one may determine the unknown model parameters by fitting the abundant experimental data. Since there is no available data, here we discuss the results with two approaches for the parametrization of the cutoffs.

\subsubsection{Common cutoff}

\begin{table}[htb]
\begin{tabular}{c|cccccc}\hline
$\Lambda_{\rm com}$ (GeV)  & 0.9   & 1.0   & 1.1   & 1.2 & 1.3&1.4\\\hline
$\Lambda_cN(^1S_0)$ [$B.E.$, $\sqrt{\langle r^2\rangle}$]&[1.24, 3.8]&[11.09, 1.5]&[27.07, 1.1]&[46.66, 0.9]&[68.45, 0.8]&[91.58, 0.7]\\
$\Sigma_cN(^1S_0)$ [$B.E.$, $\sqrt{\langle r^2\rangle}$]&$\times$&[2.22, 2.8]&[14.22, 1.3]&[31.56, 1.0]&[52.08, 0.8]&[74.62, 0.7]\\
\hline
\end{tabular}
\caption{Binding solutions for the individual channels in the $J^P=0^+$ and common-cutoff case in the OBEP model. The binding energies (B.E.) are given relative to their own thresholds. ``$\times$'' indicates that there is no binding solution. The units for the binding energy and RMS radius are MeV and fm, respectively.}\label{J0OmEP-noc}
\end{table}

\begin{table}[htb]
\begin{tabular}{c|cccc}\hline
$\Lambda_{\rm com}$ (GeV)  &0.8     & 0.9  & 1.0 & 1.1  \\\hline
$B.E. (J=0)$ (MeV)&0.12&13.60& 52.50 & 123.14   \\
$\sqrt{\langle r^2\rangle}$ (fm)& 11.2 & 1.5 & 0.9 & 0.7   \\
Prob. (\%)& (99.7/0.1/0.2)&(96.0/2.0/2.0)&(87.3/9.2/3.5)&(75.8/19.7/4.5) \\
\hline
\end{tabular}
\caption{Binding solutions for the $J^P=0^+$ and common-cutoff case with channel coupling in the OBEP model. The binding energies (B.E.) are given relative to the $\Lambda_cN$ threshold. The probabilities correspond to $\Lambda_cN(^1S_0)$, $\Sigma_cN(^1S_0)$, and $\Sigma_c^*N(^5D_0)$, respectively.}\label{J0OmEP}
\end{table}

The simplest assumption is that we use the same cutoff for different mesons $\Lambda_\pi=\Lambda_\sigma=\Lambda_{\rm vec}=\Lambda_{\rm com}$. This cutoff should be at least larger than the exchanged meson masses. As in the former subsection, we first consider the case without coupled channel effects.  In Table \ref{J0OmEP-noc}, we show the binding energy and the corresponding RMS radius for individual channels relative to each threshold, where solutions are found only in the S-wave channels. Numerically, one may obtain binding solutions for the $D$-wave channel with a larger cutoff, but the solutions are not reasonable. For example, we get $B.E.=1.99$ MeV and $r_{\rm RMS}=0.6$ fm with $\Lambda_{\rm com}=1.78$ GeV. The radius is so small for a shallow $D$- wave bound state. Because of the attractive $\sigma$ meson, $\Lambda_cN$ can be bound now. The contribution from the repulsive $\omega$ is not large in this parametrization of cutoffs. For the $\Sigma_cN$, the cancellation between the $\rho$ and $\omega$ contributions is large and the attraction comes mainly from the scalar potential.

Now we consider the coupled channel effects. Table \ref{J0OmEP} lists the results. The binding energy and the radius are more sensitive to the change of the cutoff than the OPEP case. For the same binding energy as the OPEP case, the necessary cutoff in OBEP is smaller. These features are attributed to the more attractive potentials. In order to have the molecular condition, $r_{\rm RMS}>0.7$ fm, satisfied, the binding energy should be no more than tens of MeV, as in the OPEP model calculation. As an example, we plot the wave functions of different channels with $\Lambda_{\rm com}=0.9$ GeV in Fig. \ref{WAVEu} (b).

Let us go back to the single channel $\Lambda_cN$ and the potentials. Because of the isospin conservation, direct $\pi$ and $\rho$ exchanges are forbidden. The spin-dependent $\omega$ exchange interaction also vanishes. The resulting spin-independent repulsive potential has been shown in Fig. \ref{potential-J0}. If one includes the $\delta$-functional terms in the model potentials, the $\omega$ exchange interaction is strongly attractive at short distance, which results in much deeper bound states and enhances the sensitivity of the results to the cutoff parameter. In addition, the coupled channel calculation with the $\delta$-functional terms results in unreasonable molecular bound state solutions. Thus we omit the $\delta$-functional terms in the present model construction.

\subsubsection{Scaled cutoffs}

Another possible choice of the cutoff parameters is to choose different values for the pseudoscalar, scalar, and vector mesons. To reduce the number of parameters, we adopt a parametrization used in Ref. \cite{Cheng2005}, $\Lambda_{\rm ex}=m_{\rm ex}+\alpha\Lambda_{QCD}$. Here $m_{\rm ex}$ is the mass of the exchanged meson, $\Lambda_{QCD}=220$ MeV is the scale of QCD, and $\alpha$ is a dimensionless parameter whose value is not very far from 1. We choose various values of $\alpha$ in the present investigation.
\begin{table}[htb]
\begin{tabular}{c|cccccc}\hline
$\alpha$  & 1.5   & 2.0   & 2.5   & 3.0 & 3.5&4.0\\\hline
$\Lambda_cN(^1S_0)$ [$B.E.$,$\sqrt{\langle r^2\rangle}$]&[0.12, 11.6]&[6.54, 1.9]&[20.30, 1.2]&[38.86, 0.9]&[60.56, 0.8]&[84.29, 0.7]\\
$\Sigma_cN(^1S_0)$ [$B.E.$,$\sqrt{\langle r^2\rangle}$]&$\times$&[2.33, 2.7]&[14.07, 1.3]&[31.64, 1.0]&[52.96, 0.8]&[74.75, 0.7]\\
\hline
\end{tabular}
\caption{Binding solutions for the individual channels in the $J^P=0^+$ and scaled-cutoff case in the OBEP model. The binding energies (B.E.) are given relative to their own thresholds. ``$\times$'' indicates that there is no binding solution. The units for the binding energy and RMS radius are MeV and fm, respectively.}\label{J0OmEP-noc2}
\end{table}

Table \ref{J0OmEP-noc2} shows our results for the case without channel coupling. For the $D$- wave channel $\Sigma_c^*N$, one gets a solution with the binding energy 0.92 MeV and the RMS radius 0.7 fm by using $\alpha=3.54$. A larger value $\alpha=3.6$ gives a solution with the binding energy 21.28 MeV and the radius 0.6 fm. Therefore we concern only S-wave states in that table, as in the case of common cutoff. By comparing these values with those in Table \ref{J0OmEP-noc}, one finds that these two parameterizations give similar results, especially for the $\Sigma_cN$ channel.

\begin{table}[htb]
\begin{tabular}{c|ccccc}\hline
$\alpha$                        &1.2     &1.4 & 1.6  & 1.8 &2.0  \\\hline
$B.E. (J=0)$ (MeV)& 0.11&5.26 &19.37& 43.05 & 75.65  \\
$\sqrt{\langle r^2\rangle}$ (fm)& 11.7 &2.0 &1.2 & 0.9 & 0.7    \\
Prob. (\%)& (99.6/0.4/0.0)&(95.8/4.1/0.1)&(89.7/10.1/0.2)&(76.9/22.5/0.6)& (77.0/22.5/0.5) \\
\hline
\end{tabular}
\caption{Binding solutions for the $J^P=0^+$ and scaled-cutoff case with channel coupling in the OBEP model. The binding energies (B.E.) are given relative to the $\Lambda_cN$ threshold. The probabilities correspond to $\Lambda_cN(^1S_0)$, $\Sigma_cN(^1S_0)$, and $\Sigma_c^*N(^5D_0)$, respectively.}\label{J0OmEP2}
\end{table}

\begin{figure}
\centering
\begin{tabular}{cc}
\scalebox{0.5}{\includegraphics{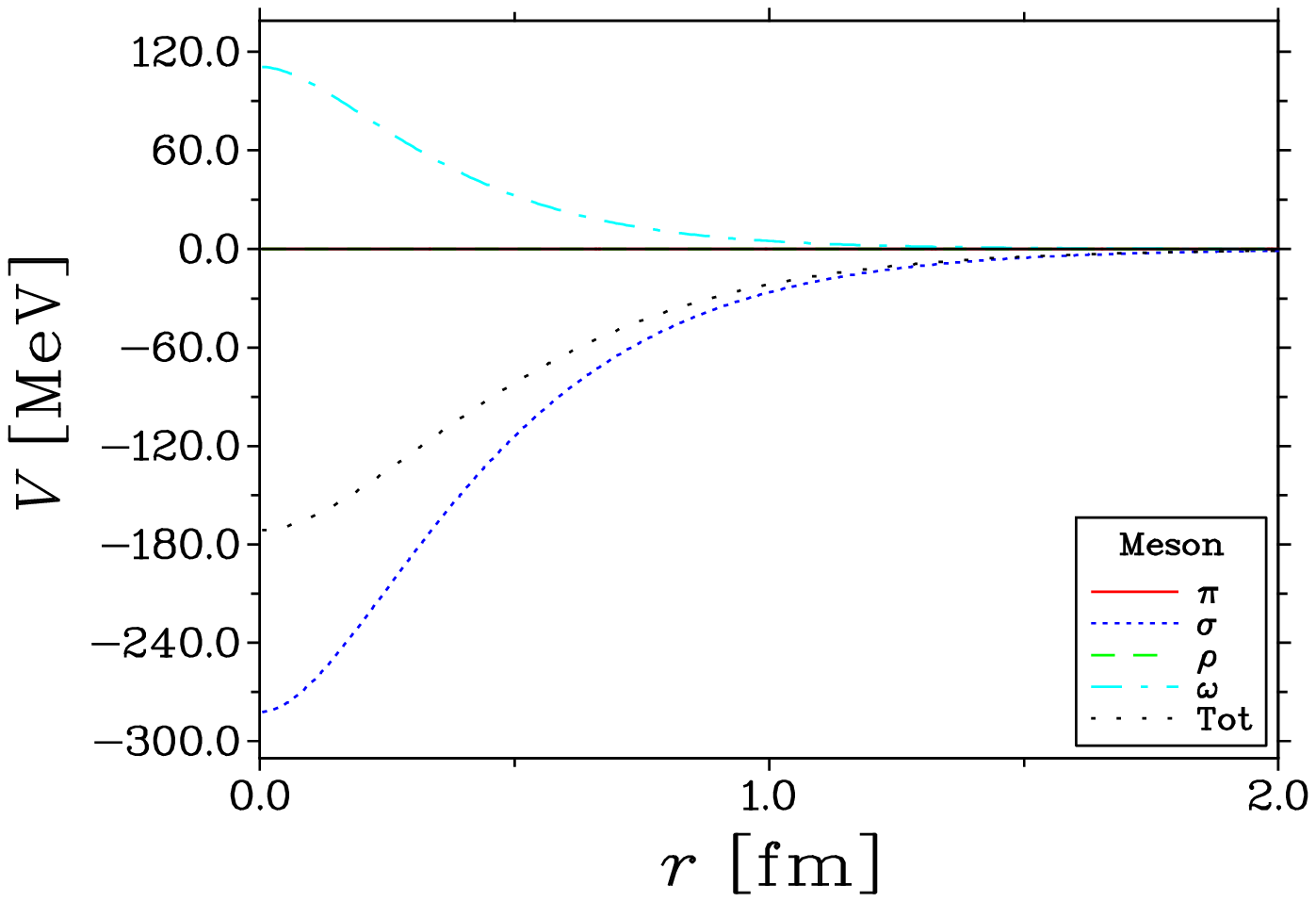}}&
\scalebox{0.5}{\includegraphics{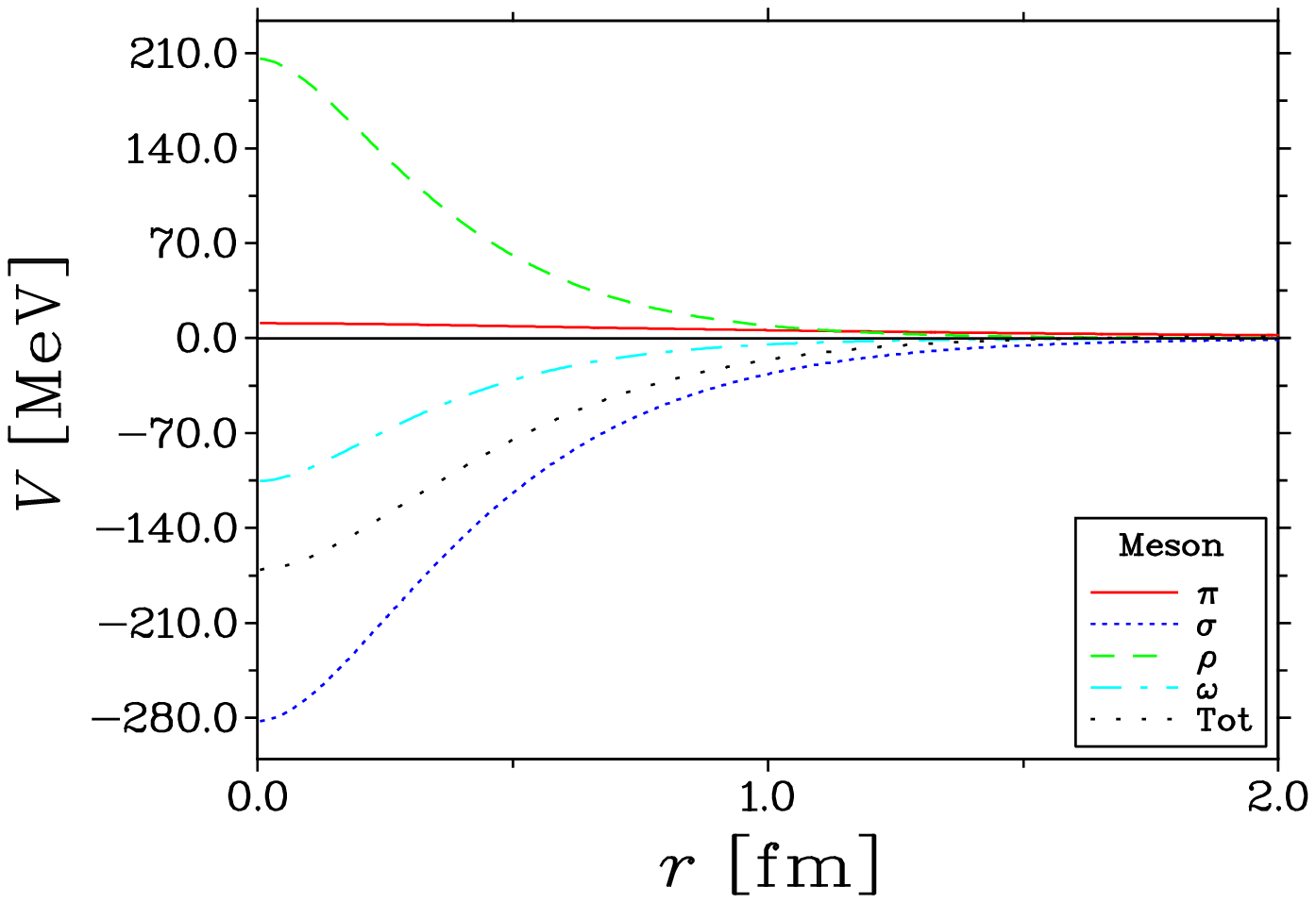}}\\
(11): $\Lambda_cN(^1S_0)\leftrightarrow\Lambda_cN(^1S_0)$&
(22): $\Sigma_cN(^1S_0)\leftrightarrow\Sigma_cN(^1S_0)$\\
\scalebox{0.5}{\includegraphics{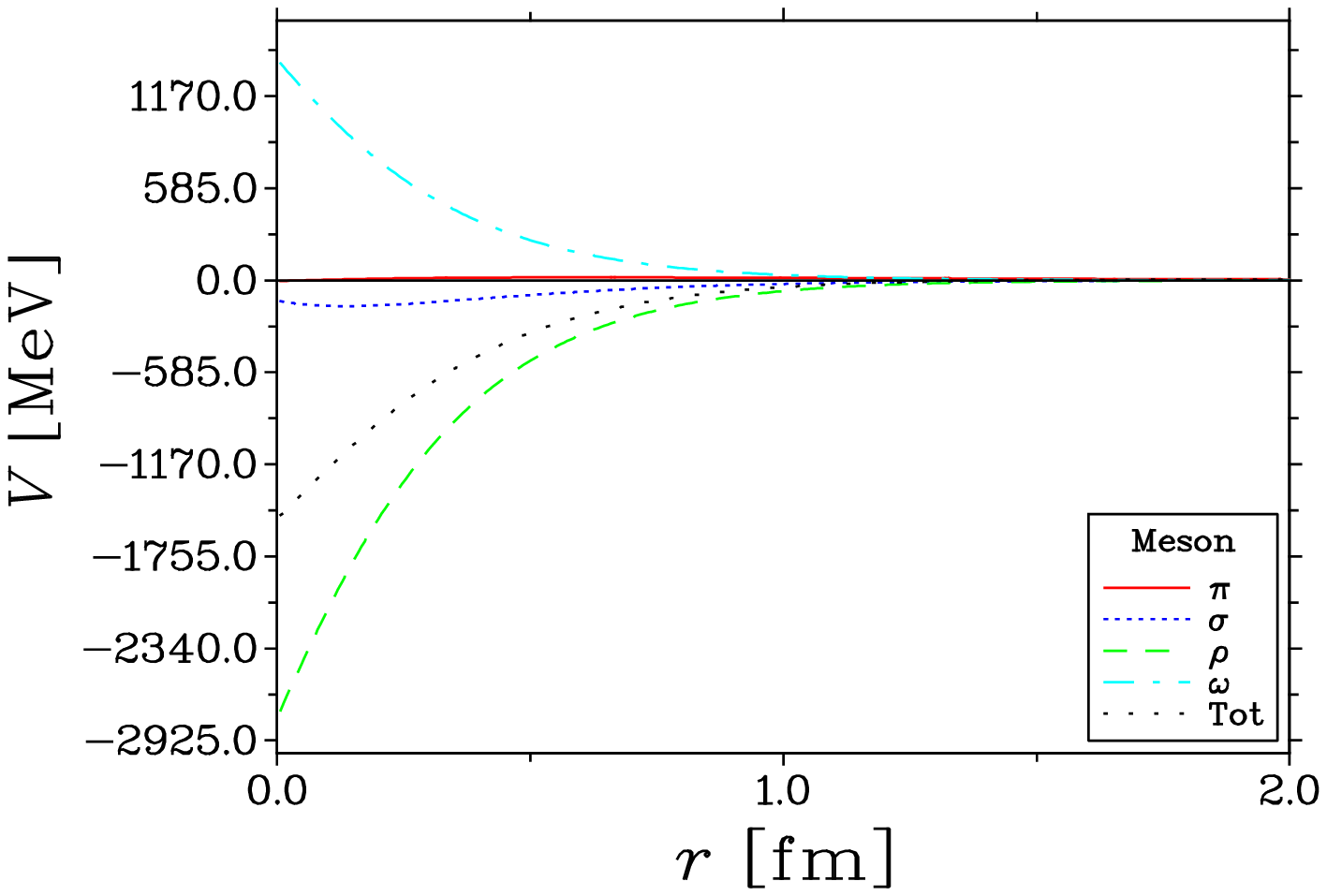}}&
\scalebox{0.5}{\includegraphics{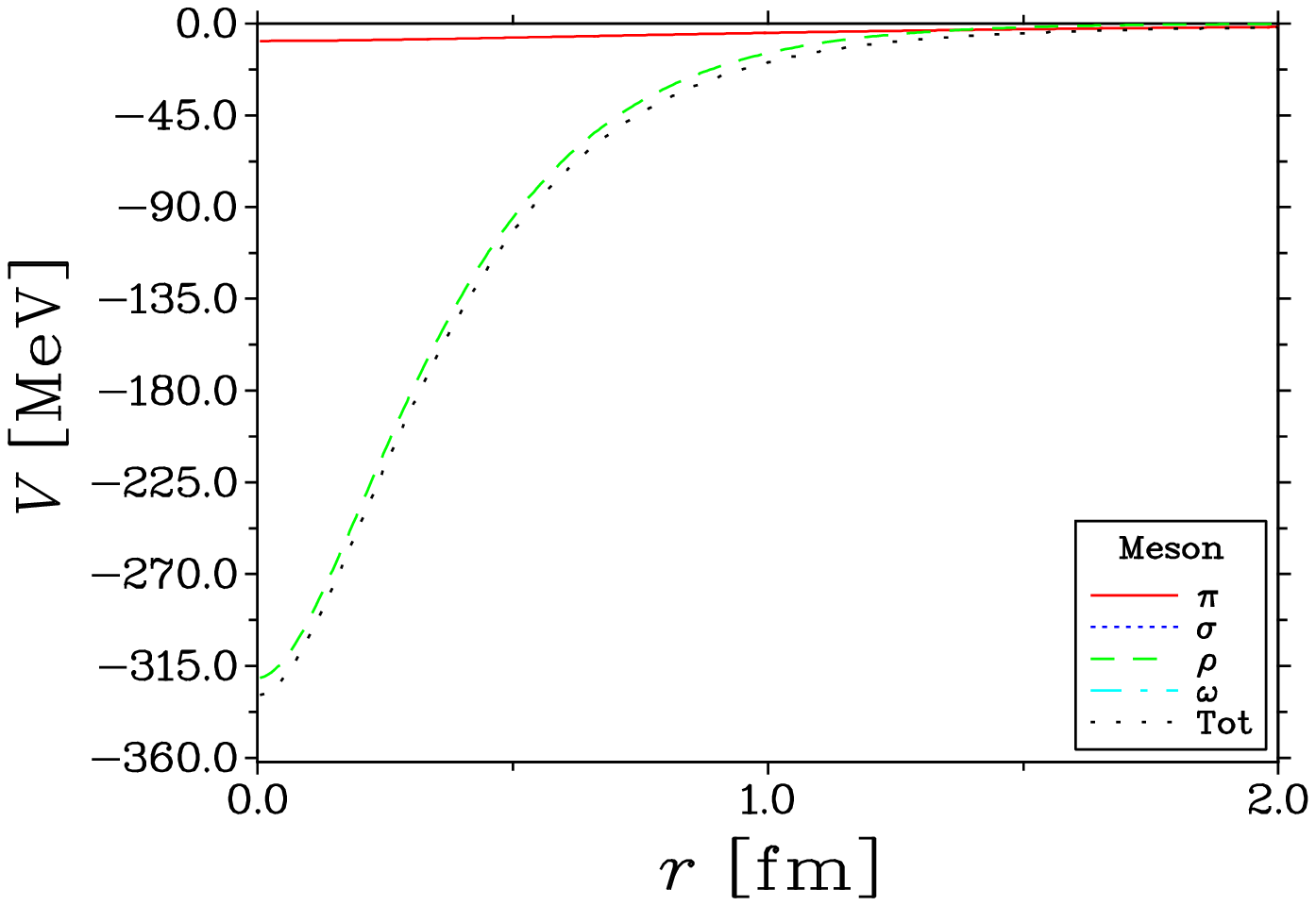}}\\
(33): $\Sigma_c^*N(^5D_0)\leftrightarrow\Sigma_c^*N(^5D_0)$&
(12): $\Lambda_cN(^1S_0)\leftrightarrow\Sigma_cN(^1S_0)$\\
\scalebox{0.5}{\includegraphics{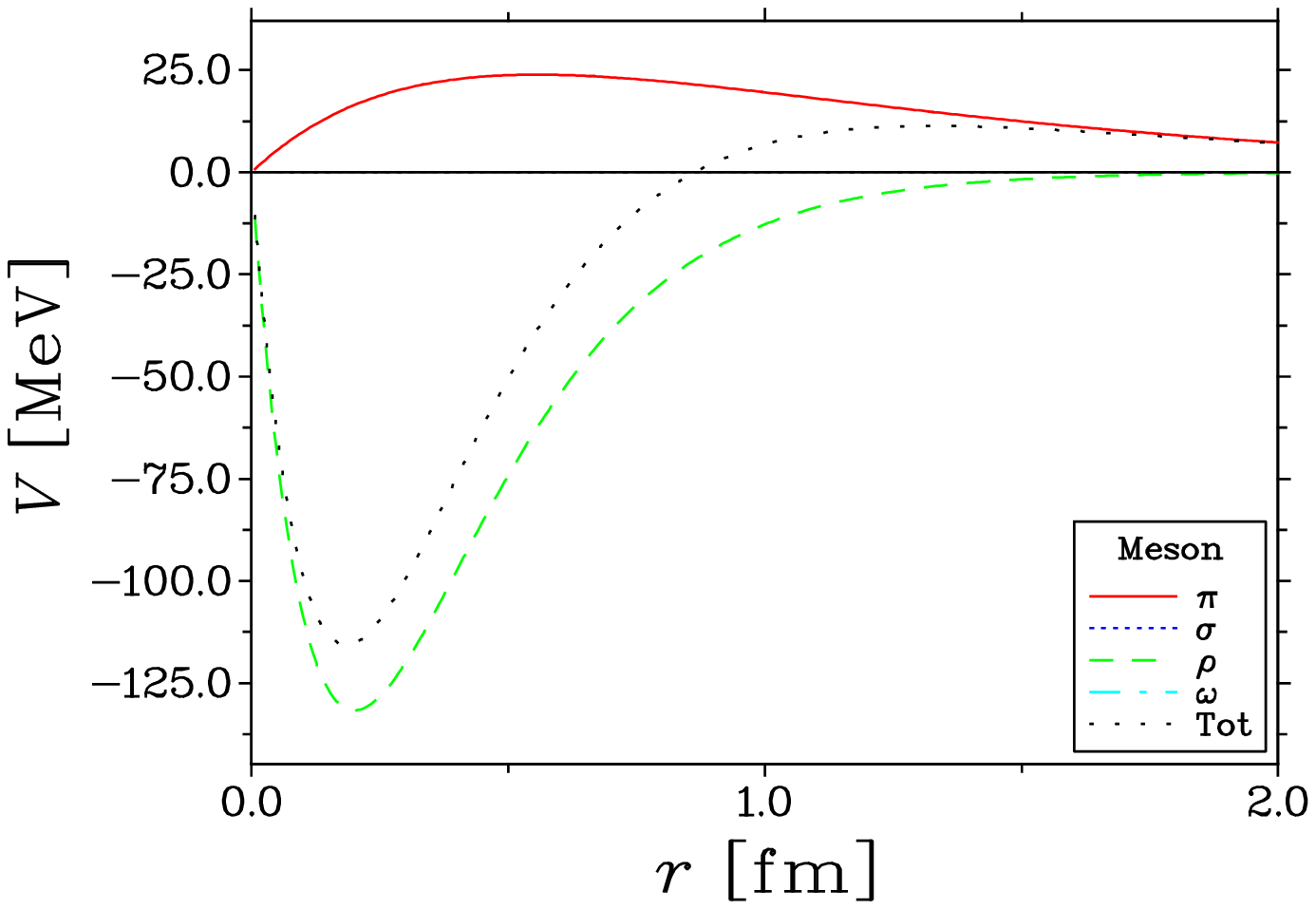}}&
\scalebox{0.5}{\includegraphics{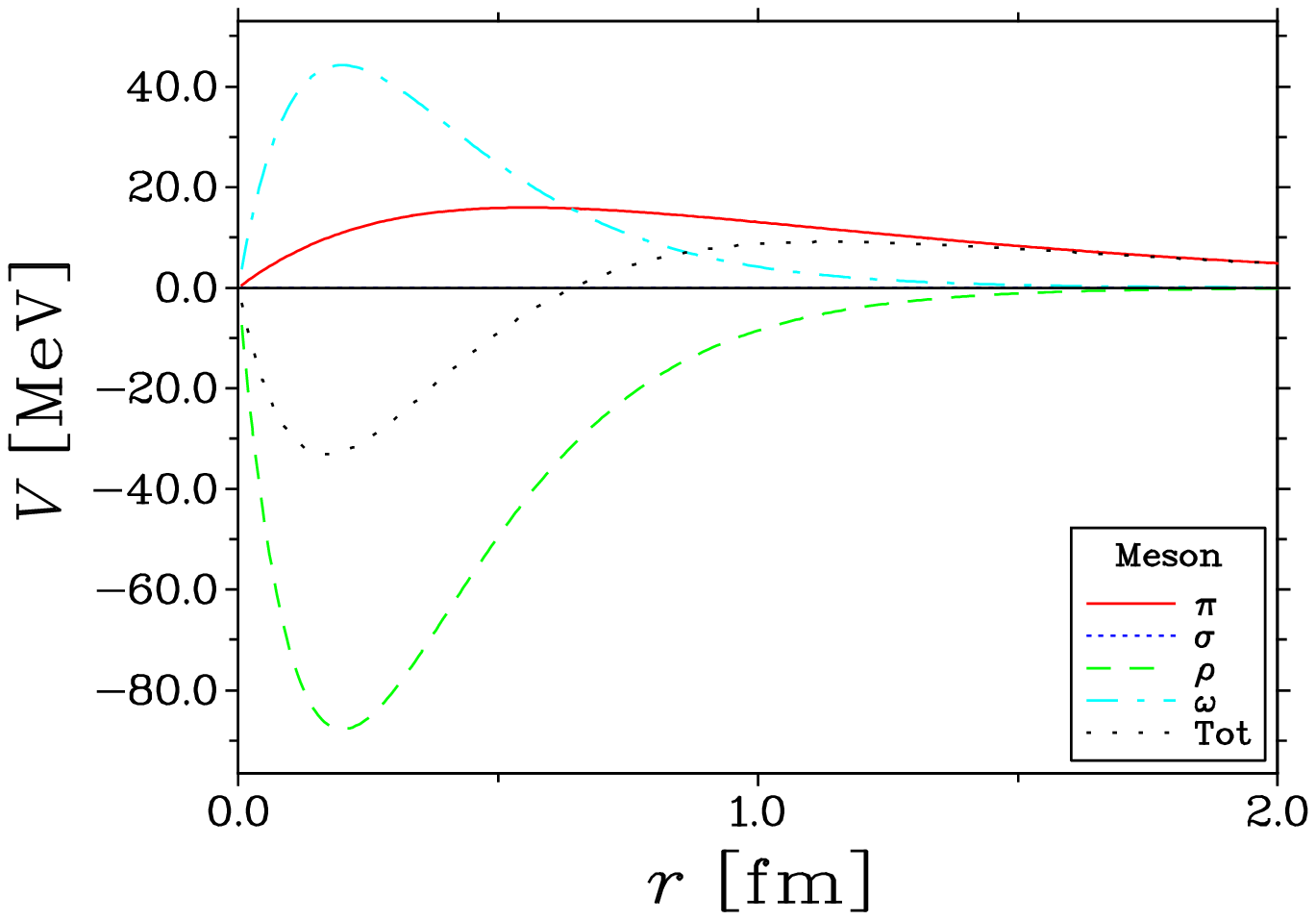}}\\
(13): $\Lambda_cN(^1S_0)\leftrightarrow\Sigma_c^*N(^5D_0)$&
(23): $\Sigma_cN(^1S_0)\leftrightarrow\Sigma_c^*N(^5D_0)$
\end{tabular}
\caption{The potentials of different channels for the $J^P=0^+$ case with $\alpha=1.4$.}\label{potential-J0alf}
\end{figure}

For the case with channel coupling, we give the numerical results in Table \ref{J0OmEP2}. From this table and Table \ref{J0OmEP}, although the binding energies and the radii in these two parametrization methods are consistent and somehow equivalent, while the probabilities of individual channels are not. The probability depends on the contributions from different exchanged mesons. It is helpful to understand this feature from the wave functions and the potentials. We show the wave functions of different channels with $\alpha=1.4$ in Fig. \ref{WAVEu} (c). There is a node in the wave function of the third channel $\Sigma_c^*N(^5D_0)$. The node does not disappear for the other $\alpha$. Actually, such a behavior results from the transition potential between the first channel and the third channel as well as that between the second channel and the third channel. These potentials have different signs in the long-range part and the short-range part. As an example, we plot the potentials of this parametrization in Fig. \ref{potential-J0alf} with $\alpha=1.4$.

\section{$J^P=1^+$ case}\label{sec6}

The spin-triplet case is a little complicated because there are four more channels than the spin-singlet case. We carry out similar analysis in this section. Before the numerical evaluation, we plot the potentials with the cutoffs $\Lambda_\pi=\Lambda_\sigma=\Lambda_{\rm vec}=1$ GeV, which are shown in Fig. \ref{potential-J1}. The potentials for the channel $\Lambda_cN(^3S_1)\leftrightarrow \Lambda_cN(^3S_1)$ and the channel $\Lambda_cN(^1S_0)\leftrightarrow \Lambda_cN(^1S_0)$ are identical. There is no transition potential $\Lambda_cN(^3S_1)\leftrightarrow \Lambda_cN(^3D_1)$, which is omitted in Fig. \ref{potential-J1}.

\begin{figure}[htb]
\centering
\subfigure{\label{potential-J1-1st}
\begin{tabular}{cc}
\scalebox{0.6}{\includegraphics{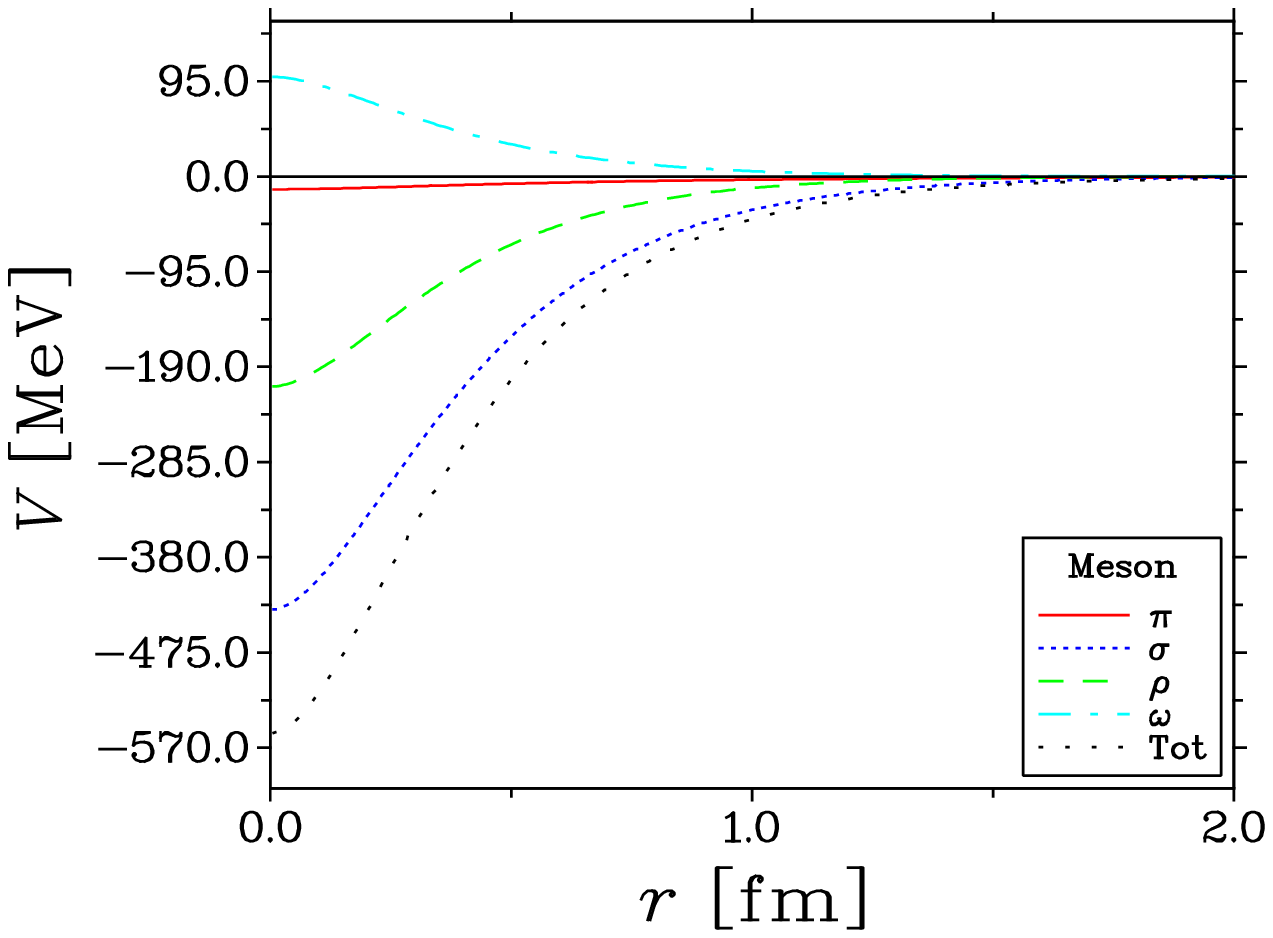}}&\scalebox{0.6}{\includegraphics{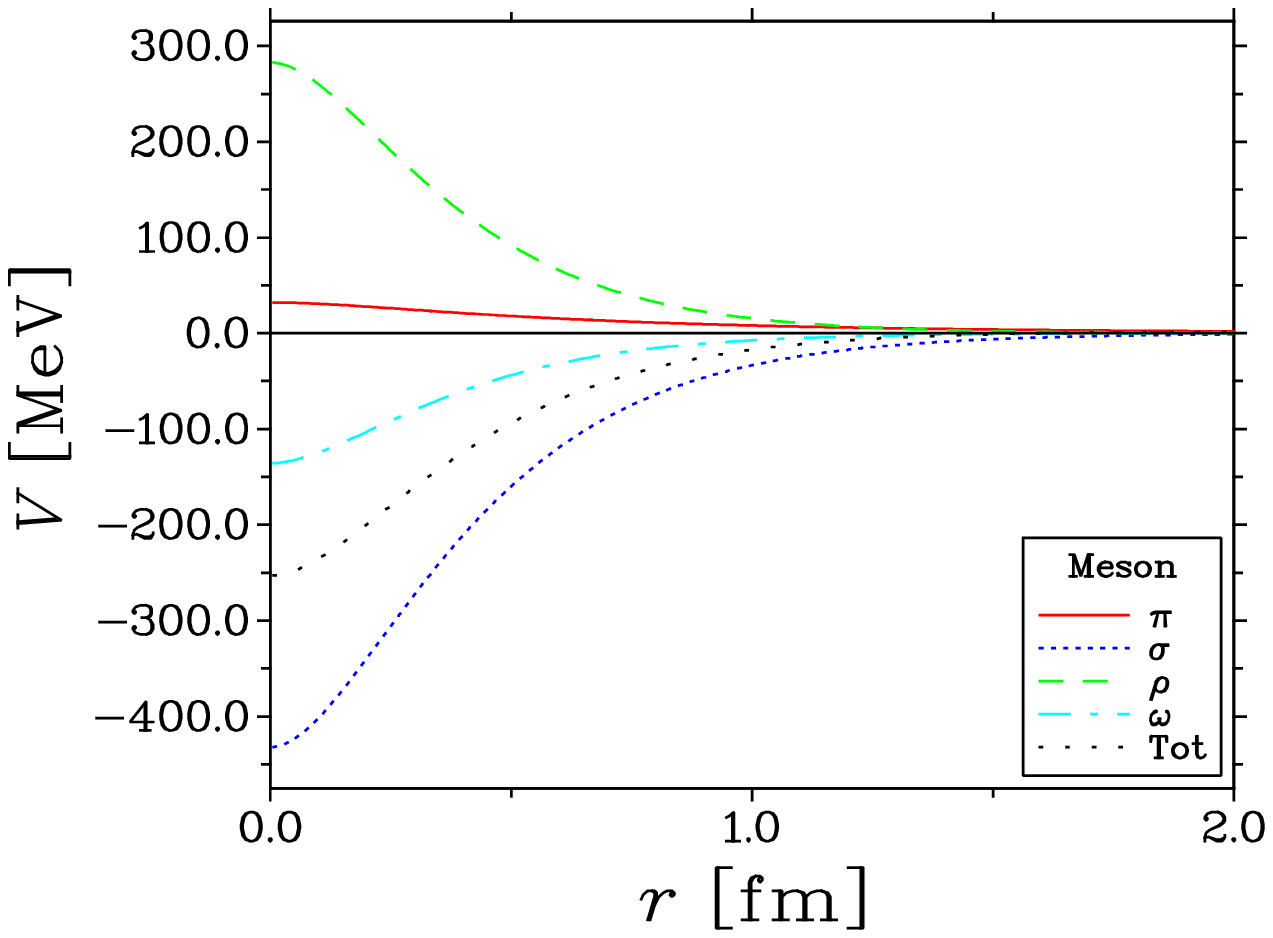}}\\
(22): $\Sigma_cN(^3S_1)\leftrightarrow\Sigma_cN(^3S_1)$&
(33): $\Sigma_c^*N(^3S_1)\leftrightarrow\Sigma_c^*N(^3S_1)$\\
\scalebox{0.6}{\includegraphics{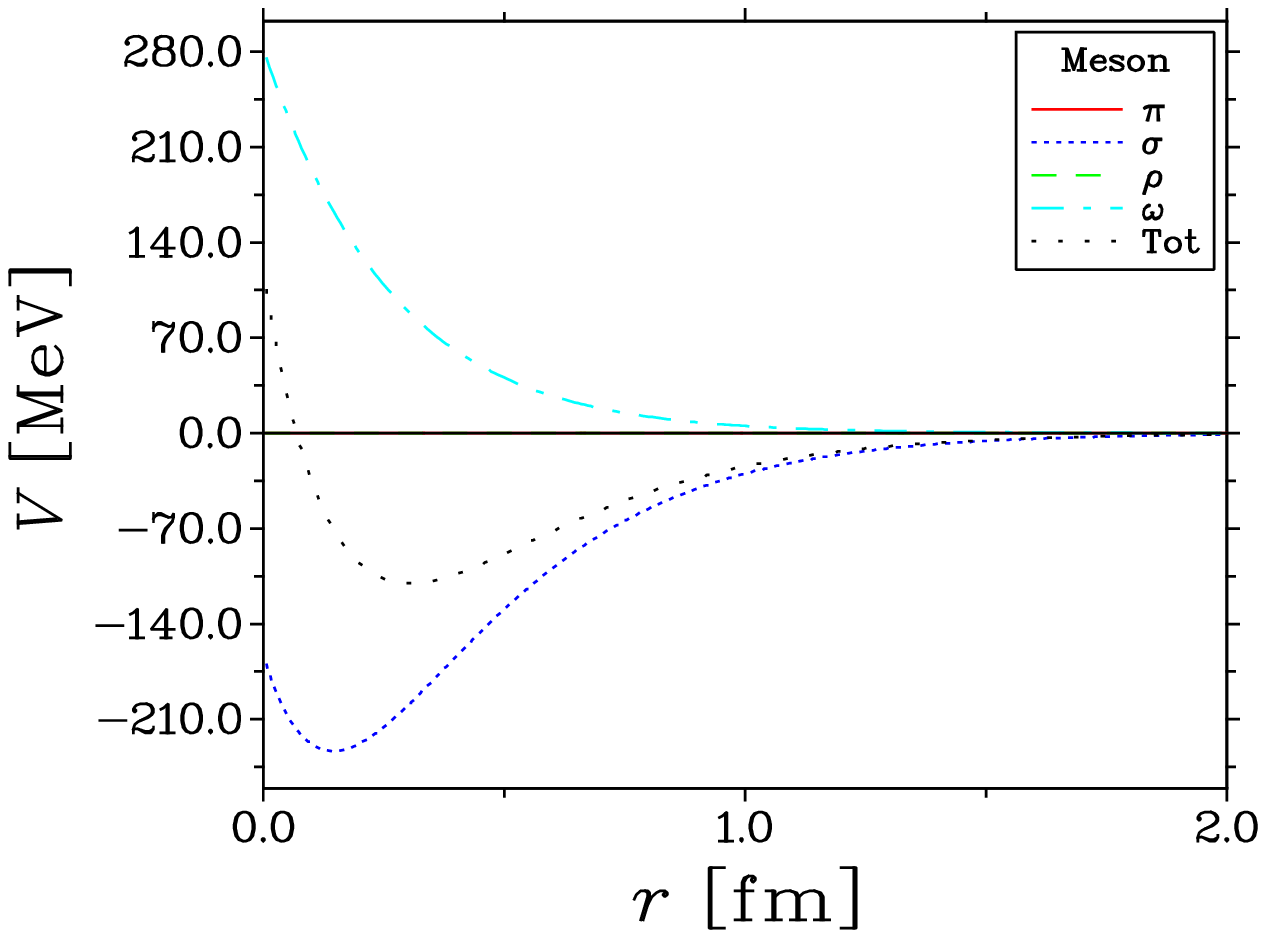}}&\scalebox{0.6}{\includegraphics{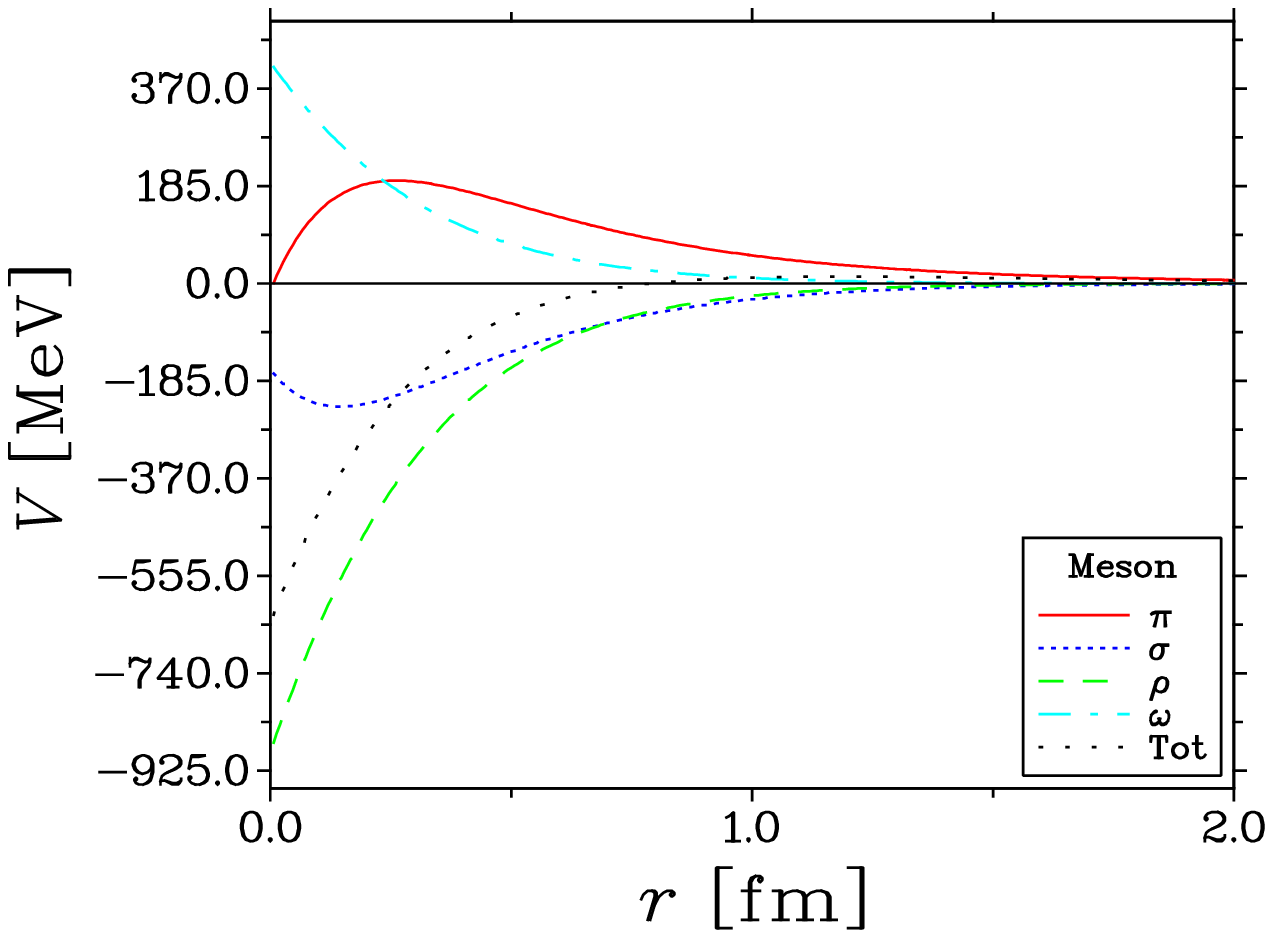}}\\
(44): $\Lambda_cN(^3D_1)\leftrightarrow\Lambda_cN(^3D_1)$&
(55): $\Sigma_cN(^3D_1)\leftrightarrow\Sigma_cN(^3D_1)$\\
\scalebox{0.6}{\includegraphics{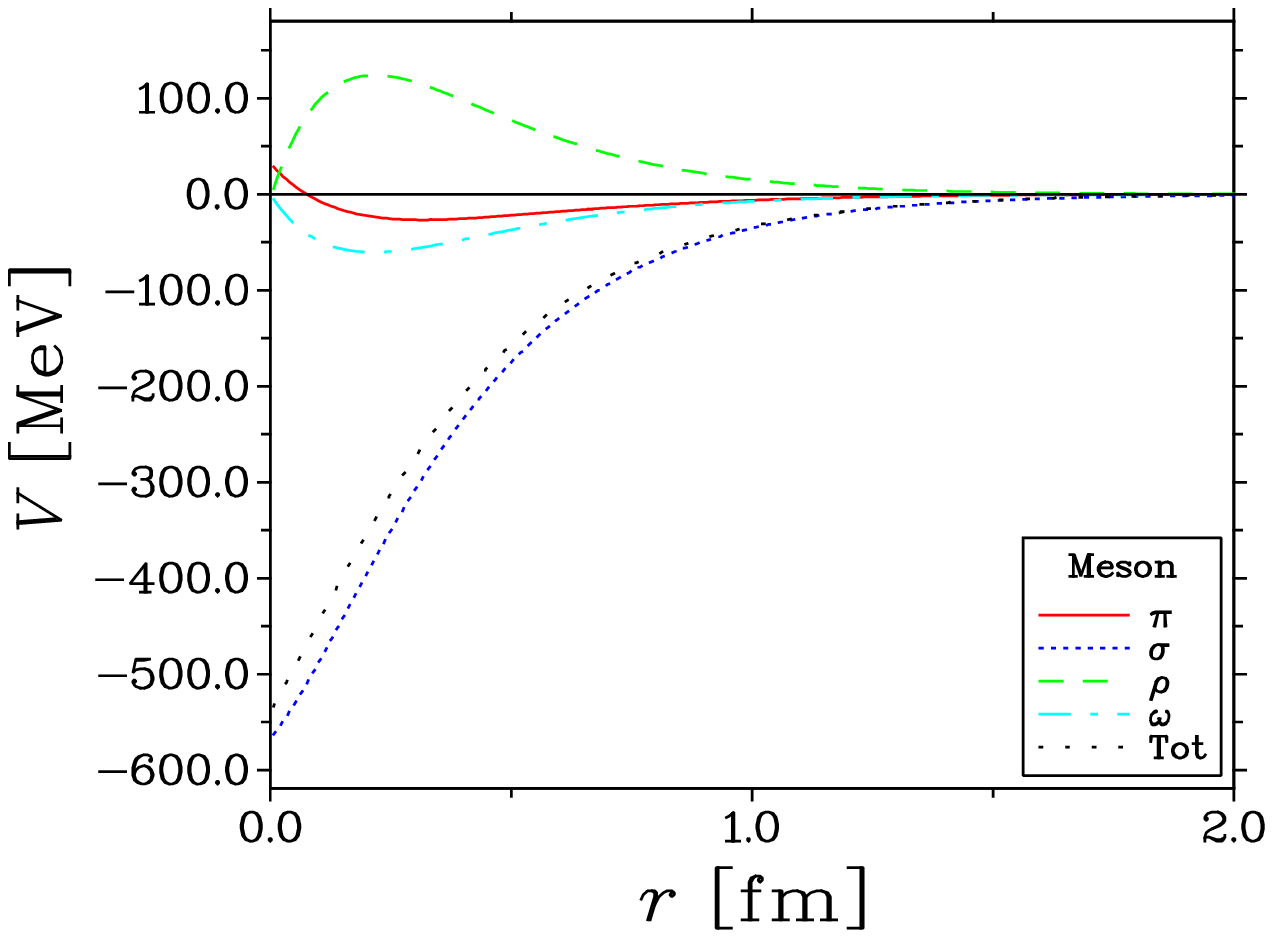}}&\scalebox{0.6}{\includegraphics{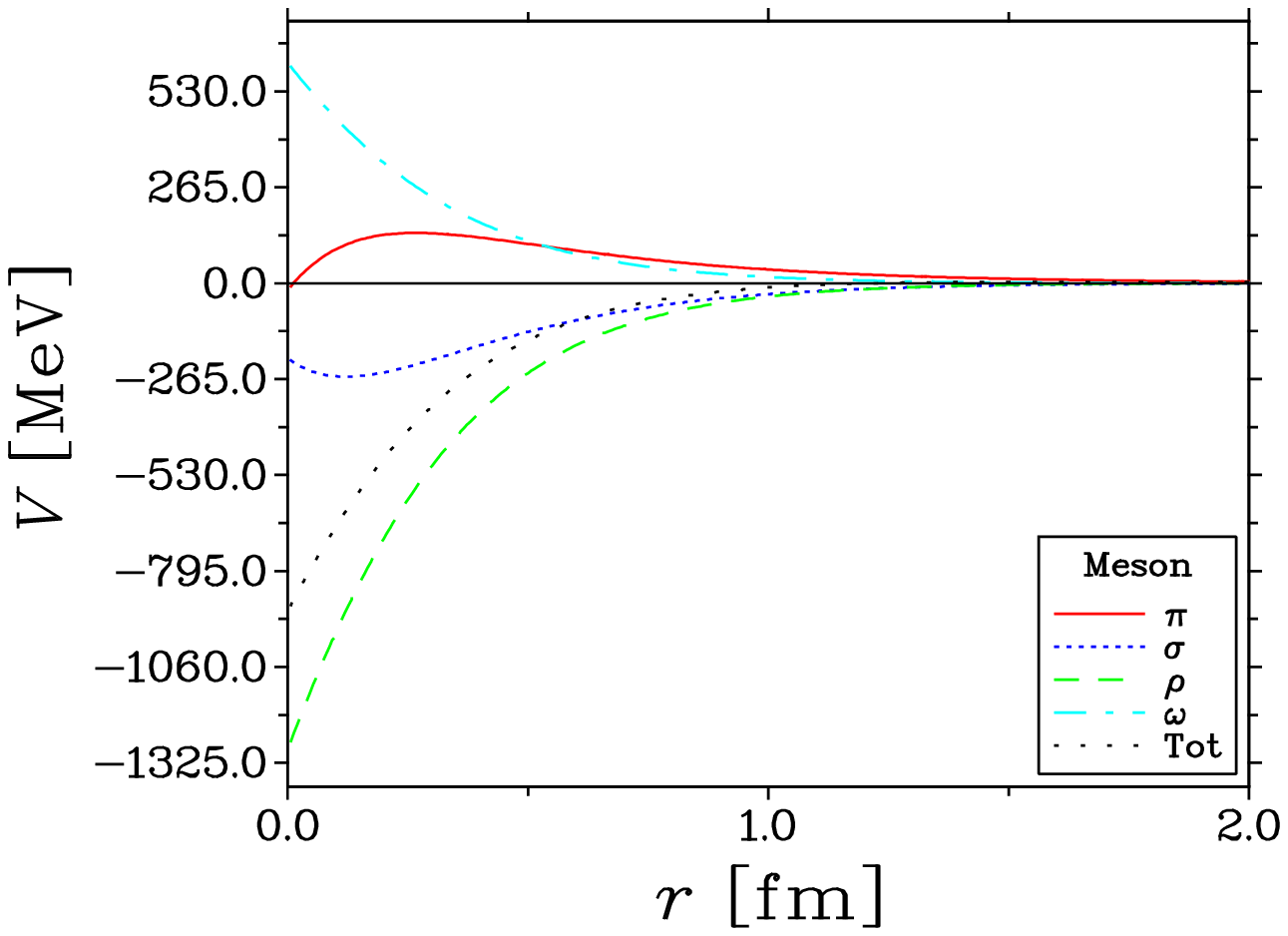}}\\
(66): $\Sigma_c^*N(^3D_1)\leftrightarrow\Sigma_c^*N(^3D_1)$&
(77): $\Sigma_c^*N(^5D_1)\leftrightarrow\Sigma_c^*N(^5D_1)$\\
\end{tabular}}
\caption{The potentials of different channels for the $J^P=1^+$ case with $\Lambda_\pi=\Lambda_\sigma=\Lambda_{\rm vec}=1$ GeV. }\label{potential-J1}
\end{figure}
\addtocounter{figure}{-1}
\begin{figure}[htb]
\addtocounter{subfigure}{1}
\centering
\subfigure{\label{potential-J1-2nd}
\begin{tabular}{cc}
\scalebox{0.6}{\includegraphics{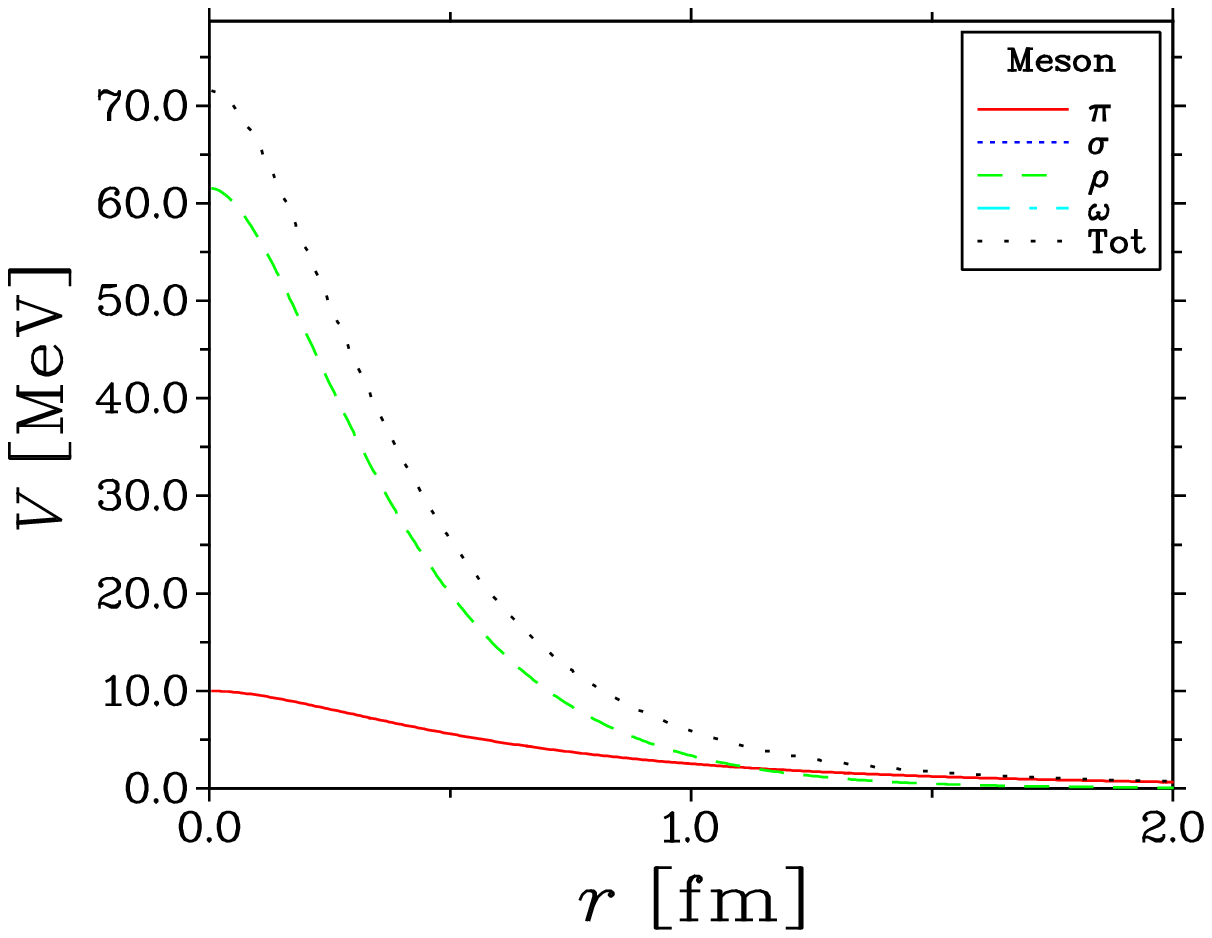}}&\scalebox{0.6}{\includegraphics{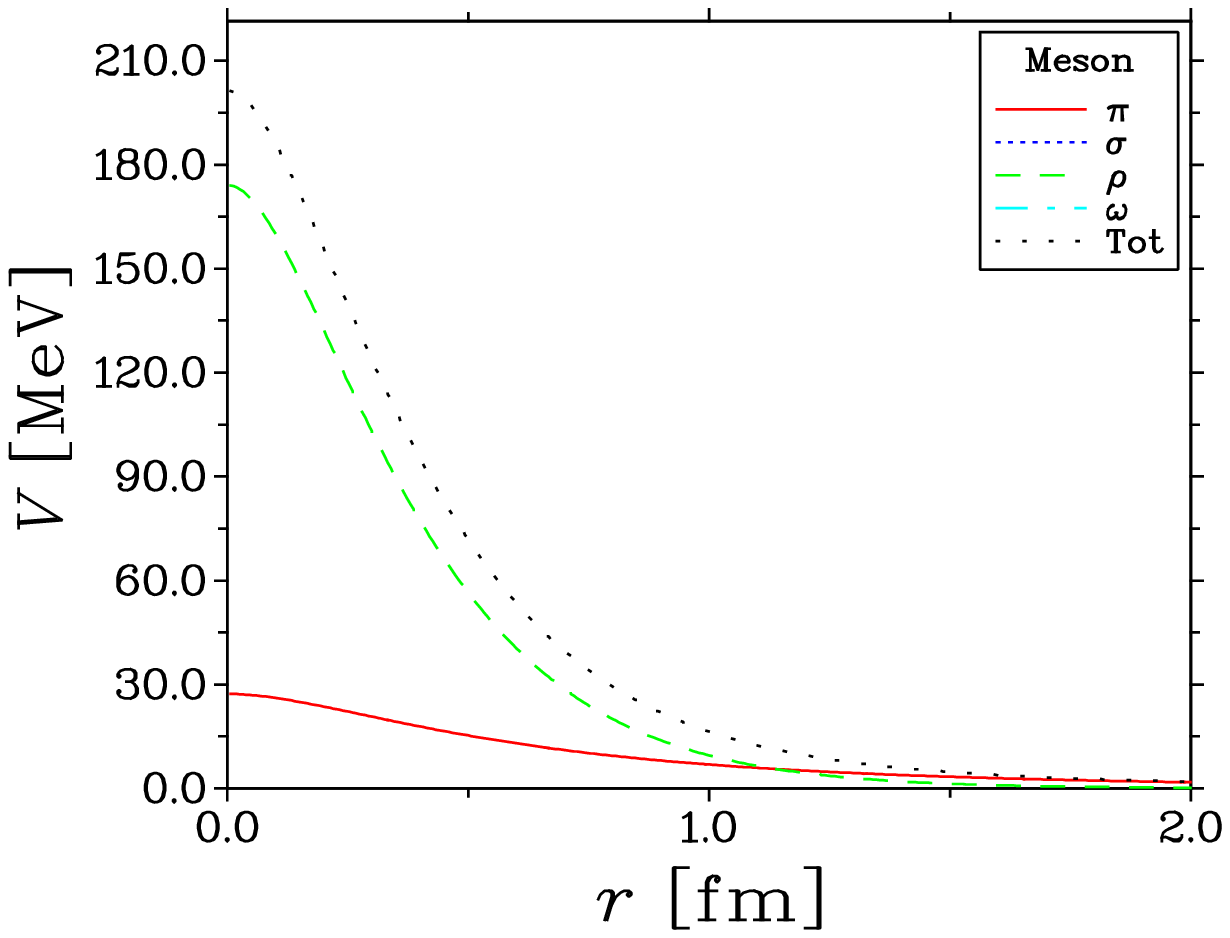}}\\
(12): $\Lambda_cN(^3S_1)\leftrightarrow\Sigma_cN(^3S_1)$&
(13): $\Lambda_cN(^3S_1)\leftrightarrow\Sigma_c^*N(^3S_1)$\\
\scalebox{0.6}{\includegraphics{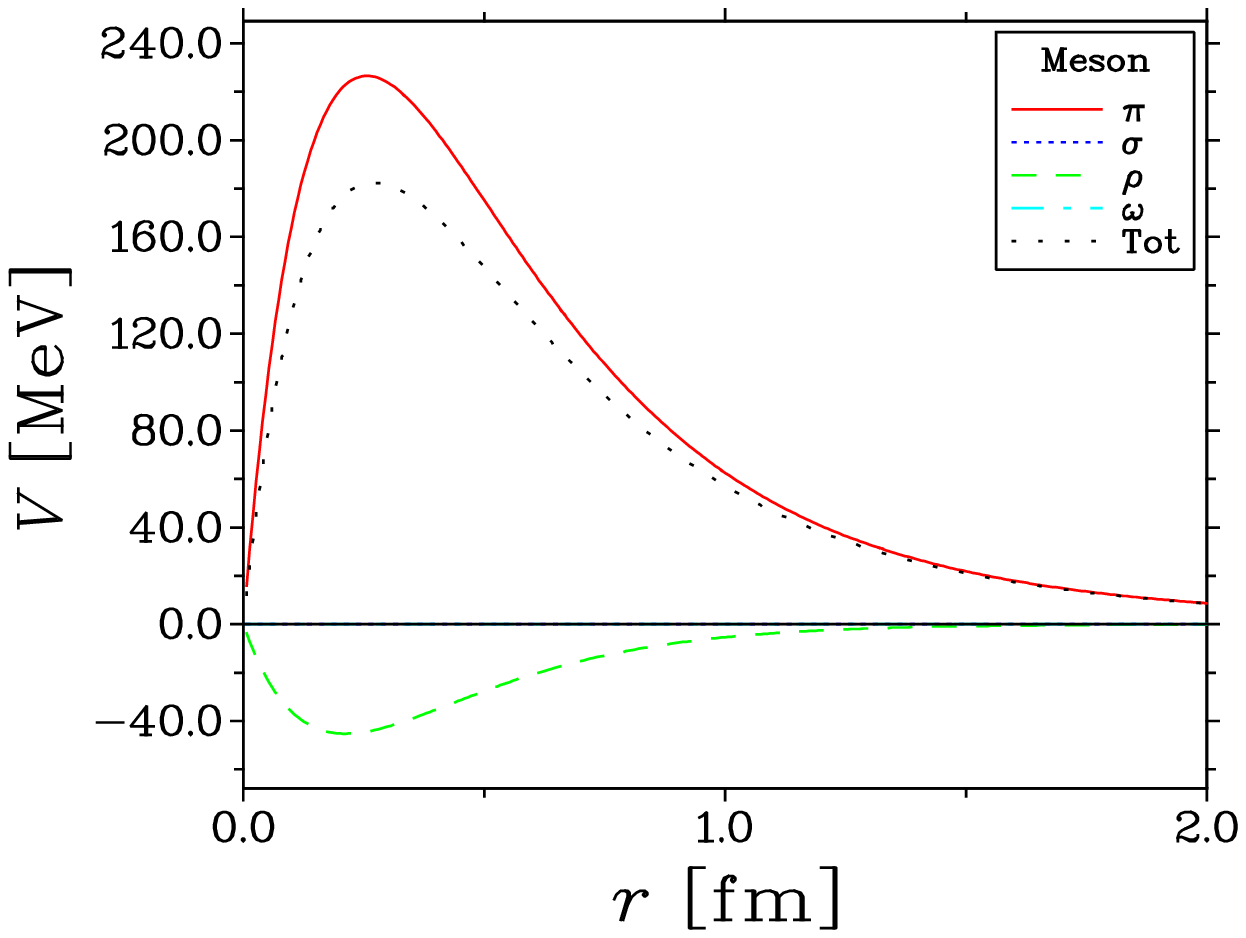}}&\scalebox{0.6}{\includegraphics{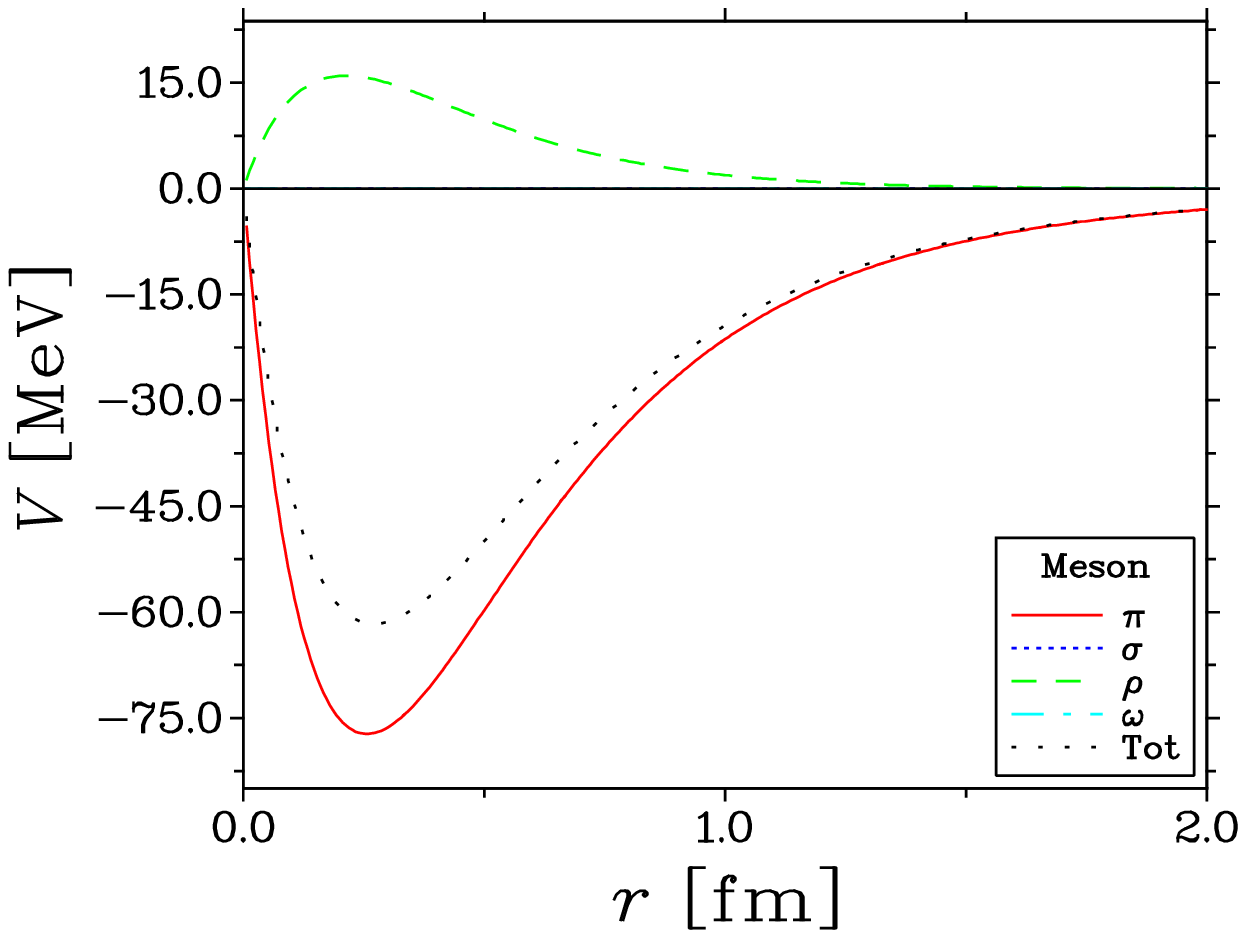}}\\
(15): $\Lambda_cN(^3S_1)\leftrightarrow\Sigma_cN(^3D_1)$&
(16): $\Lambda_cN(^3S_1)\leftrightarrow\Sigma_c^*N(^3D_1)$\\
\scalebox{0.6}{\includegraphics{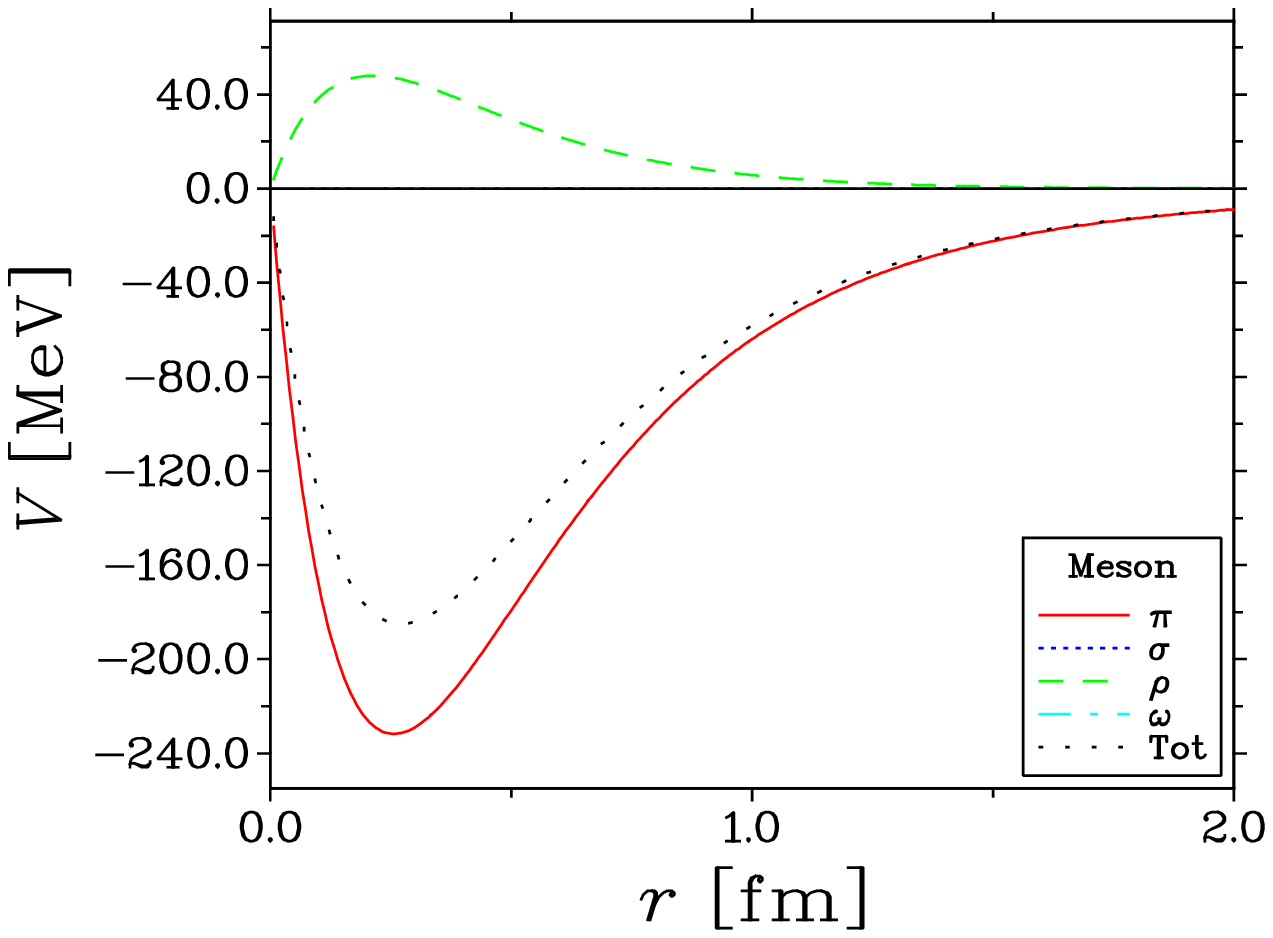}}&\scalebox{0.6}{\includegraphics{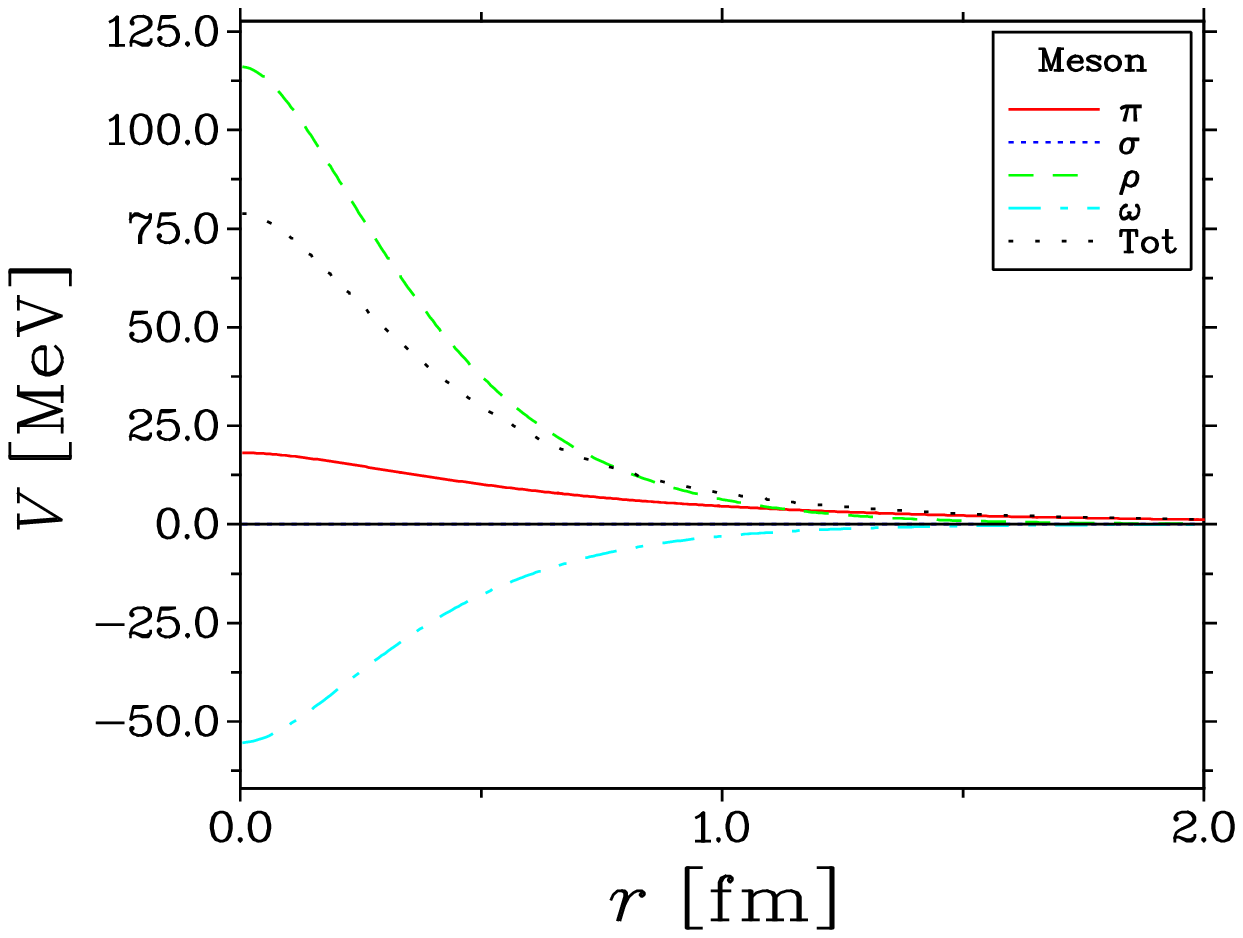}}\\
(17): $\Lambda_cN(^3S_1)\leftrightarrow\Sigma_c^*N(^5D_1)$&
(23): $\Sigma_cN(^3S_1)\leftrightarrow\Sigma_c^*N(^3S_1)$\\
\scalebox{0.6}{\includegraphics{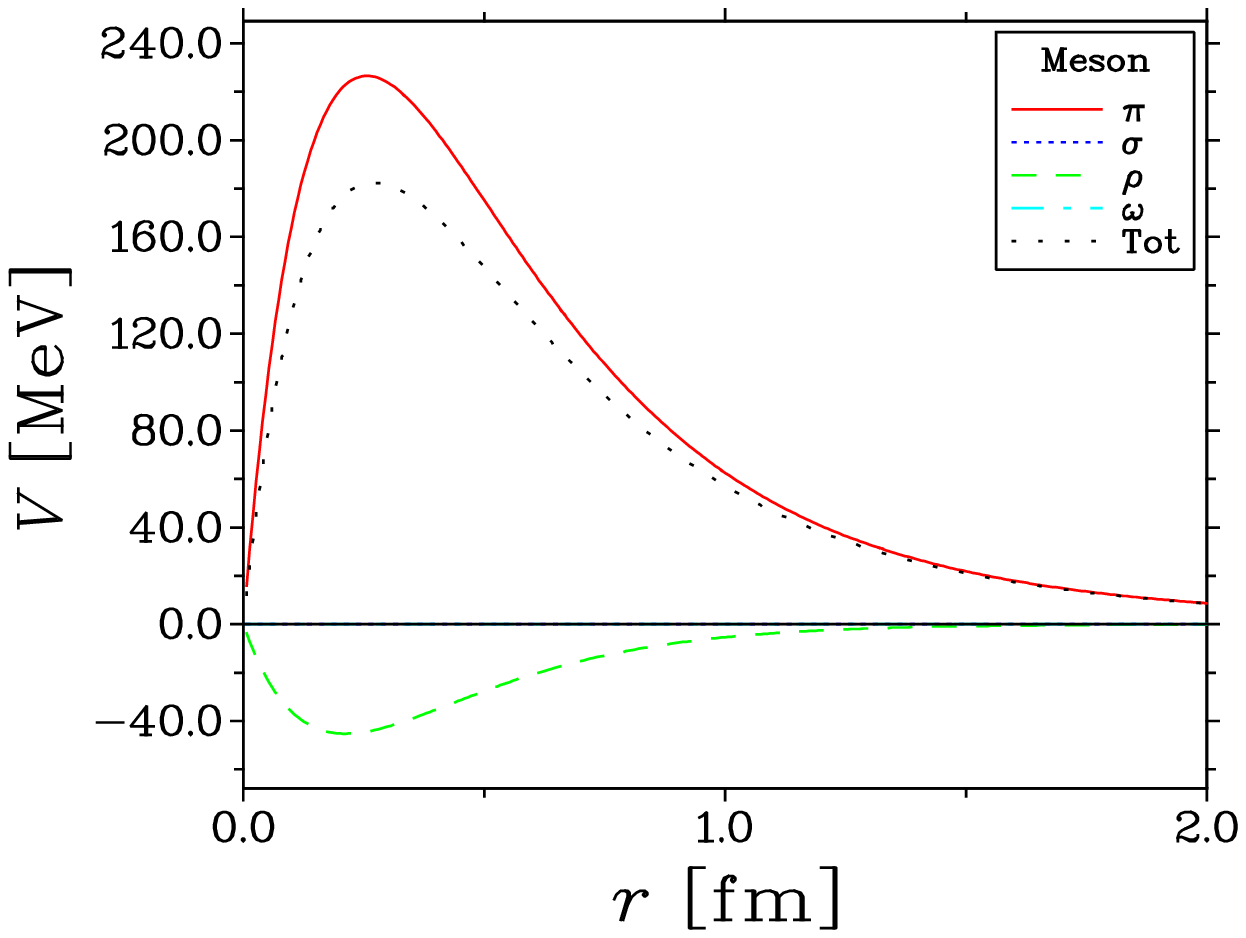}}&\scalebox{0.6}{\includegraphics{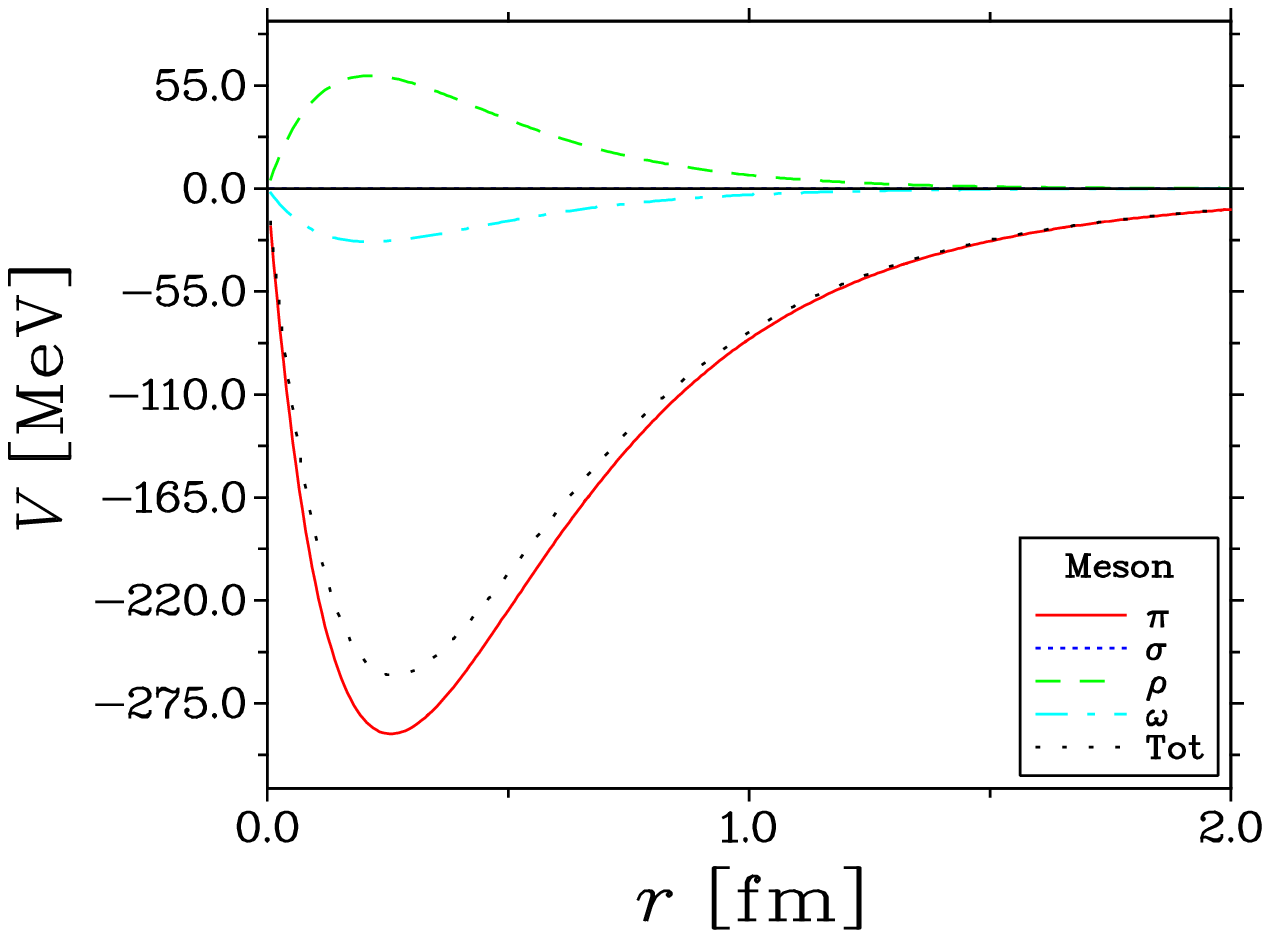}}
\\
(24): $\Sigma_cN(^3S_1)\leftrightarrow\Lambda_cN(^3D_1)$&
(25): $\Sigma_cN(^3S_1)\leftrightarrow\Sigma_cN(^3D_1)$\\
\end{tabular}}
\caption{The potentials of different channels for the $J^P=1^+$ case with $\Lambda_\pi=\Lambda_\sigma=\Lambda_{\rm vec}=1$ GeV. (cont.)}\label{potential-J1}
\end{figure}
\addtocounter{figure}{-1}
\begin{figure}[htb]
\addtocounter{subfigure}{2}
\centering
\subfigure{\label{potential-J1-3rd}
\begin{tabular}{cc}
\scalebox{0.6}{\includegraphics{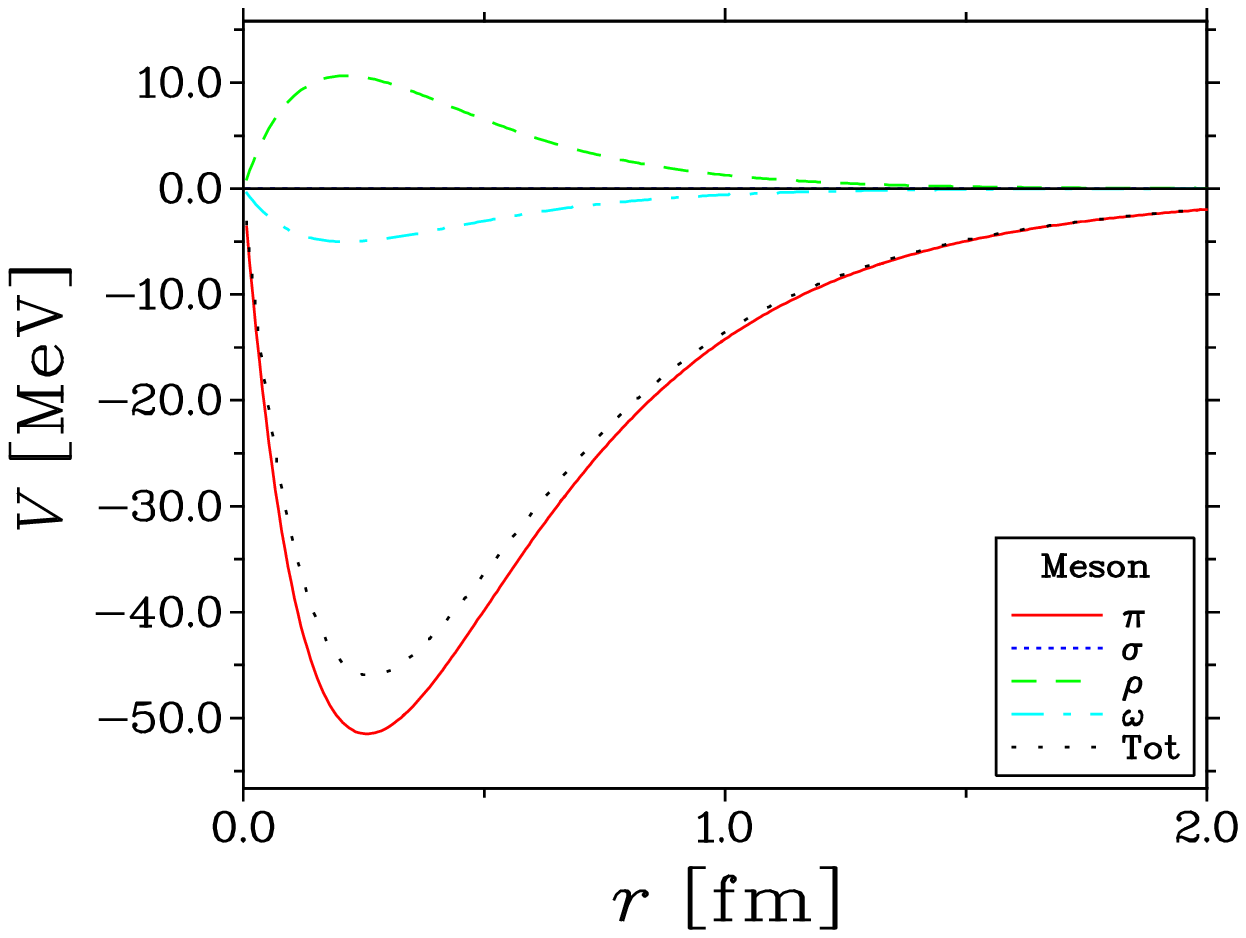}}&\scalebox{0.6}{\includegraphics{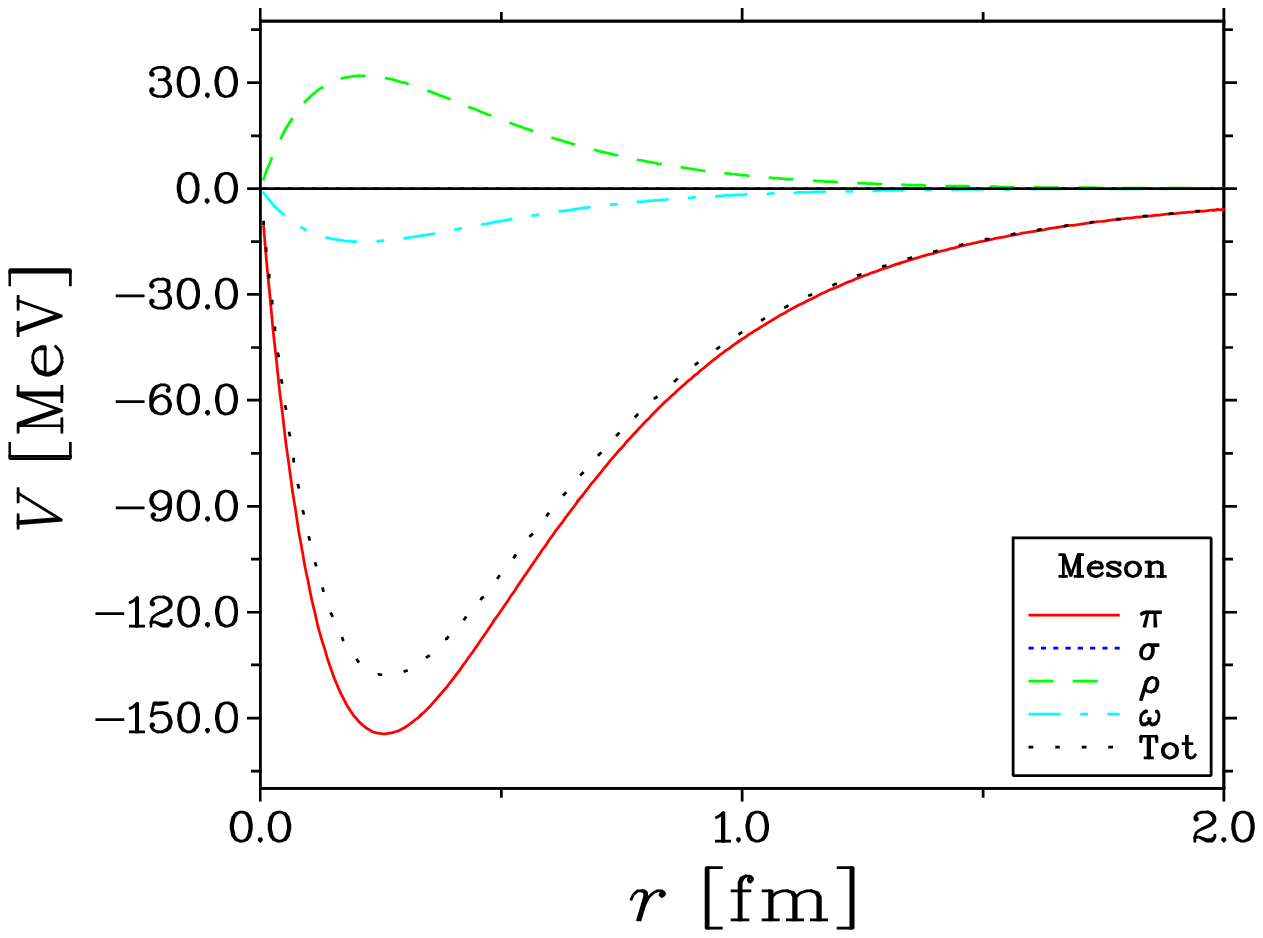}}\\
(26): $\Sigma_cN(^3S_1)\leftrightarrow\Sigma_c^*N(^3D_1)$&
(27): $\Sigma_cN(^3S_1)\leftrightarrow\Sigma_c^*N(^5D_1)$\\
\scalebox{0.6}{\includegraphics{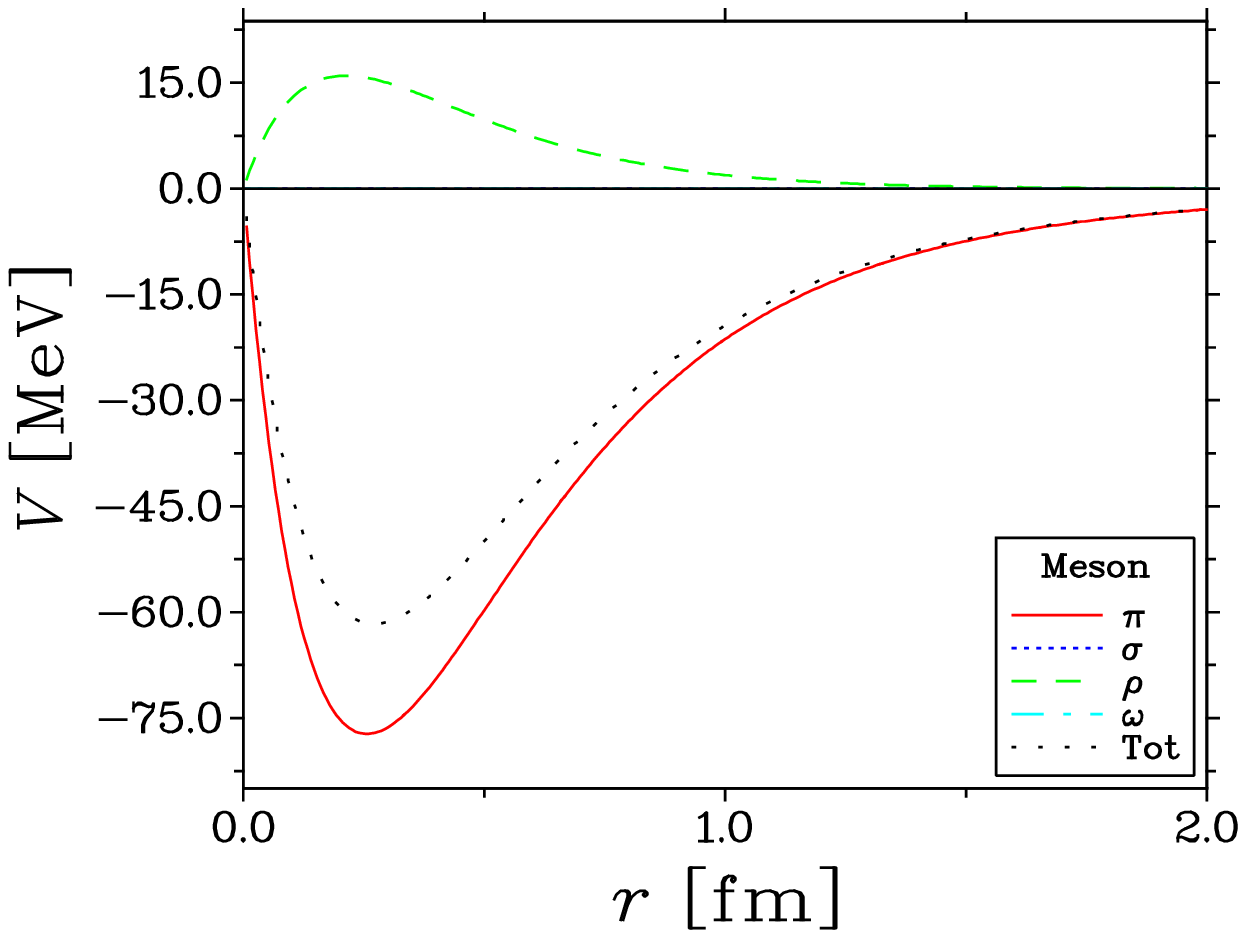}}&\scalebox{0.6}{\includegraphics{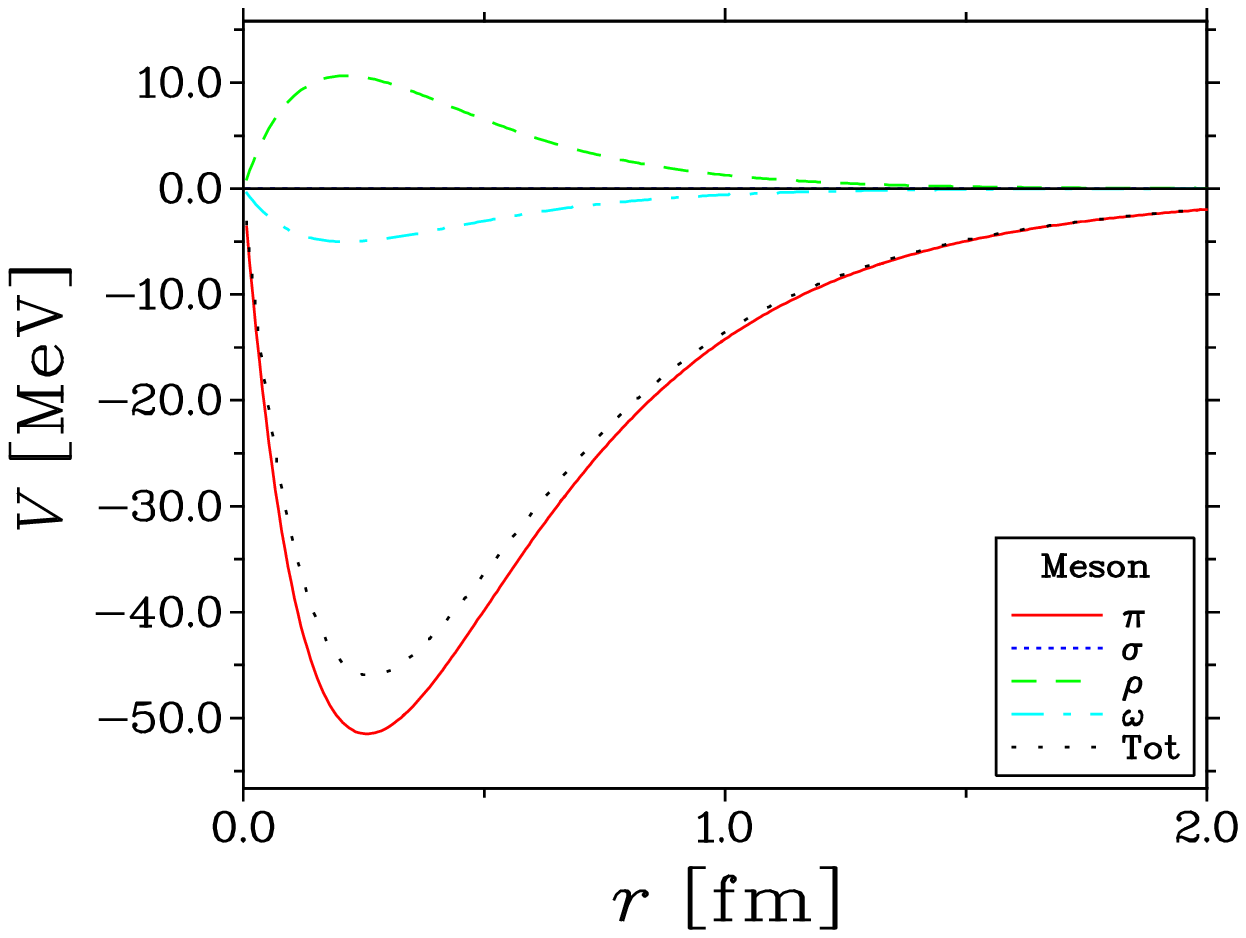}}\\
(34): $\Sigma_c^*N(^3S_1)\leftrightarrow\Lambda_cN(^3D_1)$&
(35): $\Sigma_c^*N(^3S_1)\leftrightarrow\Sigma_cN(^3D_1)$\\
\scalebox{0.6}{\includegraphics{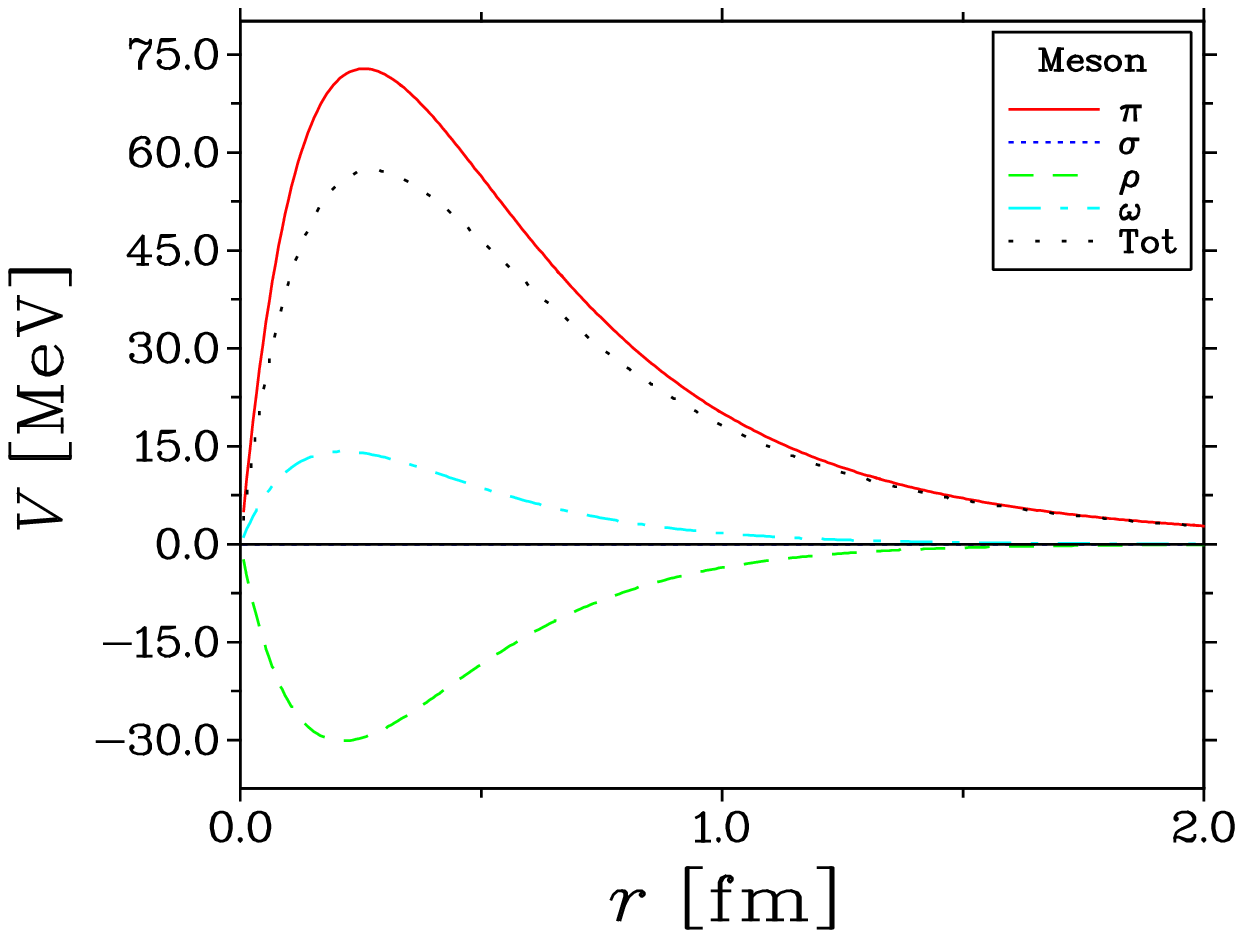}}&\scalebox{0.6}{\includegraphics{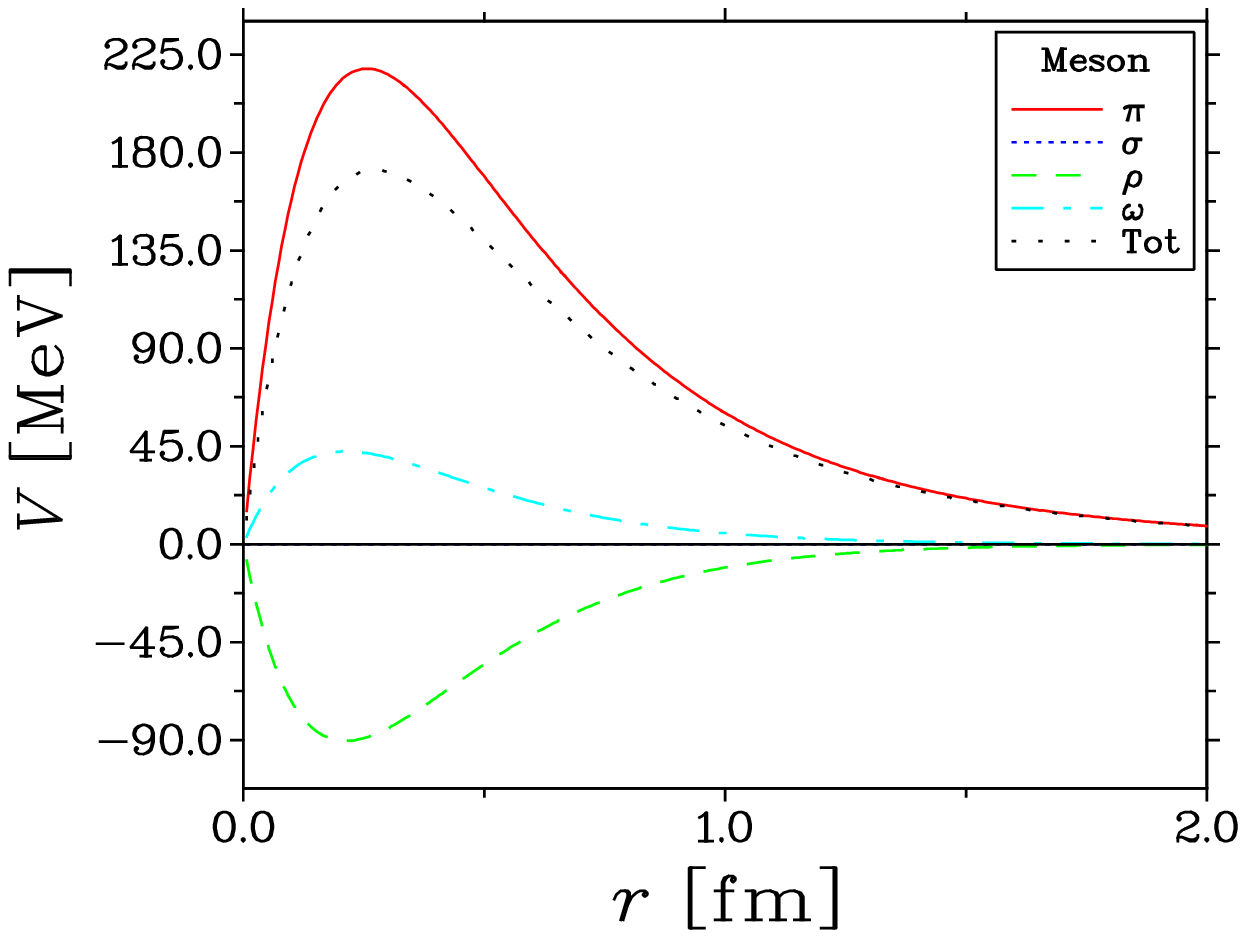}}\\
(36): $\Sigma_c^*N(^3S_1)\leftrightarrow\Sigma_c^*N(^3D_1)$&
(37): $\Sigma_c^*N(^3S_1)\leftrightarrow\Sigma_c^*N(^5D_1)$\\
\scalebox{0.6}{\includegraphics{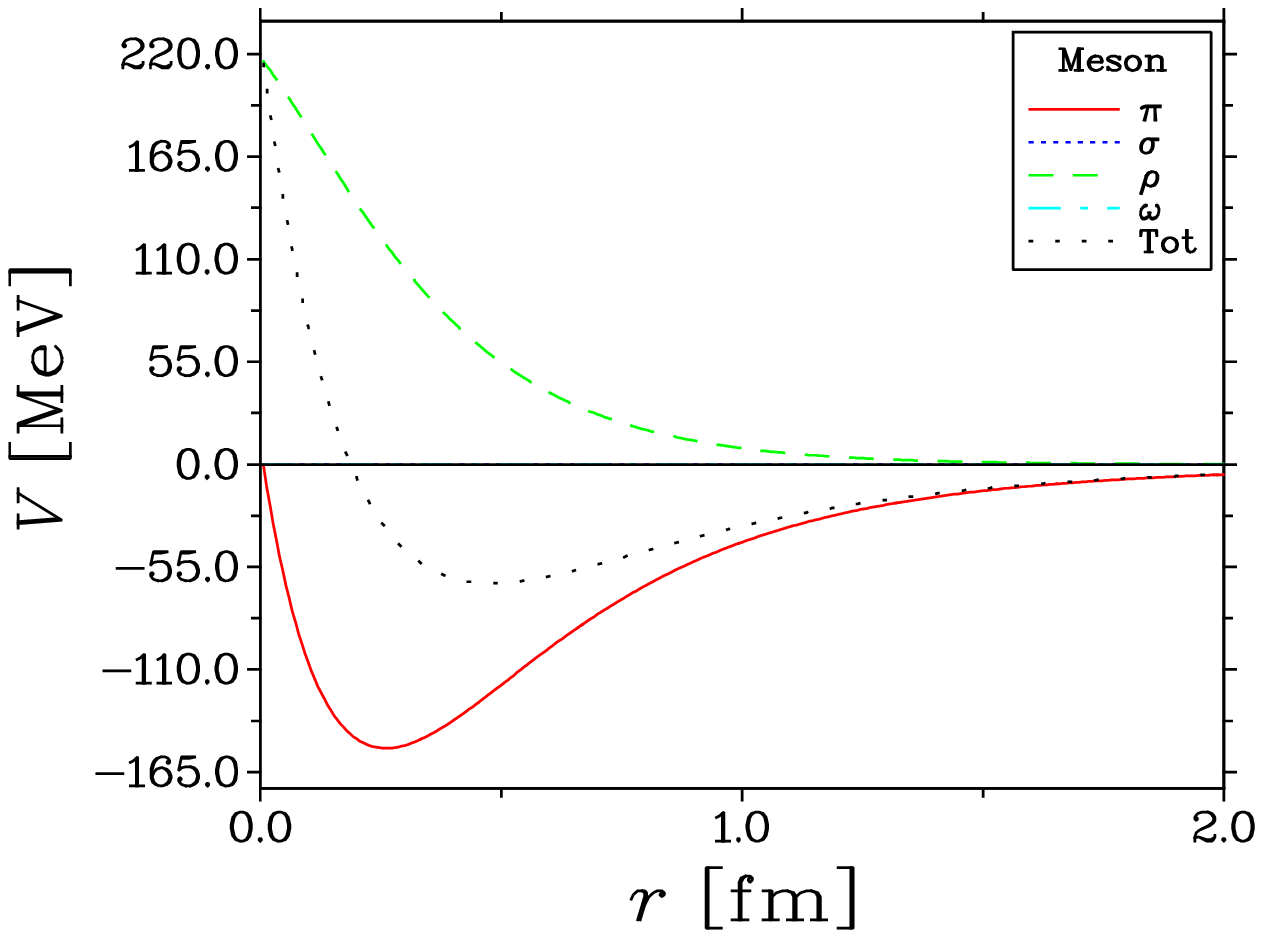}}&\scalebox{0.6}{\includegraphics{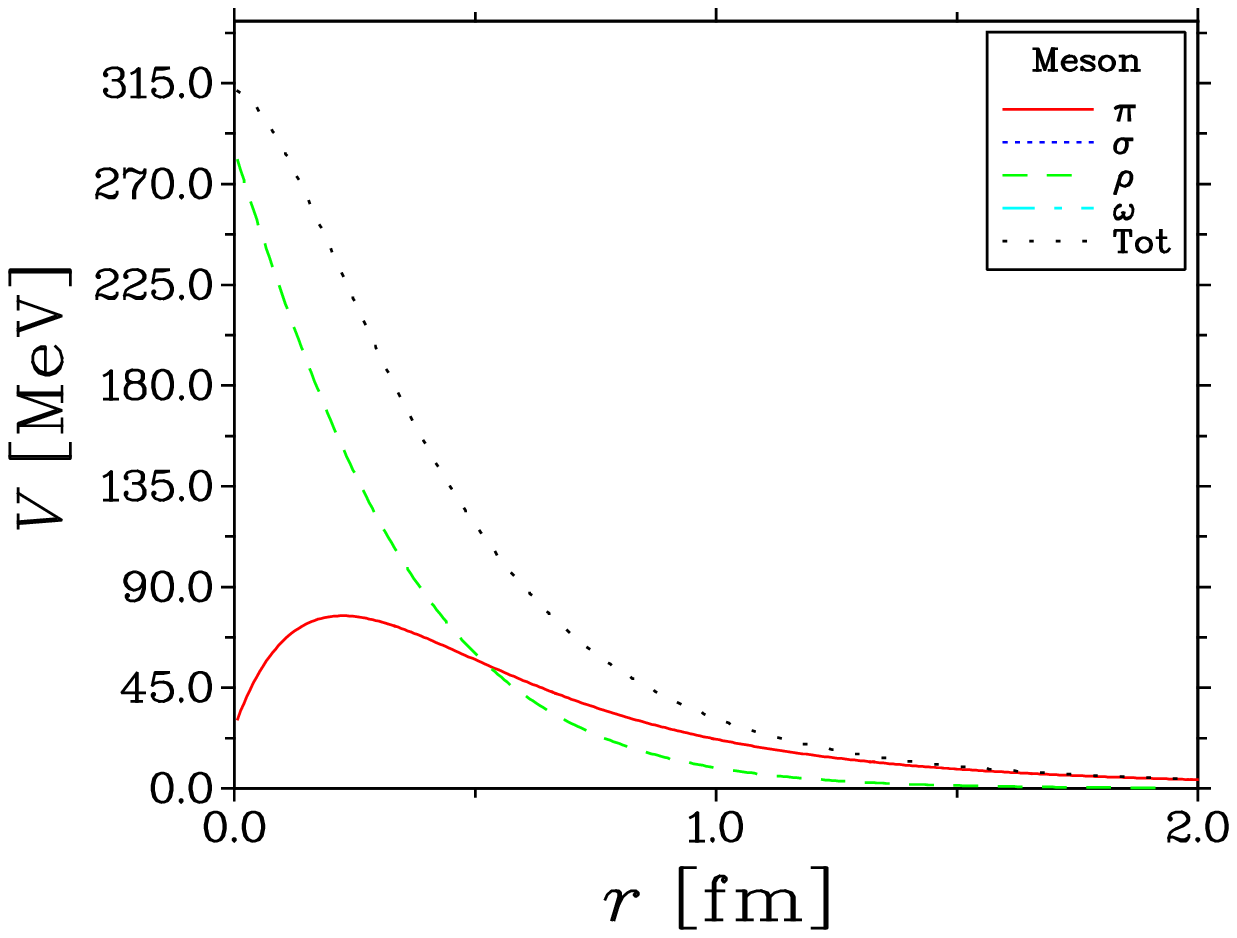}}\\
(45): $\Lambda_cN(^3D_1)\leftrightarrow\Sigma_cN(^3D_1)$&
(46): $\Lambda_cN(^3D_1)\leftrightarrow\Sigma_c^*N(^3D_1)$\\
\end{tabular}}
\caption{The potentials of different channels for the $J^P=1^+$ case with $\Lambda_\pi=\Lambda_\sigma=\Lambda_{\rm vec}=1$ GeV. (cont.)}\label{potential-J1}
\end{figure}
\addtocounter{figure}{-1}
\begin{figure}[htb]
\addtocounter{subfigure}{3}
\centering
\subfigure{\label{potential-J1-4th}
\begin{tabular}{cc}
\scalebox{0.6}{\includegraphics{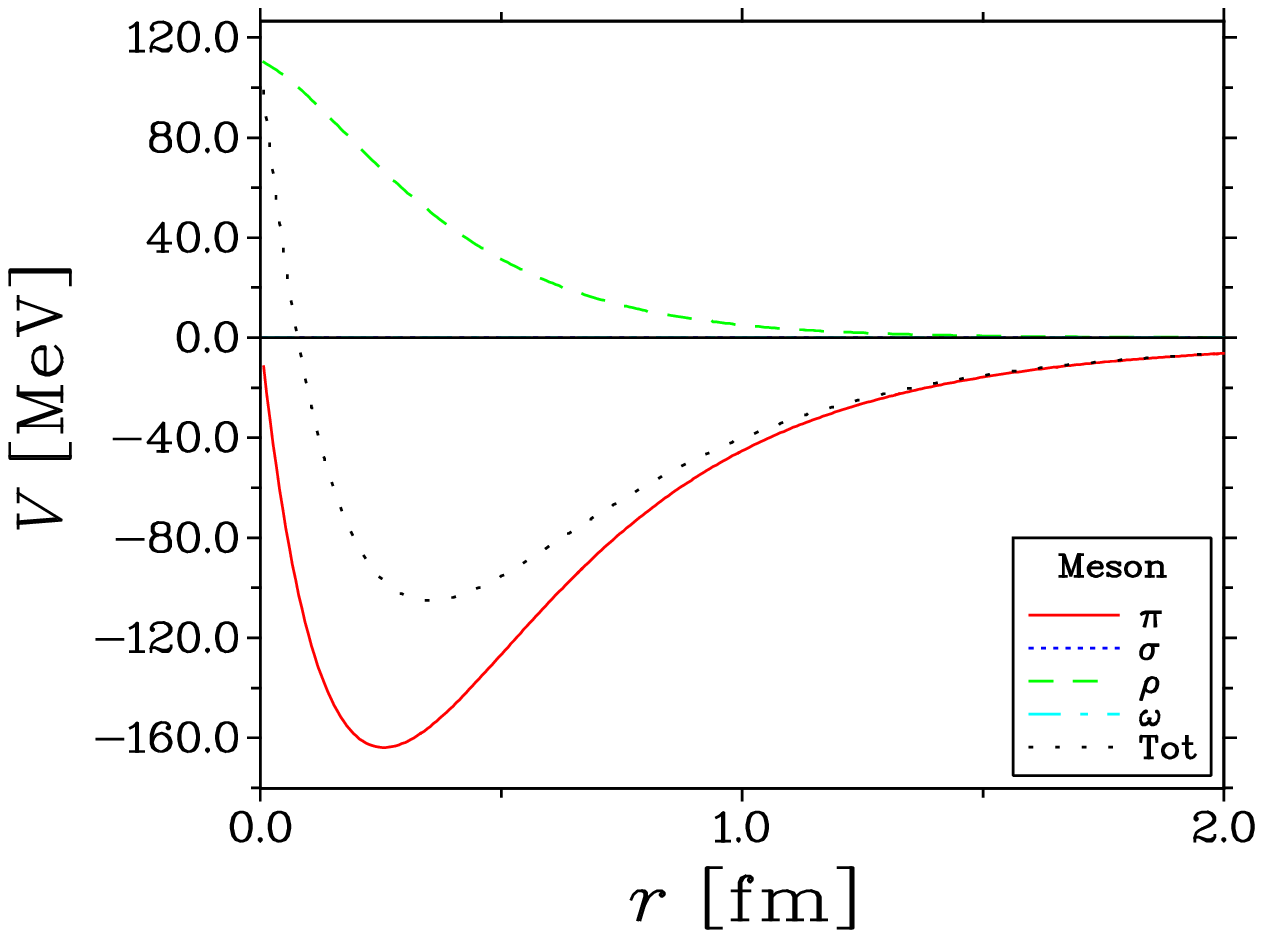}}&\scalebox{0.6}{\includegraphics{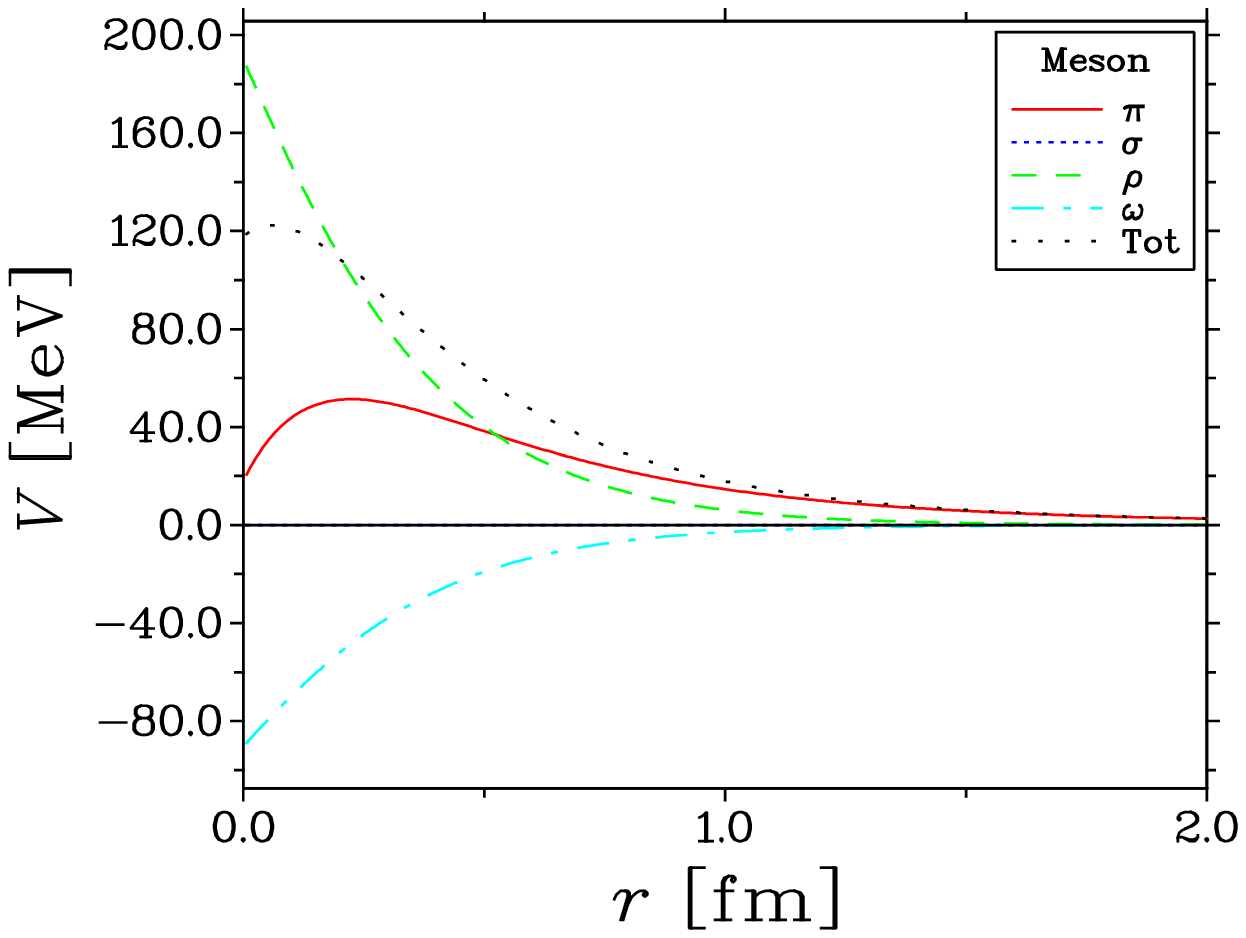}}\\
(47): $\Lambda_cN(^3D_1)\leftrightarrow\Sigma_c^*N(^5D_1)$&
(56): $\Sigma_cN(^3D_1)\leftrightarrow\Sigma_c^*N(^3D_1)$\\
\scalebox{0.6}{\includegraphics{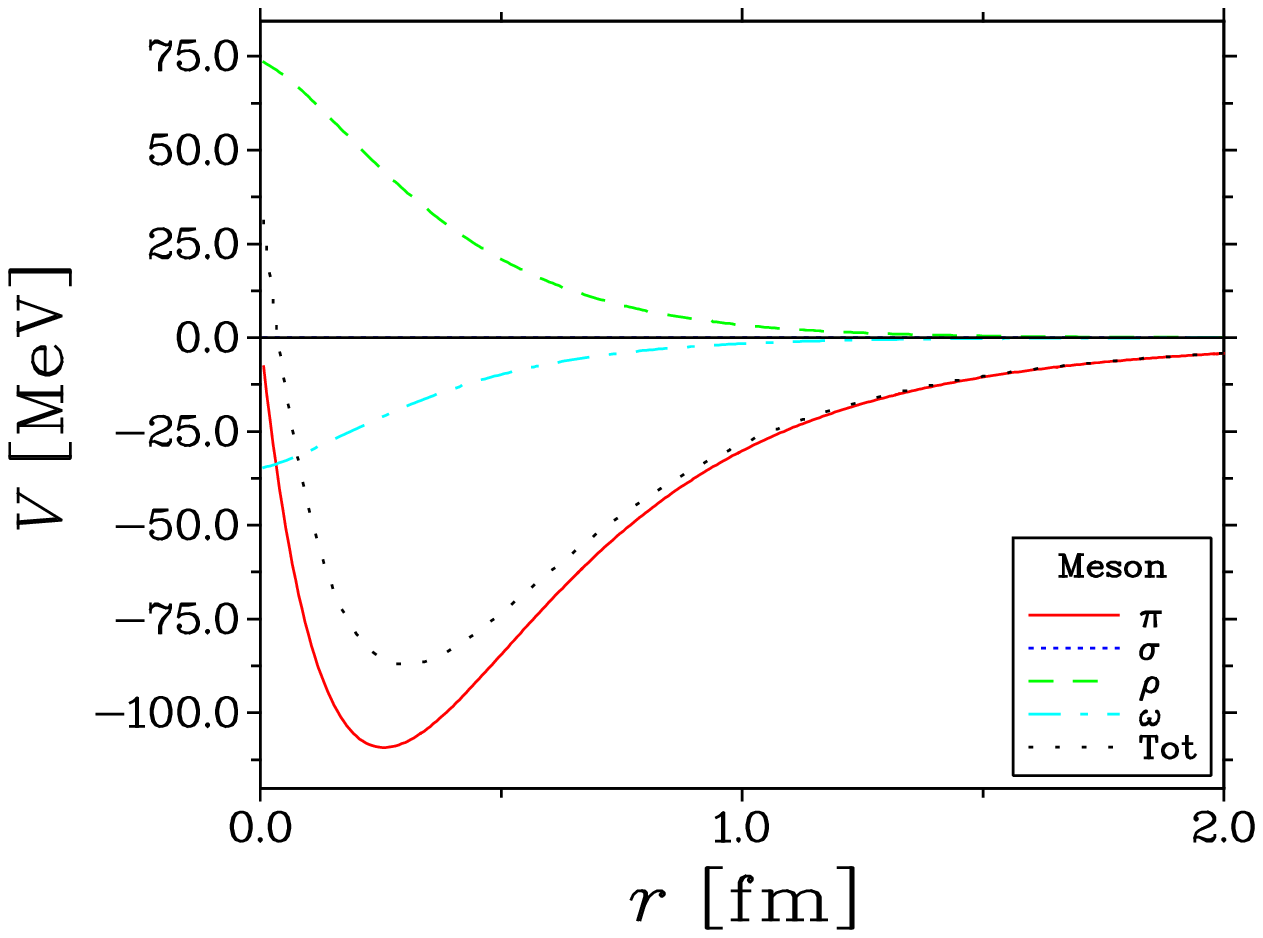}}&\scalebox{0.6}{\includegraphics{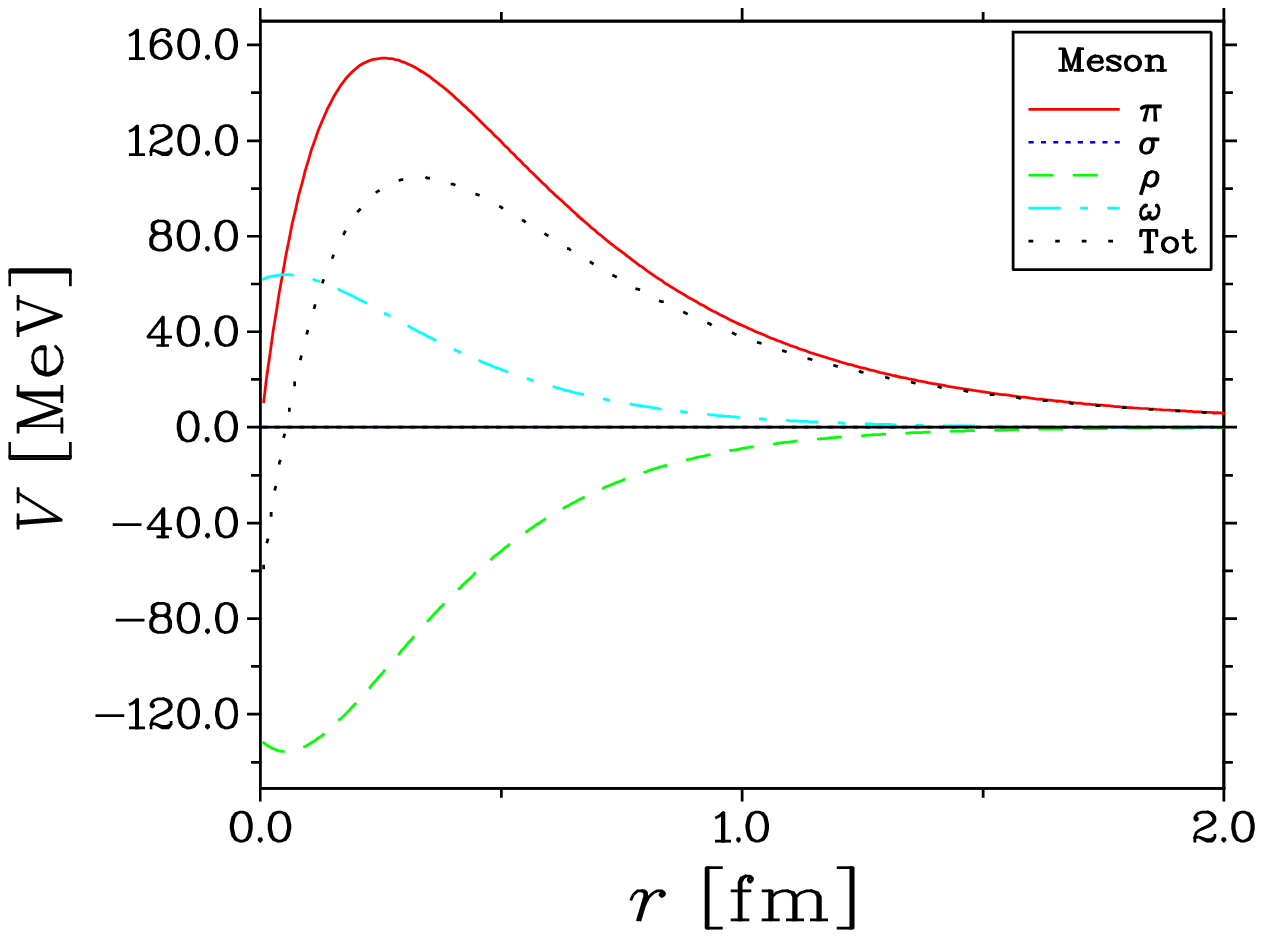}}\\
(57): $\Sigma_cN(^3D_1)\leftrightarrow\Sigma_c^*N(^5D_1)$&
(67): $\Sigma_c^*N(^3D_1)\leftrightarrow\Sigma_c^*N(^5D_1)$\\
\end{tabular}}
\caption{The potentials of different channels for the $J^P=1^+$ case with $\Lambda_\pi=\Lambda_\sigma=\Lambda_{\rm vec}=1$ GeV. (cont.)}\label{potential-J1}
\end{figure}

\subsection{OPEP model}

Now we discuss possibilities of bound states without channel coupling among $\Lambda_cN$, $\Sigma_cN$, and $\Sigma_c^*N$ in the OPEP model. For the $\Lambda_cN$ channel, there is no binding solution as before. For the $\Sigma_cN$ and $\Sigma_c^*N$ channels, S- and D- wave mixing is allowed. The former (latter) is a two-channel (three-channel) problem: $\Sigma_cN (^3S_1)$ and $\Sigma_cN (^3D_1)$ ($\Sigma_c^*N (^3S_1)$, $\Sigma_c^*N (^3D_1)$, and $\Sigma_c^*N (^5D_1)$). Numerical results are given in Table \ref{J1OpEP-noc}.

\begin{table}[htb]
\begin{tabular}{cc|ccccccccc}\hline
  &$\Lambda_\pi$ (GeV)&  1.0 & 1.1 & 1.2 & 1.3 &1.4 &1.5&1.6&1.7&1.8\\\hline
  & $B.E. (J=1)$ (MeV)& 0.08 &1.22&4.09&9.27&17.34&28.93&44.69&65.32&91.52\\
$\Sigma_cN$&$\sqrt{\langle r^2\rangle}$ (fm)&13.8&4.0 &2.4&1.7&1.4&1.1&1.0&0.8&0.7 \\
  & Prob. (\%)& (98.7/1.3)& (95.8/4.2)&(93.5/6.5)&(91.7/8.3)&(90.2/9.8)&(89.0/11.0)&(88.0/12.0)&(87.2/12.8)&(86.4/13.6)\\\hline
\end{tabular}
\begin{tabular}{cc|ccccccc}\hline
  &$\Lambda_\pi$ (GeV)&  1.6&1.7 &1.8 &1.9&2.0&2.1\\\hline
  &$B.E. (J=1)$ (MeV)&0.47&3.83&10.27&20.08&33.66&51.50\\
$\Sigma_c^*N$&$\sqrt{\langle r^2\rangle}$ (fm)&5.6&2.2&1.5&1.1&0.9&0.8\\
  & Prob. (\%)& (96.3/0.1/3.6)&(92.9/0.1/7.0)&(91.0/0.2/8.8)&(89.7/0.2/10.1)&(88.8/0.1/11.1)&(87.9/0.1/12.0)\\
 & $D$-wave prob.& 3.7\% & 7.1\% &9.0\% & 10.3\% & 11.2\% & 12.1\% \\\hline
\end{tabular}
\caption{Binding solutions for the individual channels in the $J^P=1^+$ case in the OPEP model. The binding energies (B.E.) are given relative to their own thresholds. The probabilities correspond to $^3S_1$ and $^3D_1$ for the $\Sigma_cN$ system, and $^3S_1$, $^3D_1$, and $^5D_1$ for the $\Sigma_c^*N$ system, respectively. We also present the total $D$-wave probability for $\Sigma_c^*N$.}\label{J1OpEP-noc}
\end{table}

\begin{table}[htb]
\begin{tabular}{c|ccccc}\hline
$\Lambda_\pi$ (GeV)&        1.2 & 1.3 & 1.4 &1.5&1.6\\\hline
$B.E. (J=1)$ (MeV)&1.17&7.52& 20.64&41.86&72.39\\
$\sqrt{\langle r^2\rangle}$ (fm)& 3.9 &1.8&1.2&0.9&0.8\\
Prob. (\%)&(97.6/0.2/0.4/&(93.7/0.6/1.3/&(89.3/1.4/2.5/&(84.6/2.7/3.7/& (79.6/4.6/4.9/\\
          & 0.0/1.0/0.1/0.7)      &0.0/2.5/0.2/1.7)     & 0.0/3.9/0.3/2.6)     & 0.1/5.3/0.4/3.2)     & 0.1/6.8/0.4/3.6)\\
$D$-wave prob. & 1.8\% & 4.4\% & 6.8\% & 9.0\% & 10.1\%          \\\hline
\end{tabular}
\caption{Binding solutions for the $J^P=1^+$ case with channel coupling in the OPEP model. The binding energies (B.E.) are given relative to the $\Lambda_cN$ threshold. The probabilities correspond to $\Lambda_cN(^3S_1)$, $\Sigma_cN(^3S_1)$, $\Sigma_c^*N(^3S_1)$, $\Lambda_cN(^3D_1)$, $\Sigma_cN(^3D_1)$, $\Sigma_c^*N(^3D_1)$, and $\Sigma_c^*N(^5D_1)$, respectively. We also present the total $D$-wave probability.}\label{J1OpEP}
\end{table}
After including the coupled channel effects, we obtain the results presented in Table \ref{J1OpEP}. Our calculation indicates again that the coupled channel effects are important in the OPEP model. One finds that the binding energies are slightly different between the singlet and triplet cases. The 4-th channel, $\Lambda_cN(^3D_1)$, has small contribution. Actually, if one ignores this channel, the resultant binding energy changes little. We show the wave functions of different channels with the cutoff $\Lambda_\pi=1.3$ GeV in Fig. \ref{WAVEu} (d).

\subsection{OBEP model}

In the case of the OBEP model, we again explore two cases for the parametrization of the cutoff parameters: (1) common cutoffs $\Lambda_\pi=\Lambda_\sigma=\Lambda_{\rm vec}=\Lambda_{\rm com}$, and (2) scaled cutoffs $\Lambda_{\rm ex}=m_{\rm ex}+\alpha\Lambda_{QCD}$.

\subsubsection{Common cutoff}

We first discuss the case without the channel coupling. For the $\Lambda_cN$ channel, the scalar meson and the vector meson exchanges do not lead to S- and D-wave mixing and thus one again gets the binding solutions in Table \ref{J0OmEP-noc}, i.e. the binding solution for the single channel $\Lambda_cN$ is spin independent. For the $\Sigma_cN$ and $\Sigma_c^*N$ channels, we present their results in Table \ref{J1OmEP-noc}.

\begin{table}[htb]
\begin{tabular}{cc|ccc}\hline
  &$\Lambda_{\rm com}$ (GeV)&  0.8&0.9&1.0 \\\hline
  & $B.E. (J=1)$ (MeV)& 2.84 &26.20&78.51\\
$\Sigma_cN$&$\sqrt{\langle r^2\rangle}$ (fm)&2.8&1.2 &0.8 \\
  & Prob. (\%)& (96.4/3.6)& (95.0/5.0)&(95.0/5.0)\\\hline
\end{tabular}
\begin{tabular}{cc|ccccc}\hline
  &$\Lambda_{\rm com}$ (GeV)&  0.9 &1.1 &1.3&1.5&1.7\\\hline
  &$B.E. (J=1)$ (MeV)&0.73&13.10&27.04&42.89&77.36\\
$\Sigma_c^*N$&$\sqrt{\langle r^2\rangle}$ (fm)&4.7&1.4&1.1&0.9&0.8\\
  & Prob. (\%)& (97.7/0.1/2.2)&(94.4/0.1/5.5)&(91.3/0.1/8.6)&(85.1/0.0/14.9)&(63.5/0.5/36.0)\\
  &$D$-wave prob. & 2.3\% & 5.6\% & 8.7\% & 14.9\% & 36.5\% \\\hline
\end{tabular}
\caption{Binding solutions for the individual channels in the $J^P=1^+$ and common-cutoff case in the OBEP model. The binding energies (B.E.) are given relative to their own thresholds. The probabilities correspond to $^3S_1$ and $^3D_1$ for the $\Sigma_cN$ system, and $^3S_1$, $^3D_1$, and $^5D_1$ for the $\Sigma_c^*N$ system, respectively. We also present the total $D$-wave probability for $\Sigma_c^*N$.}\label{J1OmEP-noc}
\end{table}

After considering the coupled channel effects, we get the results in Table \ref{J1OmEP}, which are consistent with the OPEP model calculation. An example of the resulting wave functions with $\Lambda_{\rm com}=0.9$ GeV is given in Fig. \ref{WAVEu} (e).

\begin{table}[htb]
\begin{tabular}{c|cccc}\hline
$\Lambda_{\rm com}$ (GeV)&        0.8 & 0.9& 1.0 &1.1 \\\hline
$B.E. (J=1)$ (MeV)&0.22&13.49& 47.50&106.16\\
$\sqrt{\langle r^2\rangle}$ (fm)&8.7 &1.5 &0.9&0.7 \\
Prob. (\%)&(99.6/0.0/0.1/&(96.6/0.3/0.9/&(90.6/2.2/3.7/&(79.1/9.9/6.4/\\
          & 0.0/0.2/0.0/0.1)&0.0/1.2/0.1/0.9)& 0.0/2.1/0.1/1.3) & 0.1/3.2/0.1/1.2)     \\
$D$-wave prob.& 0.3\% & 2.2\% & 3.5\% & 4.6\%\\\hline
\end{tabular}
\caption{Binding solutions for the $J^P=1^+$ and common-cutoff case with channel coupling in the OBEP model. The binding energies (B.E.) are given relative to the $\Lambda_cN$ threshold. The probabilities correspond to $\Lambda_cN(^3S_1)$, $\Sigma_cN(^3S_1)$, $\Sigma_c^*N(^3S_1)$, $\Lambda_cN(^3D_1)$, $\Sigma_cN(^3D_1)$, $\Sigma_c^*N(^3D_1)$, and $\Sigma_c^*N(^5D_1)$, respectively. We also present the total $D$-wave probability.}\label{J1OmEP}
\end{table}

\subsubsection{Scaled cutoffs}

\begin{table}[htb]
\begin{tabular}{cc|ccccccccc}\hline
  &$\alpha$&   0.9&1.1&1.3&1.5&1.7&1.9\\\hline
  & $B.E. (J=1)$ (MeV)&0.39&6.37&19.52&39.26&64.90&95.77\\
$\Sigma_cN$&$\sqrt{\langle r^2\rangle}$ (fm)&6.6&2.0&1.3&1.0&0.8&0.7 \\
  & Prob. (\%)& (99.8/0.2)&(99.7/0.3)&(99.8/0.2)&(99.8/0.2)&(99.9/0.1)&(99.9/0.1)\\\hline
\end{tabular}
\begin{tabular}{cc|cccccc}\hline
  &$\alpha$& 3.2& 3.3 & 3.4 & 3.5 & 3.6 & 3.7\\\hline
  &$B.E. (J=1)$ (MeV)&1.39&3.13&6.39& 12.41&22.74&38.38\\
$\Sigma_c^*N$&$\sqrt{\langle r^2\rangle}$ (fm)&3.4&2.4&1.7&1.3&0.9&0.7\\
  & Prob. (\%)& (98.0/0.1/1.9)&(96.4/0.1/3.5)&(92.8/0.2/7.0)&(85.3/0.2/14.5)&(73.7/0.2/26.1)&(61.1/0.3/38.7)\\
  & $D$-wave prob. & 2.0\% & 3.6\% & 7.2\% & 14.7\% & 26.3\% & 39.0\% \\\hline
\end{tabular}
\caption{Binding solutions for the individual channels in the $J^P=1^+$ and scaled-cutoff case in the OBEP model. The binding energies (B.E.) are given relative to their own thresholds. The probabilities correspond to $^3S_1$ and $^3D_1$ for the $\Sigma_cN$ system, and $^3S_1$, $^3D_1$, and $^5D_1$ for the $\Sigma_c^*N$ system, respectively. We also present the total $D$-wave probability for $\Sigma_c^*N$.}\label{J1OmEP-noc2}
\end{table}

We present the results without channel coupling in Table \ref{J1OmEP-noc2} and those with channel coupling in Table \ref{J1OmEP2}. If one compares Table \ref{J1OmEP2} with Table \ref{J1OmEP}, it is easy to see that the two parameterizations of the cutoffs may give consistent binding energies and radii. The wave functions for $\alpha=1.5$ in the coupled channel case are given in Fig. \ref{WAVEu} (f). As in the corresponding spin-singlet case, the wave functions also have nodes, which are due to the transition potentials. In fact, the nodes have appeared in the $D$ wave of the $\Sigma_c^*N$ channel (without the channel coupling to the $\Lambda_cN$ and $\Sigma_cN$).

\begin{table}[htb]
\begin{tabular}{c|ccccc}\hline
$\alpha$&     1.3 &1.5&1.7&1.9&2.1    \\\hline
$B.E. (J=1)$ (MeV)&0.66&6.63&19.82&40.98&70.27\\
$\sqrt{\langle r^2\rangle}$ (fm)&5.1&1.9&1.2&0.9&0.7  \\
Prob. (\%)&(99.3/0.2/0.5/&(96.7/1.0/2.2/&(92.5/2.8/4.5/&(86.9/5.8/6.8/&(80.5/10.2/8.6/\\
& 0.0/0.0/0.0/0.0)   &0.0/0.0/0.0/0.1)& 0.0/0.1/0.0/0.1)& 0.0/0.2/0.0/0.3)&0.0/0.3/0.0/0.4)\\
$D$-wave prob. & 0.0 & 0.1\% & 0.2\% & 0.5\% & 0.7\%\\\hline
\end{tabular}
\caption{Binding solutions for the $J^P=1^+$ and scaled-cutoff case with channel coupling in the OBEP model. The binding energies (B.E.) are given relative to the $\Lambda_cN$ threshold. The probabilities correspond to $\Lambda_cN(^3S_1)$, $\Sigma_cN(^3S_1)$, $\Sigma_c^*N(^3S_1)$, $\Lambda_cN(^3D_1)$, $\Sigma_cN(^3D_1)$, $\Sigma_c^*N(^3D_1)$, and $\Sigma_c^*N(^5D_1)$, respectively. We also present the total $D$-wave probability.}\label{J1OmEP2}
\end{table}

\section{Summary and conclusions}\label{sec7}

In studying the possible molecular bound states containing $\Lambda_cN$, we have constructed a potential model based on the effective Lagrangian reflecting the heavy quark symmetry, chiral symmetry, and hidden local symmetry. Solving the Schr\"odinger equation for the coupled $\Lambda_cN-\Sigma_cN-\Sigma_c^*N$ systems for $J=0$ and $J=1$ states, we obtain molecular bound states in both channels with appropriate cutoffs. By analyzing all the binding energies and the corresponding RMS radii, one observes several features: (1). the spin-singlet state and the spin-triplet state have slightly different binding energies for a given radius and the reasonable binding energy is at most tens of MeV; (2). the OPEP model and the OBEP model are somehow equivalent in getting the consistent binding energy and the corresponding RMS radius; (3). the coupled channel effects are important both in the OPEP model and in the OBEP model. To see these features, we make a comparison, which is presented in Table \ref{comparison}.

\begin{table}[htb]
\begin{tabular}{c|c|ccc|c}\hline
$J^P$&&$\Lambda_cN$ (S-wave)&$\Lambda_cN-\Sigma_cN-\Sigma_c^*N$\\\hline
$0^+$&OPEP ($\Lambda$)&$\times$             &[1.367: 13.60, 1.38]\\
     &OBEP ($\Lambda$)&[0.900: 1.24, 3.86]&[0.900: 13.60, 1.46]\\
     &OBEP ($\alpha$) &[1.533: 0.25, 8.13]&[1.533: 13.57, 1.37]\\
     \hline
$1^+$&OPEP ($\Lambda$)&$\times$             &[1.353: 13.54, 1.40]\\
     &OBEP ($\Lambda$)&[0.900: 1.24, 3.86]&[0.900: 13.49, 1.47]\\
     &OBEP ($\alpha$) &[1.618: 0.80, 4.72]&[1.618: 13.47, 1.39]\\
     \hline
\end{tabular}
\caption{Comparison between different cases. The meaning of the numbers are [cutoff $\Lambda$ in GeV or dimensionless $\alpha$: binding energy in MeV, RMS radius in fm].}\label{comparison}
\end{table}

The deuteron, the well-known $pn$ bound state with the binding energy of 2.2 MeV, is also a spin-triplet state. The $D$-wave contribution is very important and the corresponding probability of the state at $D$-wave is around 4$\sim$6\%. If the $D$-wave probability of $\Lambda_cN$ were in this range, the binding energy would be around l0 MeV in the OPEP model from Table \ref{J1OpEP}.

For the $\Lambda_cN$ single channel, only $\sigma$ and $\omega$ exchanges are allowed. There are no tensor force nor spin-dependent parts in the potentials. As a result, the coupling between the $^3S_1$ and $^3D_1$ states vanishes and one has $V_{^1S_0}=V_{^3S_1}$ for the potentials. That is, the possible $J^P=0^+$ and $J^P=1^+$ molecular states are degenerate in the heavy quark limit. After considering the contributions from the channels $\Sigma_cN$ and $\Sigma_c^*N$, the tensor force and the spin-dependent interactions enter. However, full coupled channel calculation does not give significantly different results for the two spin states. That is, the two spin states are still qualitatively degenerate. Maybe this feature indicates that the coupled channel effects do not change the degeneracy in the dominant channel. If this is the case, one expects that the states with different angular momenta of a two-body system like $\Lambda_c\Sigma_c$, $\Lambda_c\Lambda$, $\Lambda_c\Sigma$, or $\Lambda_c\Delta$ would have similar binding solutions. They may be tested in the future investigations of bound state problems and scattering problems.

The results are sensitive to the phenomenological cutoff parameter, which is a general feature for the molecule study in the meson exchange models at the hadron level. In effect, this parameter encodes the size effects of the hadrons and the information of the short-range interaction. This feature indicates that the binding energy is sensitive to the short-range interaction. At this moment, we cannot determine the values of the cutoff. So we treat it as a free parameter and discuss the results for a reasonable range of the cutoff. If the cutoff parameter around $1.2\sim1.4$ GeV in the OPEP model, $0.8\sim1.0$ GeV in the OBEP model (common cutoff case), or $1.2\sim1.7$ in the OBEP model (scaled cutoff case) is reasonable, then one obtains moleculelike bound states. Larger parameters may be beyond present models while smaller ones would result in unbound states.

\begin{figure}
\centering
\begin{tabular}{cc}
\scalebox{0.48}{\includegraphics{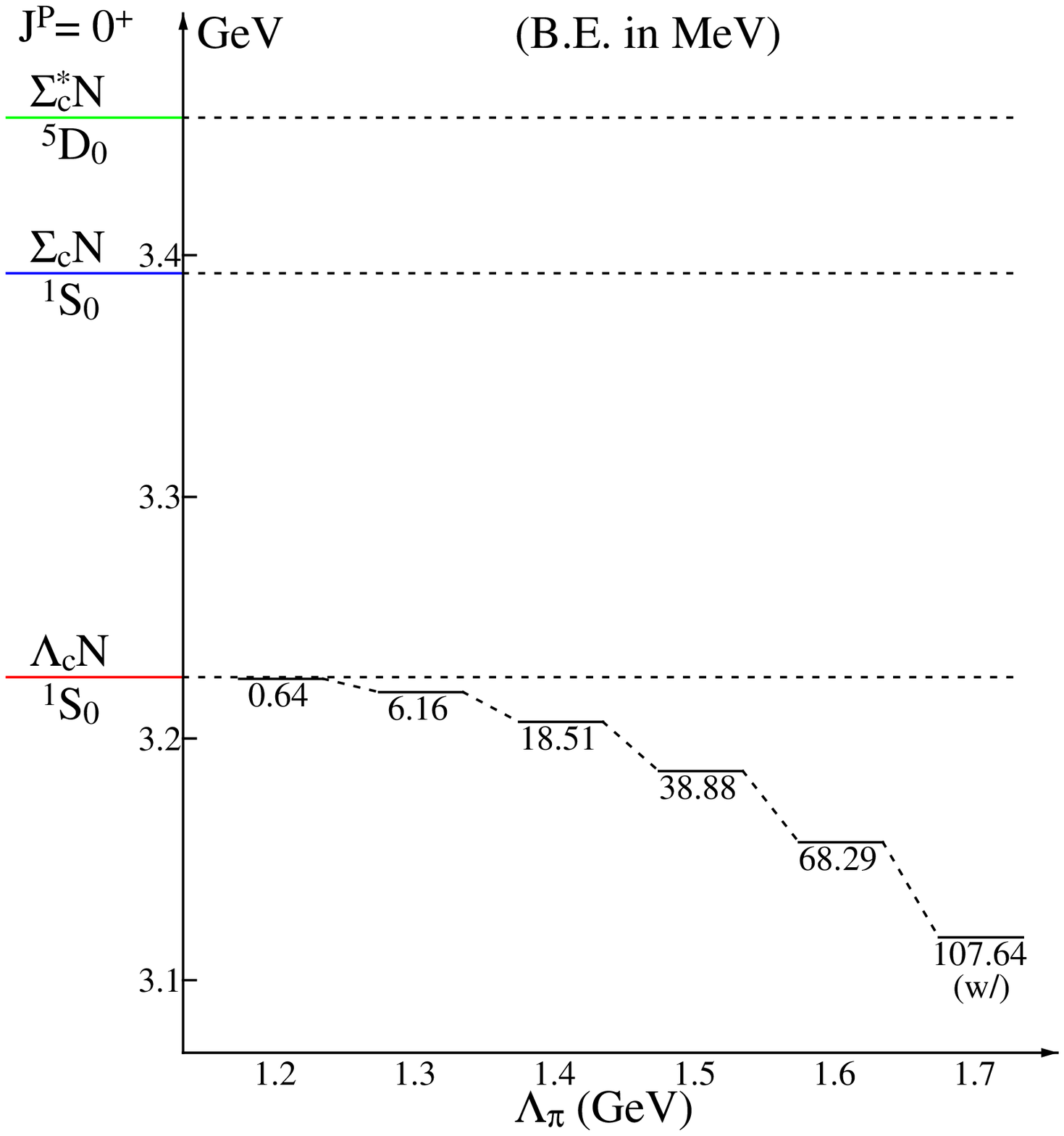}}&\scalebox{0.48}{\includegraphics{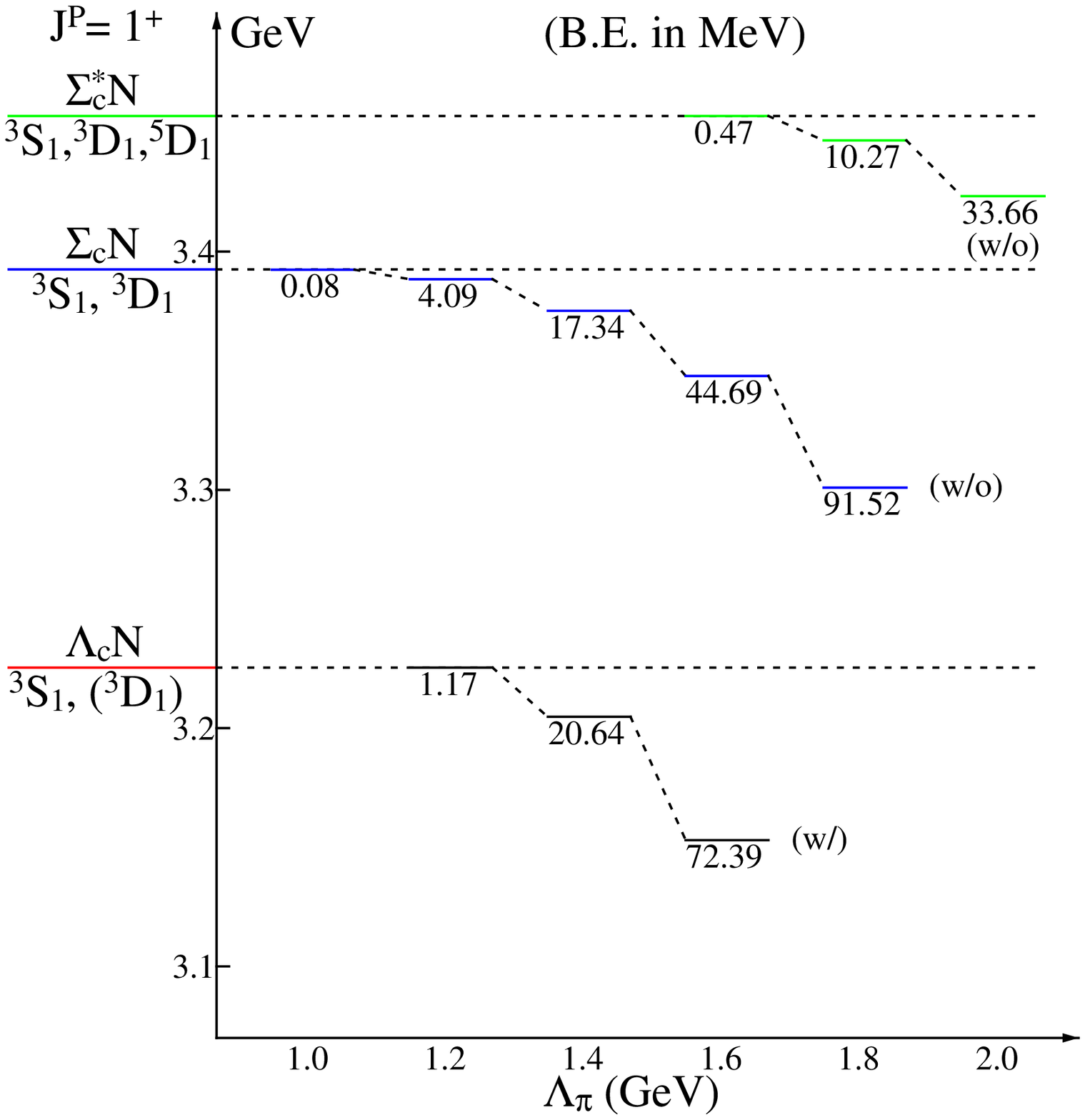}}\\
$J^P=0^+$&$J^P=1^+$
\end{tabular}
\caption{The sensitivity of the binding energy (B.E.) to the cutoff $\Lambda_\pi$ in the OPEP model for the $J^P=0^+$ and $J^P=1^+$ cases. The cases without (w/o) and with (w/) channel coupling are both shown. ($^3D_1$) means there is no $S-D$ mixing when one considers only the $\Lambda_cN$ channel. }\label{BEJ-opep}
\end{figure}

\begin{figure}
\centering
\begin{tabular}{cc}
\scalebox{0.48}{\includegraphics{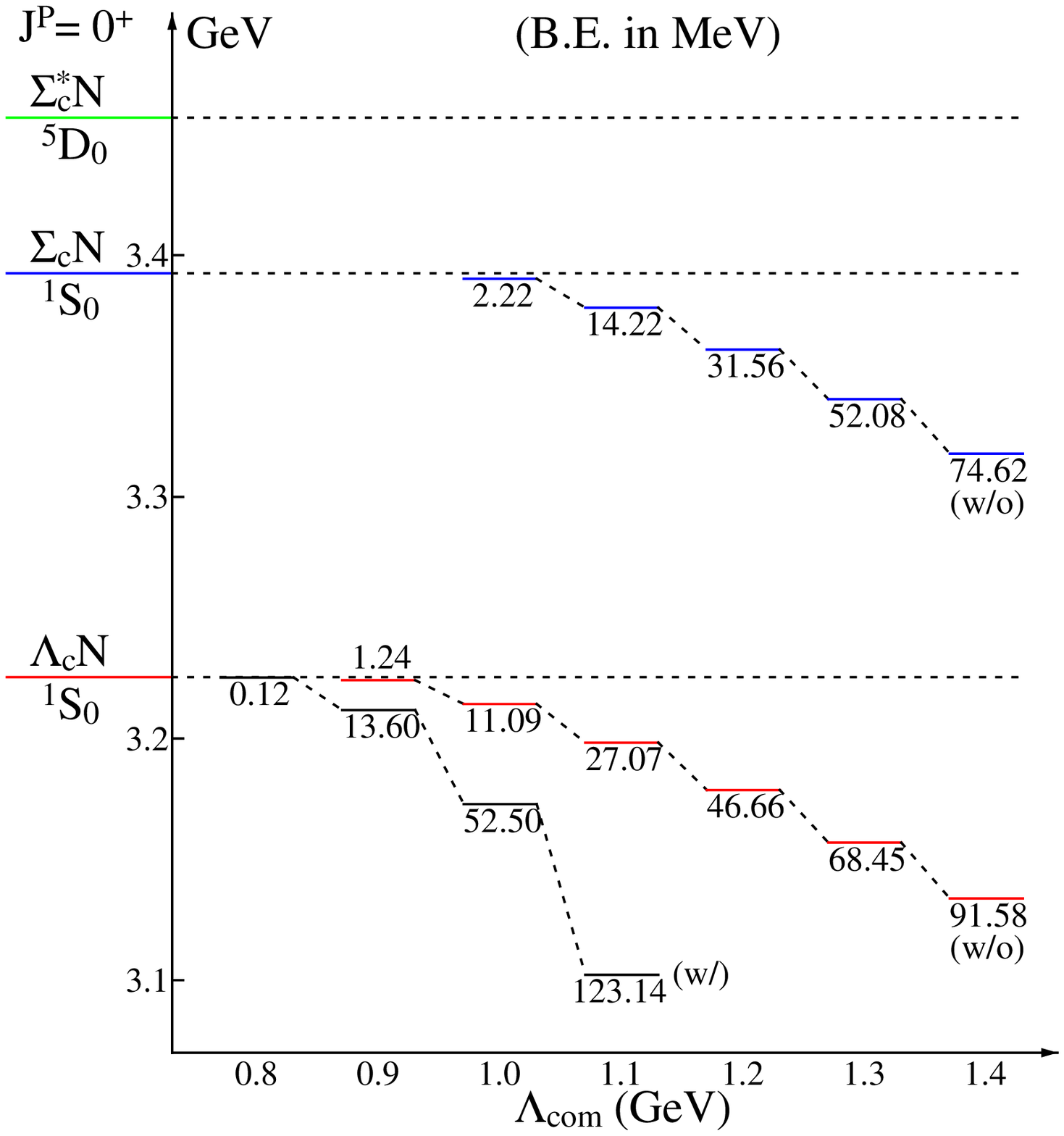}}&\scalebox{0.48}{\includegraphics{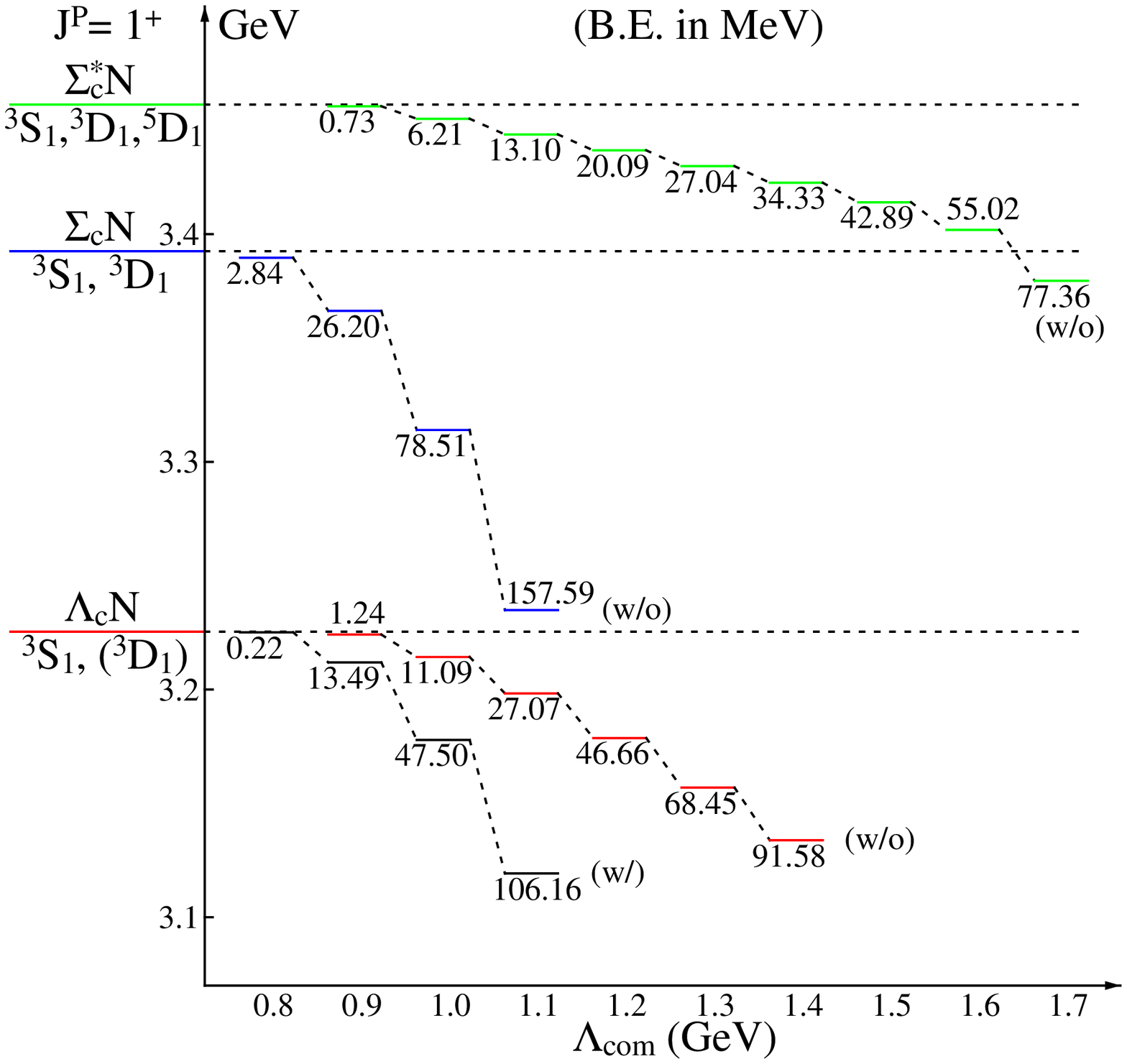}}\\
$J^P=0^+$&$J^P=1^+$
\end{tabular}
\caption{The sensitivity of the binding energy (B.E.) to the cutoff $\Lambda_\pi=\Lambda_\sigma=\Lambda_{\rm vec}=\Lambda_{\rm com}$ in the OBEP model for the $J^P=0^+$ and $J^P=1^+$ cases. The cases without (w/o) and with (w/) channel coupling are both shown. ($^3D_1$) means there is no $S-D$ mixing when one considers only the $\Lambda_cN$ channel.}\label{BEJ-omepL}
\end{figure}

\begin{figure}
\centering
\begin{tabular}{cc}
\scalebox{0.46}{\includegraphics{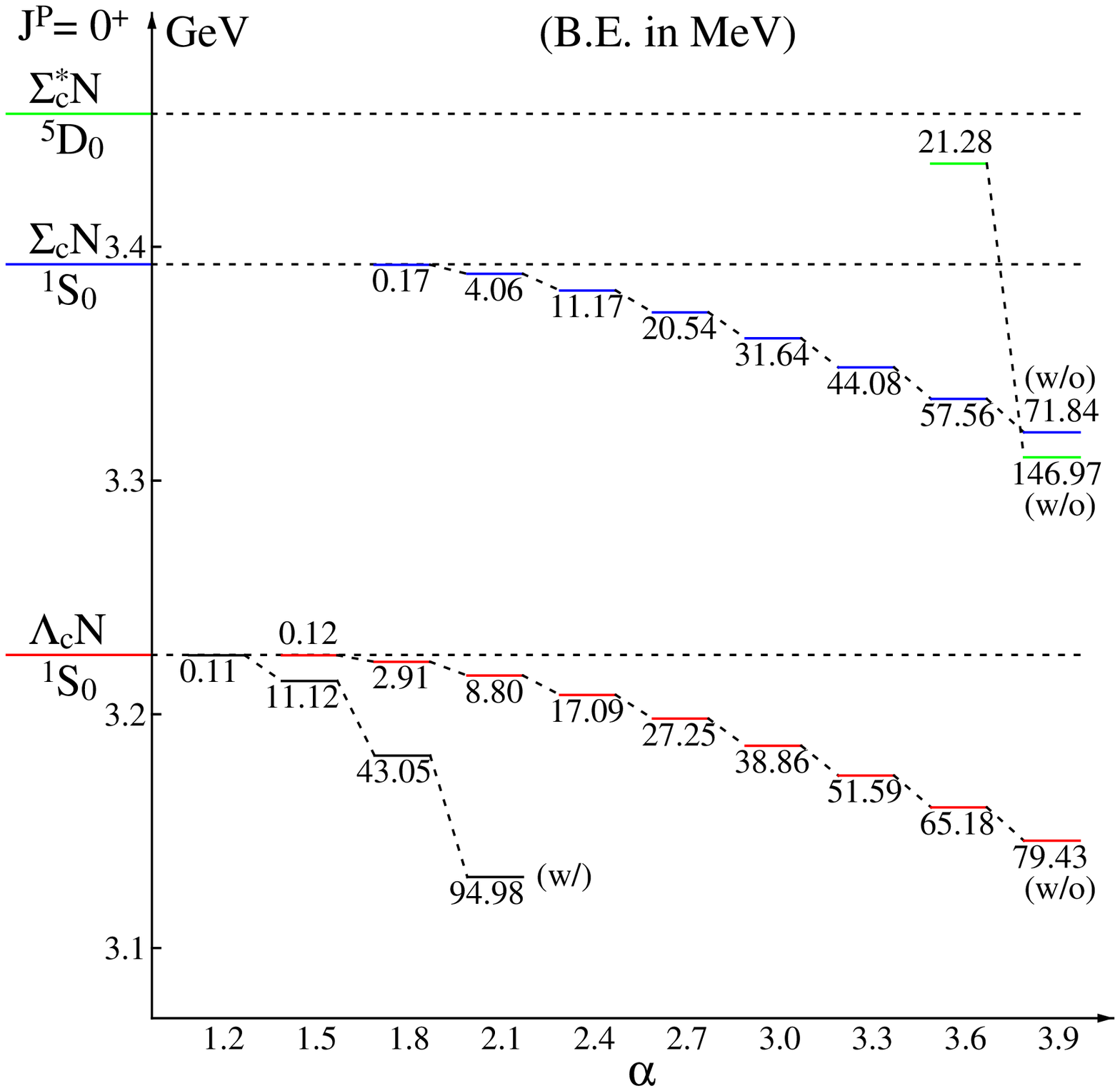}}&\scalebox{0.46}{\includegraphics{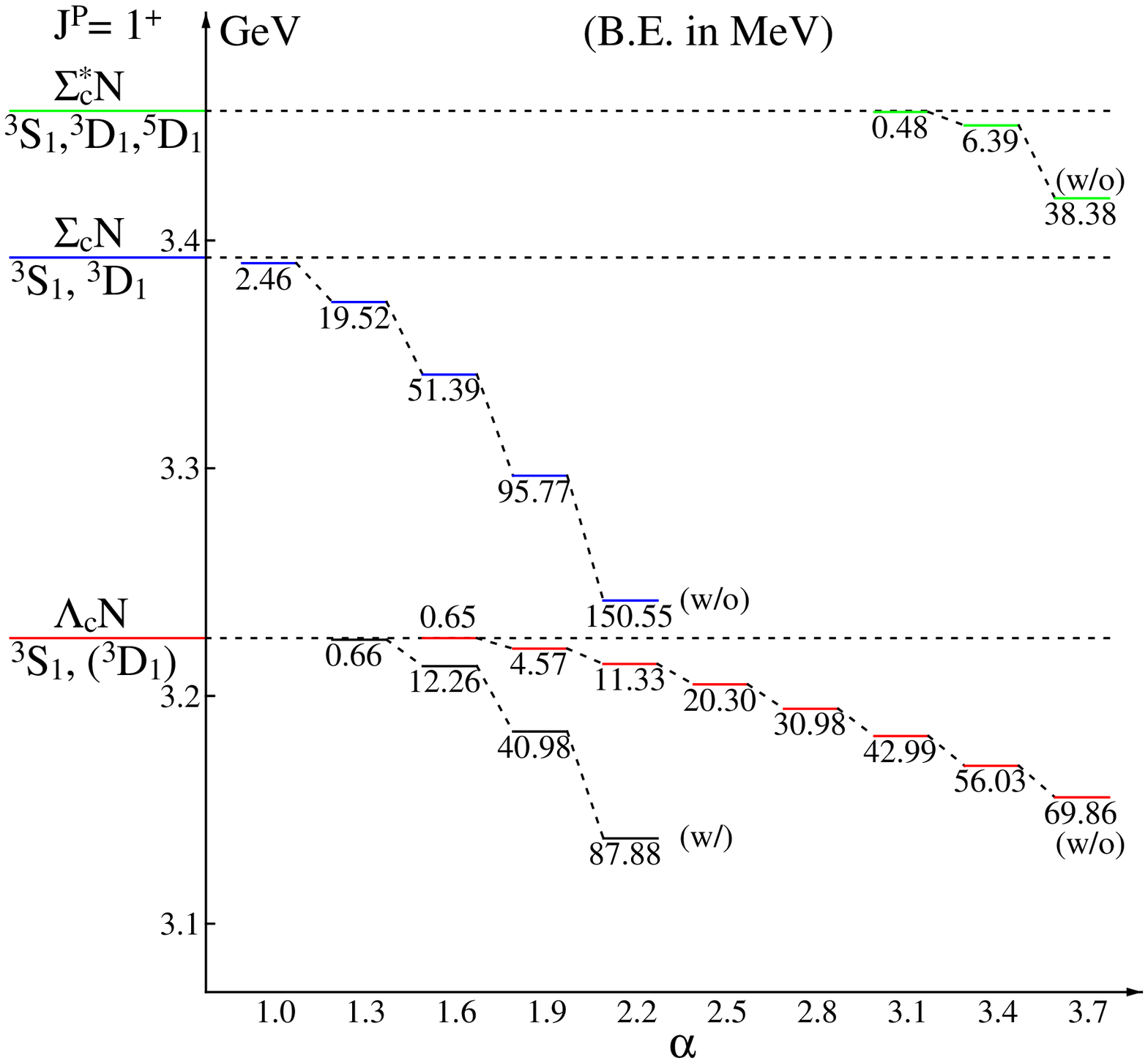}}\\
$J^P=0^+$&$J^P=1^+$
\end{tabular}
\caption{The sensitivity of the binding energy (B.E.) to the parameter $\alpha$ in the OBEP model for the $J^P=0^+$ and $J^P=1^+$ cases. The cases without (w/o) and with (w/) channel coupling are both shown. ($^3D_1$) means there is no $S-D$ mixing when one considers only the $\Lambda_cN$ channel.}\label{BEJ-omepalf}
\end{figure}

To see the sensitivity of the binding energy to the cutoff parameter $\Lambda$ or $\alpha$, we show the results in a diagrammatic form. In Fig. \ref{BEJ-opep}, we present the dependence of the binding energy on $\Lambda_\pi$ for the $J^P=0^+$ and $J^P=1^+$ cases in the OPEP model. The dependence of the binding energy on the common cutoff $\Lambda_{\rm com}$ is illustrated in Fig. \ref{BEJ-omepL}. That on the parameter $\alpha$ is shown in Fig. \ref{BEJ-omepalf}. The binding energies in these figures are not all given in the former tables. Here, we have included some additional binding energies whose corresponding RMS radii are smaller than 0.7 fm. From these figures, it is obvious that the coupling among the channels $\Lambda_cN$, $\Sigma_cN$, and $\Sigma_c^*N$ may result in bound states even if there are no binding solutions within individual channels.

In the OPEP model, the attraction for the formation of the $\Lambda_cN$ bound states comes only through the $\Sigma_cN$ and $\Sigma_c^*N$ channels indirectly. So the coupled channel effects are essential. The same observation is made in the $\Lambda_c\Lambda_c$ case \cite{Meguro2011}. In the OBEP model, the single channel $\Lambda_cN$ itself may have binding solutions. Then the effects of the channel coupling make the binding energy larger with a significant value.

In the above discussions, we do not consider the uncertainty of the coupling constants in Eq. (31). To investigate the sensitivity of the results to them, we vary separately $g_2$, $g_4$, $\ell_B$, $(\beta_Bg_V)$, $(\beta_Sg_V)$, $(\lambda_Sg_V)$, $h_\sigma$, $h_V$, and $h_T$ and compare the binding energies. As a well-known value, $g_A$ is not varied. We still adopt the relations in Eqs. (31) in determining other coupling constants. Table \ref{sen-coup} illustrates the sensitivity when we use the common cutoff parameter $\Lambda_{com}=0.9$ GeV. Larger values of the coupling constants (except $(\beta_Bg_V)$ and $h_V$) may result in deeper bound states. From the numerical results, one finds that the results are sensitive to $g_4$, $\ell_B$, and $h_\sigma$. If one uses the set of coupling constants $g_2=-0.5$, $g_4=0.7$, $\ell_B=-2.0$, $(\beta_Bg_V)=-7.0$, $(\lambda_Sg_V)=13.0$, $h_\sigma=8.0$, $h_V=4.0$, and $h_T=5.0$, one may also get binding solutions with a slightly larger cutoff $\Lambda_{com}=1.1$ GeV. Since this set is not helpful to the formation of bound states, the main conclusion that the molecular bound states are possible does not change.

\begin{table}
\begin{tabular}{cc|cc|cc|cc}\hline
$g_2$ &B.E.&$g_4$&B.E.&$\ell_B$&B.E.&$(\beta_Bg_V)$&B.E.\\
-0.50&13.17, 11.36&0.7&6.93, 9.54&-2.0&2.09, 4.15&-3.0&14.75, 14.67\\
-0.55&13.38, 12.38&0.8&8.77, 10.67&-2.5&6.24, 9.15&-4.0&14.36, 14.27\\
-0.60&13.61, 13.54&0.9&10.99, 11.99&-3.0&12.21, 15.75&-5.0&13.98, 13.88\\
-0.65&13.83, 14.84&1.0&13.62, 13.51&-3.5&19.75, 23.76&-6.0&13.60, 13.49\\
-0.70&14.07, 16.29&1.1&16.68, 15.26&-4.0&28.68, 33.03&-7.0&13.22, 13.11\\\hline\hline

$(\lambda_Sg_V)$&B.E.&$h_\sigma$&B.E.&$h_V$&B.E.&$h_T$&B.E.\\
13.0&13.03, 13.21& 8.0& 4.04, 4.10&2.0&13.92, 13.89&5.0&13.48, 13.53\\
15.0&13.20, 13.29&10.0&10.00, 9.96&2.5&13.76, 13.69&6.0&13.56, 13.50\\
17.0&13.39, 13.39&12.0&18.08, 17.90&3.0&13.60, 13.49&7.0&13.66, 13.48\\
19.0&13.58, 13.48&14.0&27.98, 27.63&3.5&13.44, 13.30&8.0&13.77, 13.47\\
21.0&13.78, 13.59&16.0&39.46, 38.93&4.0&13.28, 13.10&9.0&13.89, 13.47\\\hline
\end{tabular}
\caption{$J=0$ and $J=1$ binding energies (MeV) corresponding to the variation of only one coupling constant. The unchanged coupling constants are the same as in Eq. (31). The units of $(\lambda_Sg_V)$ and $h_T$ are GeV$^{-1}$. The common cutoff $\Lambda_{com}=0.9$ GeV is used.}\label{sen-coup}
\end{table}

Theoretically, the formation of these $J^P=0^+,1^+$ $\Lambda_cN$ bound states occurs after the production of $\Lambda_c$ and through the coalescence mechanism, similar to that of X(3872) in the molecular picture \cite{Braaten2004}. Once the molecules are formed, they would be rather stable because they do not decay through strong interactions. Possible places to search for them are GSI-FAIR, J-PARC, and RHIC. It is also possible to look for them at BELLE.

In summary, we have obtained effective Lagrangians describing the charmed baryon interactions with light mesons according to the heavy quark symmetry, chiral symmetry, and hidden local symmetry. We estimate the coupling constants with various methods and get consistent results. Based on these interaction Lagrangians, one-boson exchange potentials for the $\Lambda_cN-\Sigma_cN-\Sigma_c^*N$ systems are derived. With the cutoffs around 1 GeV, it is possible to get binding solutions by solving the bound state problem with the variational method. One finds that the channel coupling has very important effects for the possible molecular bound states. These effects do not significantly change the feature that the binding solutions are spin-independent. If these states exist, the reasonable binding energies should be at most tens of MeV.

\section*{Acknowledgments}

This project was supported by the
Japan Society for the Promotion of Science under Contract No.
P09027; KAKENHI under Contract Nos. 19540275, 20540281, 22105503, and 21$\cdot$09027.

\appendix

\section{Phase convention in the superfield $S_\mu$}\label{app1}

If one reduces Eq. (3.12) in Ref. \cite{Yan1992} to the relevant terms in Eq. (\ref{totLag}) in this paper with the superfield $S_\mu$ defined in Eq. (\ref{superfield}), one gets
\begin{eqnarray}
g_3=-\frac{\sqrt3}{2}\delta g_1,\quad g_5=-\frac32 g_1, \quad g_4=\sqrt3 \delta g_2, \quad g_6=0.
\end{eqnarray}
To determine the phase $\delta$, we resort to the quark model and the heavy quark symmetry. The spin wave functions of $\Sigma_c$ and $\Sigma_c^*$ ($S_z=\frac12$) are
\begin{eqnarray}
\Sigma^*(\uparrow)&\equiv&\sqrt{\frac13}(11)\downarrow+\sqrt{\frac23}(10)\uparrow,\nonumber\\
\Sigma(\uparrow)&\equiv&\epsilon\left[\sqrt{\frac23}(11)\downarrow-\sqrt{\frac13}(10)\uparrow\right],
\end{eqnarray}
where $\epsilon$ is an arbitrary relative phase, $\downarrow$ or $\uparrow$ is the heavy quark spin, and $(11)$ or $(10)$ represents the diquark spin and its $z$ component. Let $\sigma_h^z$ denotes the spin operator acting only on the heavy quark, and
\begin{eqnarray}
\sigma_h^z\Sigma^*(\uparrow)&=&a\Sigma^*(\uparrow)+b\Sigma(\uparrow),\nonumber\\
\sigma_h^z\Sigma(\uparrow)&=&c\Sigma^*(\uparrow)+d\Sigma(\uparrow).
\end{eqnarray}
One gets $a=\frac13$, $b=-\frac{4}{3\sqrt2}\epsilon^*$, $c=-\frac{4}{3\sqrt2}\epsilon$, and $d=-\frac13$. Now we demand the superfield $S_\mu$ does not change under the heavy quark transformation. In the static limit, the superfield $S_\mu$ ($\mu=3$) is
\begin{eqnarray}
S_\mu&=&S_{t\mu}\Phi+\delta\frac{1}{\sqrt3}(\gamma_\mu+v_\mu)\gamma^5S_t\chi\to S_{t3}\Phi+\delta\frac{1}{\sqrt3}\sigma_3S_t\chi\to \frac{1}{\sqrt3}(\sqrt2+\delta)\left(\begin{array}{c}1\\0\end{array}\right),
\end{eqnarray}
where $S_{t\mu}$ and $S_t$ are transition spin operators of Rarita-Schwinger field and Dirac field, respectively, $\Phi=(0,1,0,0)^T$ is the wave function of $\Sigma_c^*$, and $\chi=(0,1,0,0)^T$ is the wave function of $\Sigma_c$. With $\sigma_h^zS_3=S_3$, one may get $(\delta\epsilon)=-1$. A natural choice is $\epsilon=1$ and then $\delta=-1$. Therefore, we have
\begin{eqnarray}
&S_\mu=B_{6\mu}^*-\frac{1}{\sqrt3}(\gamma_\mu+v_\mu)\gamma^5 B_6,&\\
&g_3=\frac{\sqrt3}{2} g_1,\quad g_5=-\frac32 g_1, \quad g_4=-\sqrt3 g_2, \quad g_6=0.&
\end{eqnarray}
The relations among the coupling constants have been presented in Ref. \cite{Yan1992}.

\end{document}